%% file: 2021_MicrostructuralVariability.tex
\documentclass[
               DIV=15, font size=10.5pt,
                 onecolumn,
               ]{scrartcl}  

\input{template/packages}
\input{template/pgfplotsettings}

\input{template/commonCommands}
\input{template/titleAndAuthors}
\input{template/journalFooter}

\crefname{figure}{Fig.}{Fig.}
\crefname{equation}{Eq.}{Eq.}
\crefname{table}{Tab.}{Tab.}

\drawboundingboxtrue
\drawboundingboxfalse 

\newcommand*{\figref}[2][]{%
	\hyperref[{fig:#2}]{%
		Fig.~\ref*{fig:#2}%
		\ifx\\#1\\%
		\else
		\,#1%
		\fi
	}%
}
\definecolor{changes}{RGB}{0,0,0}
\definecolor{changez}{RGB}{0,0,0}

\begin{document}  

\normalem
\maketitle  
  
\vspace{-1.5cm} 
\hrule 
\input{template/abstract}
 \vspace{.2cm} 
\vspace{0.25cm}\\
\noindent \textit{Keywords:} \input{template/keywords} 
\vspace{0.35cm}
\hrule 
\vspace{0.15cm}
\captionsetup[figure]{labelfont={bf},name={Fig.},labelsep=colon}
\captionsetup[table]{labelfont={bf},name={Tab.},labelsep=colon}
\tableofcontents
\vspace{0.5cm}
\hrule 
	\input{sections/introduction/introduction}
	\input{sections/model/model}
	\input{sections/parameters/parameters}
	\input{sections/realization/realization}
	\input{sections/MLMC/MLMC}
	\input{sections/numericalExperiments/numericalExperiments}
	\input{sections/conclusion/conclusion}
	\begin{appendices}
	\addtocontents{toc}{\protect\setcounter{tocdepth}{0}}
	\crefalias{section}{appendix}
	\input{sections/appendix}
    \end{appendices}
\section*{Acknowledgements} 
\input{template/acknowledgements}
\newpage
\bibliographystyle{apalike}
 \bibliography{library}

\end{document}

%% file: template/packages.tex
\usepackage[dvipsnames]{xcolor}
\usepackage{psfrag} 
\usepackage{subfig}
\usepackage{hyperref}
\usepackage{amsmath,amssymb}
\usepackage{graphics}
\usepackage{mathtools}
 \usepackage{pstool} 
\usepackage{todonotes}
\usepackage{gensymb}
\usepackage{bm}
\usepackage{multirow}
\usepackage{cancel}
\usepackage{float}
\usepackage{textcomp}
          
\usepackage[toc,page]{appendix}
\usepackage[square, numbers,sort&compress]{natbib}
 
\usepackage{pgfplots}
\usepackage{pgfplotstable}
\usepgfplotslibrary{statistics}
\usepgfplotslibrary{groupplots}
\usepackage{filecontents}
\usepackage{algorithm2e}
\usepackage{algorithmic}
\usepackage{siunitx}
\usepackage{tikz}
\usepackage{tikzscale}
\usepackage{tabularx} 

\usetikzlibrary{spy}
\usetikzlibrary{snakes,arrows,shapes}    
\usetikzlibrary{spy}
\usetikzlibrary{backgrounds}  
\usetikzlibrary{decorations}
\usetikzlibrary{positioning}
\usetikzlibrary{patterns}
\usetikzlibrary{calc} 

\usepackage{booktabs}   
\usepackage{makecell} 
\usepackage{colortbl}   

\usepackage{xspace}   
\usepackage{setspace}   
 
\usepackage{soul}
\usepackage{ulem}
\usepackage{currfile}
\usepackage{import}

\usepackage{authoraftertitle} 
\usepackage{authblk} 
\usepackage{cleveref}

\usepackage{breqn}

%% file: template/pgfplotsettings.tex
\pgfplotsset{compat=1.8}

\newcommand{\findmax}[3]{
    \pgfplotstablesort[sort key={#2},sort cmp={float >}]{\sorted}{#1}%
    \pgfplotstablegetelem{0}{#2}\of{\sorted}%
    \let #3=\pgfplotsretval%
}

\pgfplotscreateplotcyclelist{myCycleListBlackAndWhite}
{%
  {black, mark=o},
  {black, mark=square},
  {black, mark=triangle},
  {black, mark=diamond}, 
  {black, mark=pentagon}, 
  {black, mark=*},
  {black, mark=square*},
  {black, mark=triangle*},
  {black, mark=diamond*}, 
  {black, mark=pentagon*}, 
}

\definecolor{darkgreen}{rgb}{0,0.4,0} 
\definecolor{darkbrown}{rgb}{0.5, 0.396, 0.09}

\definecolor{c1}{rgb}{0.0, 0.4196078431372549, 0.6431372549019608}
\definecolor{c2}{rgb}{1.0, 0.5019607843137255, 0.054901960784313725}
\definecolor{c3}{rgb}{0.6705882352941176, 0.6705882352941176,
0.6705882352941176} \definecolor{c}{rgb}{0.34901960784313724, 0.34901960784313724, 0.34901960784313724}
\definecolor{c4}{rgb}{0.37254901960784315, 0.6196078431372549,
0.8196078431372549} \definecolor{c}{rgb}{0.7843137254901961, 0.3215686274509804, 0.0}
\definecolor{c5}{rgb}{0.5372549019607843, 0.5372549019607843,
0.5372549019607843} \definecolor{c}{rgb}{0.6352941176470588, 0.7843137254901961, 0.9254901960784314}
\definecolor{c6}{rgb}{1.0, 0.7372549019607844, 0.4745098039215686}
\definecolor{c7}{rgb}{0.8117647058823529, 0.8117647058823529,
0.8117647058823529}

\pgfplotscreateplotcyclelist{myCycleListColor}
{%
  {blue, mark=o},
  {red, mark=square},
  {darkbrown, mark=triangle},
  {darkgreen, mark=diamond},
  {black, mark=pentagon},
  {blue, mark=*},
  {red, mark=square*},
  {darkbrown, mark=triangle*},
  {darkgreen, mark=diamond*},
  {black, mark=pentagon*},
}

\pgfplotsset{every axis/.append style= 
              {
                font=\small,
                mark size=2,
                line width = 0.1,
                legend style={font=\small, mark size=3, draw=none, fill=none},
                legend cell align=left,
                cycle list name=myCycleListColor,
              }
            }
            
\pgfplotstableset
{
  every odd row/.style={before row={\rowcolor[gray]{0.9}}},
  every head row/.style=
  {
    before row=
    {%
      \toprule
    },
    after row=\midrule
  },
  every last row/.style=
  {
    after row=\bottomrule
  }
}

\newif\ifdrawboundingbox

\usetikzlibrary{external} 
\tikzset{external/system call={pdflatex \tikzexternalcheckshellescape
-halt-on-error -interaction=batchmode -jobname "\image" "\texsource"}} 

\pgfkeys
{ 
  /pgf/number format/.cd,
  fixed,
  set thousands separator={\,},
  1000 sep in fractionals,
}

%% file: template/commonCommands.tex
\newcolumntype{C}[1]{>{\centering\arraybackslash}m{#1}}
\newcolumntype{R}[1]{>{\raggedright\arraybackslash}m{#1}}
\newcolumntype{L}[1]{>{\raggedleft\arraybackslash}m{#1}}

\newcommand{\delete}[1]{\xspace}

\definecolor{Reviewer1}{rgb}{0.0, 0.0, 1.0}
\definecolor{Reviewer2}{rgb}{0.0, 0.5, 0.0}

%% file: template/titleAndAuthors.tex
\title{Uncertainty quantification of microstructure variability and mechanical behaviour of additively manufactured lattice structures}

\author[1]{N. Korshunova\thanks{\href{mailto:nina.korshunova@tum.de}{\texttt{nina.korshunova@tum.de}},
    Corresponding author}}
\author[2]{I. Papaioannou}
\author[1]{S. Kollmannsberger}
\author[2]{D. Straub}
\author[3]{E. Rank}
 \affil[1]{Chair of Computational Modeling and Simulation,
	Technische Universit\"at M\"unchen, Germany}
 \affil[2]{Associate Professorship of Engineering Risk Analysis, Technische Universit\"at M\"unchen, Germany}
\affil[3]{Institute for Advanced Study, Technische Universit\"at M\"unchen, Germany}

\newcommand{\journal}{Computer Methods in Applied Mechanics and Engineering}
\newcommand{\publicationDate}{March 17, 2021}

%% file: template/journalFooter.tex
\usepackage{fancyhdr}
\date{}
\fancypagestyle{plain}{%
  \fancyhf{}%
  \fancyfoot[L]{
  \vspace{-1.25cm} 
  \footnotesize{Preprint submitted to \textit{\journal{}}  \hfill
  \publicationDate
  \\
  }} }

%% file: template/abstract.tex
\section*{Abstract}
{
	Process-induced defects are the leading cause of discrepancies between as-designed and as-manufactured additive manufacturing (AM) product behavior. Especially for metal lattices, the variations in the printed geometry cannot be neglected. Therefore, the evaluation of the influence of microstructural variability on their mechanical behavior is crucial for the quality assessment of the produced structures. Commonly, the as-manufactured geometry can be obtained by computed tomography (CT). However, to incorporate all process-induced defects into the numerical analysis is often computationally demanding. Thus, commonly this task is limited to a predefined set of considered variations, such as strut size or strut diameter. In this work, a CT-based binary random field is proposed to generate statistically equivalent geometries of periodic metal lattices. The proposed random field model in combination with the Finite Cell Method (FCM), an immersed boundary method, allows to efficiently evaluate the influence of the underlying microstructure on the variability of the  mechanical behavior of AM products. Numerical analysis of two lattices manufactured at different scales shows an excellent agreement with experimental data. Furthermore, it provides a unique insight into the effects of the process on the occurring geometrical variations and final mechanical behavior.
}

%% file: template/keywords.tex
additive manufacturing, metal lattice structures, uncertainty quantification, process-induced defects, 
computed tomography, statistical model, numerical analysis, Finite Cell method

%% file: sections/introduction/introduction.tex
\section{Introduction}
\label{sec:introduction}
{ 	
	Lattice structures are generating considerable interest due to their lightweight and superior mechanical, acoustic or dielectric properties~\cite{Barba2020,Christensen2012,Fang2006}. However, their production by traditional manufacturing methods is rather limited to specific architectures~\cite{Tao2016}. Recent advances in Additive Manufacturing (AM) allow to fully exploit the potential of possible lattice designs~\cite{Plocher2019,Zhang2018a}. The lattice microstructures can now be produced at very small scales. Especially, metal lattice structures receive much attention due to their applicability in biomedical~\cite{Ahmadi2014, Alabort2019,Shidid2016}, aerospace~\cite{Hasan2010,Shen2013} or automotive industry~\cite{Beyer2016}. 
	
	Additively manufactured lattices often exhibit geometrical deviations with respect to their nominal geometries~\cite{Bagheri2017,Dallago2018,Maconachie2019,Vayre2012a}. Such imperfections in the geometry of as-manufactured lattices lead to strong deviations from the designed mechanical behavior~\cite{Dallago2019, Liu2017, Maconachie2019}. This has motivated a number of experimental investigations aiming at evaluating the mechanical properties of AM final parts~\cite{Gumruk2013,Tancogne-Dejean2016,Xiao2020}. Since the defects are rather specific for different designs, materials, and process parameters, a strong interest in the incorporation of as-manufactured geometries into the numerical analysis has emerged. Several research works have indicated that this is essential for an accurate prediction of the lattice behavior~\cite{duPlessis2020,Geng2019, Lei2019, Liu2017, Lozanovski2019}. Yet, this task is not trivial and leads to high computational costs. The first challenge is to obtain the as-manufactured geometry. Commonly, this is achieved through computed tomography (CT)~\cite{Echeta2020,Yan2012}. It provides comprehensive information about the internal geometrical structure and allows to account for multiple process-induced defects depending on the scan resolution. The second challenge is to make such a non-standard geometrical model suitable for the mechanical analysis. When CT geometries are used, the geometry reconstruction and mesh generation can be highly demanding and sometimes close to impossible for the whole object. Furthermore, small geometrical features present in the lattices require a high CT resolution to obtain a reliable geometrical approximation leading to large data sets. Despite such computational demands the CT-based numerical analysis is widely used to predict the mechanical behavior of imperfect lattices~\cite{Dallago2019, Geng2019, Korshunova2020a,Wang2019a}.

	While such predictive evaluations of the as-manufactured geometries capture the observed experimental behavior quite accurately, they do not evaluate the effects of the geometrical variability introduced by the AM process. This, however, is an important research question to improve the quality of the AM products and predict the possible variability of the obtained mechanical characteristics.  Furthermore, performing CT scans on large amount of samples is practically infeasible. Hence, an alternative approach to perform uncertainty quantification of geometrical variability on the mechanical response of the structures is required. At present, the most common approach to perform such characterization is to use statistical models. To this end, a set of considered defects, such as, e.g., strut waviness, diameter, or a cross-sectional shape change, is usually defined. Then, a statistical model is developed based on their occurrence in the acquired CT or optical microscopy images. Finally, these process-induced defects are incorporated into the CAD model. The challenging task is then to generate corresponding random geometrical structures to estimate the effects of the induced variations on the mechanical response.
	
	 A group of methods performs numerical analysis using beam elements, where such defects as strut diameter or strut center axis deviations can be easily incorporated. Campoli et al.~\cite{Campoli2013} accounted for the uncertainty on the strut diameter into the beam elements using the Gaussian distribution and concluded that their integration in a numerical analysis can significantly improve the accuracy of predicted linear elastic mechanical properties. Liu et al.~\cite{Liu2017} modeled both the strut waviness and thickness via a continuous generic probability density function and studied the mechanical behavior under compression. Similar statistical models have been applied to estimate the energy absorption variations caused by the strut waviness and diameter variations~\cite{Cao2020,Lei2019}.
	 
	 A second category of methods aims to simulate the effects of geometrical uncertainties using 3D solid elements. This approach allows to further extend the range of possible defects and include, e.g., internal porosity of the struts. However, such approaches challenge the modeling procedure of imperfect CAD geometries. Karamooz Ravari et al.~\cite{KaramoozRavari2014} proposed a generation of a spline curve describing the strut diameter variation. The method was extended to incorporate the defects as a union of spheres in the direction of the struts allowing for strut waviness and spherical pores within the strut volumes~\cite{KaramoozRavari2015}. Lozanovski et al.~\cite{Lozanovski2019} proposed a statistical model to incorporate the change in the cross-sectional shape using solid loft techniques for the generation of elliptical cross-sections.
	 
	 In light of the above literature review, existing research has focused on the incorporation of the geometrical imperfections into a CAD model for facilitating the subsequent numerical analysis. A flexible description of all occurring geometrical and topological variations can be rather challenging within this approach. Thus, the main novelty of the present contribution is to provide a CT-based approach to incorporate process-induced defects in lattice structures into an uncertainty analysis of the final product's mechanical behavior. The presented method does not require an ideal CAD model of the considered lattice and is easy to incorporate in an image-to-material-characterization workflow~\cite{Korshunova2020, Korshunova2020a}. It employs a non-homogeneous binary random field model to efficiently generate three-dimensional CT-based statistically equivalent realizations of the underlying lattices without limitations on the occurring type of geometrical imperfections. 
	 
	 Binary random field models are extensively used for the simulation of random two-phase media (e.g.,~\cite{Koutsourelakis2006,Khristenko2020,Fu2021, Su2021}).
     Such random fields are typically constructed by a level-cut of an underlying Gaussian random field. 
     Most existing studies employ a homogeneous threshold level, which is suitable for representing materials with random porous microstructure (e.g.,~\cite{Pigarin2004, Ilango2017, Khristenko2020,Fu2021}).
     In our study, we aim at modeling random process-induced defects in periodic AM lattice structures.
     Hence, the periodic pattern of the underlying microstructure needs to be retained.
     We achieve this through introducing a non-homogeneous cut-off level.

	We note that an alternative equivalent approach for defininig the non-homogeneous binary random field is to apply a homogeneous cut-off level to a Gaussian random field with non-homogeneous mean value (e.g.,~\cite{Chib1998,DeOliveira2000}). In~\cite{DeOliveira2000}, the parameters of the model are learned by application of a sampling-based Bayesian approach. Such an approach is computationally challenging for large-scale models such as the AM lattice structures studied in this paper. Therefore, we identify the parameters of the proposed binary random field model based on matching sample estimates of its first and second moment functions obtained from the CT images of sample structures. 
	 
	 As the output of the proposed model are generated statistically equivalent CT images, the Finite Cell Method, an immersed boundary method~\cite{Duster2017}, is used to evaluate the mechanical properties in a computationally efficient and fully automatic way.
	 Due to the large scale separation in the considered images, the binary random field has to be parameterized in terms of a large number of input random variables. Thus, we resort to Monte Carlo-based uncertainty quantification, whose efficiency does not depend on the dimension of the input random variable space.
	 To reduce the computational costs, we employ a multilevel Monte Carlo approach~\cite{Cliffe2011, Krumscheid2020} for efficient evaluation of the effect of the random microstructure on the mechanical properties of the final parts.
	 
	 The paper begins by introducing a non-homogeneous random field model used to reproduce a lattice geometry in~\cref{sec:binaryRandomField}. \Cref{sec:parameterIdentification} shows how a given CT image can be used to identify the design parameters of the random field model to obtain geometrical realizations, which are statistically similar to the original lattice geometry. Having identified these model parameters, an efficient technique to generate statistically similar lattice structures is outlined in~\cref{sec:numericalgeneration}. These artificially generated models are then used to evaluate the first two moments of the considered quantity of interest. A multilevel Monte Carlo method is applied for these computations, which is described in~\cref{sec:MLMC}. In~\cref{sec:numerics}, the proposed workflow is applied to square and octet-truss lattice structures manufactured with laser powder bed fusion. Finally, in~\cref{sec:conclusions} we draw the main conclusion before an outlook is presented.

}

%% file: sections/model/model.tex
\section{Binary random field model for geometrical description of periodic structures}
\label{sec:binaryRandomField}
{ 
	Computed tomographic images of additively manufactured products provide an unique opportunity to obtain an as-manufactured representation of the microstructure. To simplify these images for the numerical analysis, segmentation is traditionally performed to locate the object boundaries. The focus of this work is metal lattices. In this case, the final segmented CT images consist only of values $0$ and $1$, with $1$ indicating metal and $0$ indicating void. Representative slices of two lattice structures are shown in~\cref{fig:2DSlice}. 
	\newsavebox{\tempbox}
	\begin{figure}[H]%
	\centering
	\captionsetup[subfigure]{oneside,margin={0cm,0cm},labelformat=empty}
	\centering\sbox{\tempbox}{\includegraphics[scale=0.26]{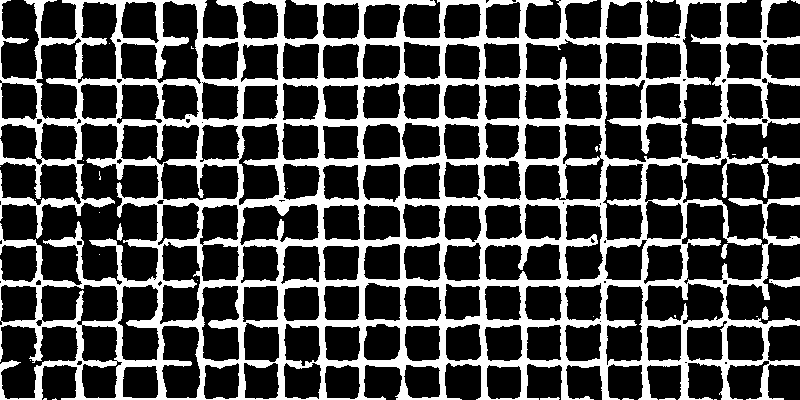}}%
	\subfloat[(a) Square grid lattice]{\usebox{\tempbox}}%
	\hspace*{-3cm}\subfloat[(b) Octet-truss lattice]{%
	\vbox to \ht\tempbox{%
	\vfil\includegraphics[scale=0.26, trim={8.5cm 15cm 8.5cm 15cm},clip]{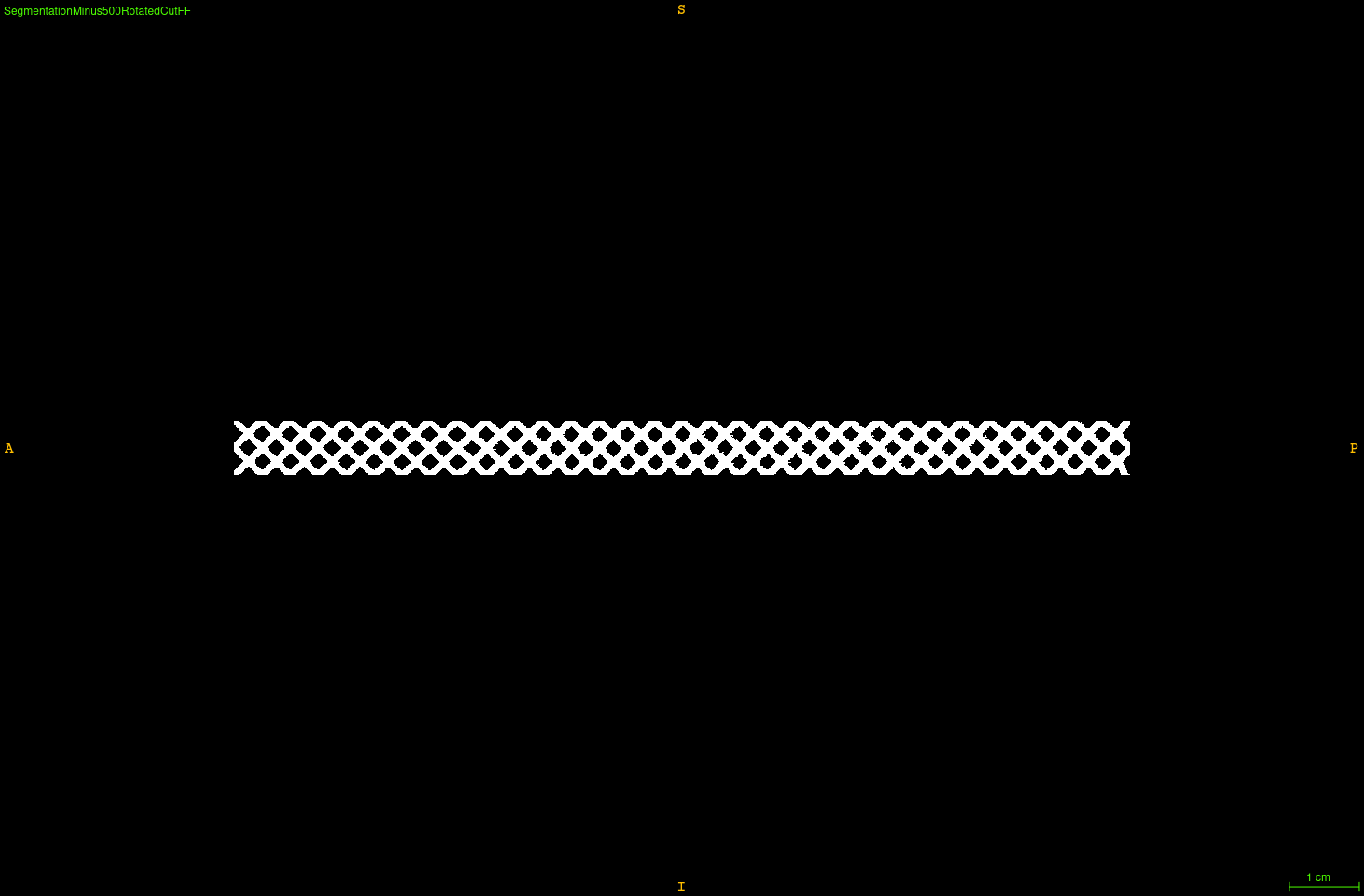}\vfil}}%
	\caption{2D slices of the segmented CT scans of AM periodic lattice structures (white pixels represent material and black pixels represent void).}
	\label{fig:2DSlice}
	\end{figure}		
	
	A natural choice to describe such geometrical models is through a binary random field model.
	A binary random field $Y(\bm{x})$ represents an infinite collection of binary random variables, i.e., variables taking outcomes in $\{0,1 \}$, indexed by a spatial coordinate $\bm{x} \in \Omega$, where $\Omega \subset \mathbb{R}^p$ denotes a spatial domain~\cite{Vanmarcke2007}. This model is then used to generate realizations, i.e., artificial geometries which are statistically similar to a physical structure recorded through CT images. The properties of the random field model should be such that the generated realizations retain certain important geometrical features while their remaining characteristics are varying. In particular, the lattices considered in this work are periodic as depicted in~\cref{fig:2DSlice}. Thus, the proposed model should preserve a general repetitive periodic cell structure, while the individual small features within a cell, e.g., surface roughness, internal porosity, strut sizes, should vary.
	
	A common way to formulate the binary random field is to define a parent (or latent) underlying continuous random field, which is clipped in the consequent step with the pre-defined threshold $d$. The parent field is assumed to be a real-valued homogeneous Gaussian random field $U(\bm{x})$ with zero mean value $\mu_U=0$, unit standard deviation $\sigma_{U} = 1$, and auto-correlation function $\rho_{UU}(\bm{x}_1,\bm{x}_2)$. Commonly, the cut-off level $d$ is assumed to be homogeneous, i.e., constant with respect to the spatial coordinate (see, e.g.,~\cite{Lin2005, Khristenko2020, Koutsourelakis2006, Ogorodnikov2018, Pigarin2004, Vanmarcke2007}). However, the use of homogeneous threshold would lead to a random porous geometrical model. The generation of periodic structures as shown in~\cref{fig:2DSlice} would not be possible. Thus, in this work, we choose the threshold to vary with respect to the spatial location, i.e., $d(\bm{x})$, which allows us to control the overall periodic structure of the lattices.  Hence, the binary random field $Y(\bm{x})$ can be expressed in terms of the Gaussian field $U(\bm{x})$ as
	\begin{equation}
	Y(\bm{x}) = 
	\begin{cases}
	0 & \text{ for } U(\bm{x}) \in \left(-\infty, d(\bm{x})\right)\\
	1 & \text{ for } U(\bm{x}) \in \left[d(\bm{x}), \infty\right)
	\label{eq:binaryDefinition}
	\end{cases}
	\end{equation}
	where $d(\bm{x})$ is the truncation threshold depending on the spatial location. As the Gaussian random field $U(\bm{x})$ has zero mean and unit variance, $U(\bm{x})$ is completely characterized by its auto-correlation function $\rho_{UU} (\bm{x}_1,\bm{x}_2)=\mathrm{E}[U(\bm{x}_1)U(\bm{x}_2)]$. Hence, the binary random field $Y(\bm{x})$ is defined by the threshold function $d(\bm{x})$ and the function $\rho_{UU}(\bm{x}_1,\bm{x}_2)$. These functions should be identified such that the generated geometrical models attain a similar lattice structure. A graphical illustration of the proposed model is given in~\cref{fig:BayesianNetwork}. The figure shows the process of generation of a random realization from the binary random field $Y(\bm{x})$ for a single unit cell of lattice structure. Given the function $\rho_{UU}(\bm{x}_1,\bm{x}_2)$, realizations from the Gaussian field $U(\bm{x})$ can be generated, which can be transformed to realizations of the field $Y(\bm{x})$ based on the threshold function $d(\bm{x})$. The parameters $\bm{l}$ and $\bm{\nu}$ shown in~\cref{fig:BayesianNetwork} are parameters of the function $\rho_{UU}(\bm{x}_1,\bm{x}_2)$, which will be discussed in detail further below.
	
	The design parameter space of the binary random field model, including the threshold level $d(\bm{x})$ and the auto-correlation function $\rho_{UU}(\bm{x}_1,\bm{x}_2)$, can be identified based on matching sample estimates of the first and second moment functions of $Y(\bm{x})$ obtained from the CT images of sample structures. 
	In the following, we provide expressions for these functions in terms of the model parameters. We remark that for the case of a homogeneous threshold, expressions for the second moment function of the binary field as a function of $\rho_{UU} (\bm{x}_1,\bm{x}_2)$ and the (homogeneous) threshold $d$ are given, e.g., in \cite{Teubner1991,Kendall1994,Pigarin2004,Lin2005}.
	
	\begin{figure}[H]
		\centering
		\includegraphics[scale=0.22]{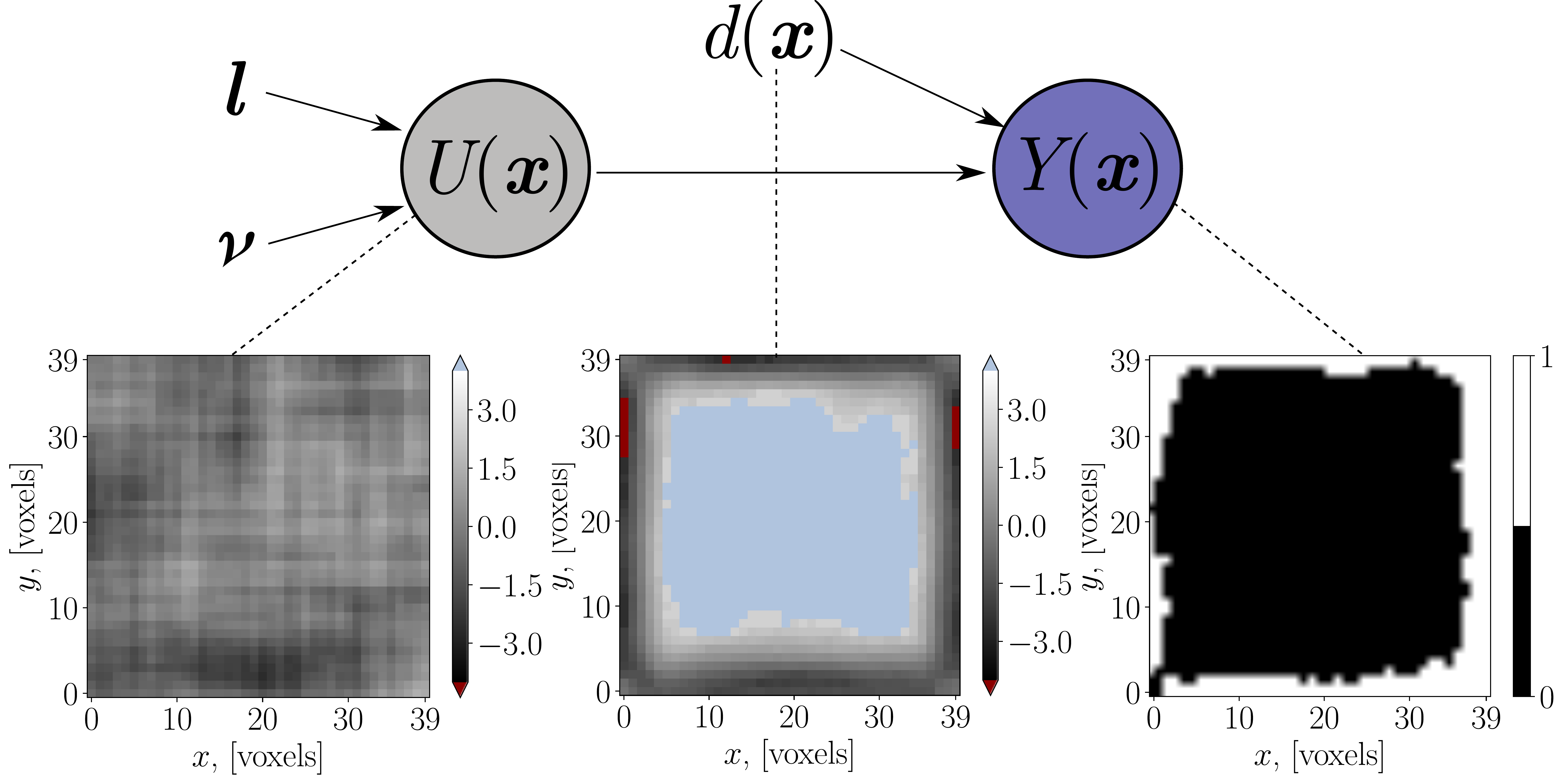} 
		\caption{Graphical illustration of the proposed non-homogeneous binary random field model and the process of generation of a random realization from the model (in dark red: $-\infty$ values, light blue: $\infty$ values). Lower left: one realization of the Gaussian random field $U(\bm{x})$; lower middle: threshold function $d(\bm{x})$ acquired from CT image of a sample; lower right: one realization of a binary random field $Y(\bm{x})$.}
		\label{fig:BayesianNetwork}
	\end{figure}		
	
	The marginal probability mass function (PMF) of $Y(\bm{x})$ is
	\begin{equation}
	p_Y (y,\bm{x}) = 
	\begin{cases}
	\Phi(d(\bm{x})) & \text{ for } y = 0\\
	1 - \Phi(d(\bm{x})) & \text{ for } y = 1
	\label{eq:marginalPMF}
	\end{cases}
	\end{equation}
	where $\Phi(\cdot)$ is the standard normal cumulative distribution function (CDF). Hence, the mean of $Y(\bm{x})$ is given as follows:
	\begin{equation}
	\mu_{Y} (\bm{x}) = 1-\Phi(d(\bm{x}))
	\label{eq:meanDefinition}
	\end{equation}
	and the variance function is:
	\begin{equation}
	\sigma_{Y}^2 (\bm{x}) = \left[1-\Phi(d(\bm{x}))\right]\Phi(d(\bm{x}))
	\label{eq:variaceDefinition}
	\end{equation}
	
	The covariance of the binary random field with spatially varying thresholds can be written as
	\begin{equation}
	\Gamma_{YY}(\bm{x}_1, \bm{x}_2)=\int_{0}^{\rho_{UU}(\bm{x}_1,\bm{x}_2)}\frac{1}{2\pi \sqrt{1-z^2}}\exp\left[-\frac{d(\bm{x}_1)^2 + d(\bm{x}_2)^2 - 2d(\bm{x}_1)d(\bm{x}_2)z}{2(1-z^2)}\right]dz
	\label{eq:covarianceBinaryFinal2}
	\end{equation}
	A full derivation of this result can be found in~\cref{sec:appendix}. The auto-correlation function of $Y(\bm{x})$ is then obtained as:
	\begin{equation}
	R_{YY}(\bm{x}_1, \bm{x}_2) = \frac{\Gamma_{YY}(\bm{x}_1, \bm{x}_2)}{ \sigma_{Y}(\bm{x}_1) \sigma_{Y}(\bm{x}_2) }
	\end{equation}
	with $\sigma_{Y} (\bm{x})$ as defined in~\cref{eq:variaceDefinition}.
	
	The functions $d(\bm{x})$ and $\rho_{UU}(\bm{x}_1,\bm{x}_2)$, which define the binary field $Y(\bm{x})$, can be estimated by comparing sample estimates of the mean and auto-covariance functions of $Y(\bm{x})$ with the expressions in \cref{eq:meanDefinition,eq:covarianceBinaryFinal2}. Such estimates can be obtained based on CT images of manufactured AM products. 
	Before describing this identification procedure, we further simplify the modeling process by introducing a parametric correlation model for the function $\rho_{UU}(\bm{x}_1,\bm{x}_2)$, based on the Mat\'ern model.
	The one-dimensional Mat\'ern correlation model reads~\cite{Handcock1994}:
	\begin{equation}
	\rho_M\left(x_1,x_2\right)= \frac{2^{1-\nu}}{\Gamma(\nu)}\left( \sqrt{2\nu}\frac{\Delta x}{l}\right)^\nu K_{\nu}\left(\sqrt{2\nu}\frac{\Delta x}{l}\right)
	\label{eq:correlationMatern}
	\end{equation} 
	where $\Delta x=|x_1-x_2|$ indicates the spatial lag, $l$ is the correlation length parameter, $\nu$ is a non-negative smoothness parameter, $\Gamma$ is the gamma function, and $K_\nu(\cdot)$ is the modified Bessel function of the second kind. 
	The design space for the one-dimensional Mat\'ern correlation model includes two parameters, namely the correlation length $l$ and smoothness parameter $\nu$. The smoothness parameter provides great flexibility to describe spatial correlations. When $\nu$ tends to zero, the spatial variation is rather rough, while large values of $\nu$ lead to a smooth spatial process. Furthermore, it combines a wide range of other parametric correlation models. For example, for $\nu=0.5$ the exponential model is recovered, while for $\nu \rightarrow \infty$ the Gaussian correlation kernel is obtained. 
	
	We introduce the separability assumption on the correlation function, which, as discussed in~\cref{sec:numericalgeneration}, simplifies the generation of the random field $U(\bm{x})$. The product family of the correlation kernels can be formulated as follows \cite{Sacks1989}:
	\begin{equation}
	\rho_{UU}(\bm{x}_1,\bm{x}_2) =\rho_M\left(x_1,x_2\right)\rho_M\left(y_1,y_2\right)\rho_M\left(z_1,z_2\right)
	\label{eq:separability}
	\end{equation}
	Thus, the design space for the binary model in~\cref{eq:binaryDefinition} with~\cref{eq:correlationMatern,eq:separability} consists of the following set of parameters: the threshold function $d(\bm{x})$, the correlation length vector $\bm{l}=(l_x,l_y,l_z)$ and the smoothness vector $\bm{\nu}=(\nu_x,\nu_y,\nu_z)$ as depicted in~\cref{fig:BayesianNetwork}. Given CT scans of the produced lattice structures, these parameters can be determined directly from the segmented images. The procedure for the parameter identification will be described in the following section.
	
}

%% file: sections/parameters/parameters.tex
\section{CT-based model parameter identification}
\label{sec:parameterIdentification}
{ 
	We want to select the design parameters $d(\bm{x})$, $\bm{l}$, and $\bm{\nu}$ of the proposed random binary field model that lead to geometric models that are statistically similar to the structures depicted in~\cref{fig:2DSlice}. Thus, we establish a procedure to identify these parameters such that the overall macroscopic periodic structure, together with the process-induced defects, are retained by the model in a statistical sense. The available CT images of already produced lattices serve as a basis to learn these features because they contain all necessary data to proceed with the model generation.
	
	First, we describe how the threshold $d(\bm{x})$ can be identified from available CT images. As mentioned in~\cref{sec:binaryRandomField}, this function primarily controls a general macroscopic shape of the structure. As an example, if the threshold level $d(\bm{x})=\text{const}$, it regulates a macroscopic volume fraction of the voids or inclusions in the generated geometry. In contrast, when it is dependent on the spatial coordinate, structure-specific information can be incorporated. In the considered case, we aim at retaining the periodicity of the lattice cells. Thus, the threshold level can be identified by taking advantage of the repetition of the unit cells. 
	
	We introduce a local unit cell with the attached local coordinate system $\bm{x}^l=(x^l,y^l,z^l)$. This is shown in Step 1 of~\cref{fig:schemeForThresholds} on a two-dimensional slice. The size of this cell depends on the design geometry of the structure. The mapping between the local and the global coordinate system is defined inherently due to the existing periodicity. Then, we exploit the underlying voxel structure of the CT scan (Step 2 in~\cref{fig:schemeForThresholds}). In particular, as the segmented values are constant within one voxel, the threshold $d(\bm{x})$ becomes piece-wise constant. It can be evaluated at a discrete set of local coordinates $\bm{X}^l=[\bm{x}^l_{1};\ldots;\bm{x}^l_{n_l}]$ located at the center of every voxel (see step 2 in~\cref{fig:schemeForThresholds}). 
\pdfliteral direct {/Interpolate false}

	\begin{figure}[H]
		\centering
		\includegraphics[scale=0.15]{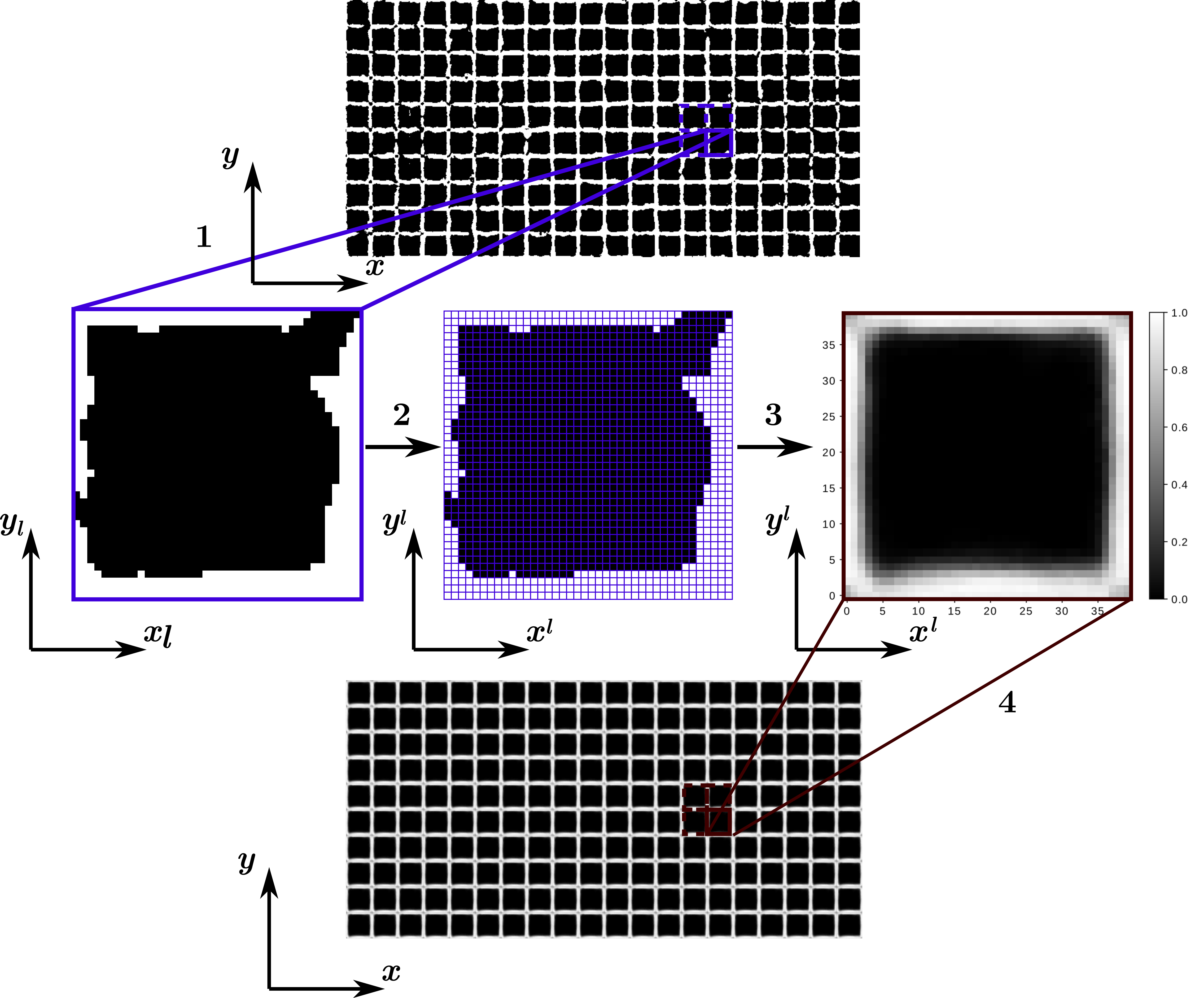}
		\caption{Threshold workflow identification.}    
		\label{fig:schemeForThresholds}
	\end{figure}	
\pdfliteral{/Interpolate false}

	Then, as the CT image consists of many local unit cells $N_{cells}$, all of them can be collected in a pool of realizations of the binary random field defined at the voxels of the local cells $\{ y^{l}_{i}({\bm{X}}^l),i=1,\ldots,N_{\text{cells}} \}$. In this way, the mean function of $Y$ at the discrete grid points ${\bm{X}}^l$ in the local coordinate system can be estimated as follows (step 3 in~\cref{fig:schemeForThresholds}):
	\begin{equation}
	\hat{\mu}_Y({\bm{X}}^l)= \frac{1}{N_{\text{cells}}}\sum_{i=1}^{N_{\text{cells}}} y^{l}_{i}({\bm{X}}^l)
	\label{eq:meanBinary}
	\end{equation}
	 Note that the computed mean values coincide with the estimates of the probabilities $\Pr(Y({\bm{X}^l})=1)$. In particular, the black voxels in step 3 in~\cref{fig:schemeForThresholds} indicate that there is always a void at these locations in all periodic cells. By contrast, the white-coloured voxels are always filled with the material in the whole structure. This grey value distribution is considered representative of the AM process for the considered lattice structure itself. 
	
	The global mean values $\hat{\mu}_Y({\bm{X}})$, are then obtained through performing the inverse mapping of the local coordinate system to the global one (step 4 in~\cref{fig:schemeForThresholds}). Using~\cref{eq:meanDefinition}, the thresholds $d(\bm{x})$ then are piece-wise constant with the values evaluated at the set of coordinates ${\bm{X}}$ as follows:
	\begin{equation}
	\hat{d}({\bm{X}}) = \Phi^{-1}(1 - \hat{\mu}_Y({\bm{X}}))
	\end{equation}
	where $\Phi^{-1}$ is the inverse of the standard normal CDF.

    Second, a correlation structure similar to the one occurring in the manufactured lattice should be identified. The correlation parameters $(\bm{l},\bm{\nu})$ control an overall smoothness of the lattice grid. For example, the combination of a large correlation length with large smoothness values would lead to a smooth surface, while low values would be associated with rough lattice surfaces. Following~\cref{eq:separability}, the correlation kernel is assumed to be separable. Thus, the correlation fit can be performed separately in every spatial direction. In the following, we explain the procedure only for spatial direction $x$, where the design parameters to be determined are $(l_x,\nu_x)$. The other directions are fitted analogously.
	
	The covariance function of the binary random field can be estimated from the CT image for each local coordinate pair $(\bm{x}^l_i,\bm{x}^l_j)$ having a certain spatial lag $|\bm{x}^l_i-\bm{x}^l_j|=(\Delta x_{(ij)},0,0)$ . The spatial lag in x direction, $\Delta x_{(ij)}$, can attain the minimum value of one voxel and runs until the width of the local unit cell. Since the threshold values are repeated in each unit cell, an estimate of the covariance for a given lag $\Delta x_{(ij)}$ is obtained as follows:
	\begin{equation}
		\hat{\Gamma}_{YY}(\bm{x}^l_i,\bm{x}^l_j) =  \frac{1}{N_{\text{cells}}}\sum_{k=1}^{N_{\text{cells}}}\left(y^l_{k}({\bm{x}}_i^l)-\hat{\mu}_{Y}({\bm{x}}_i^l)\right)\left(y^l_{k}({\bm{x}}_j^l)-\hat{\mu}_{Y}({\bm{x}}_j^l)\right)
		\label{eq:covBinaryFromImage}
	\end{equation}
	The auto-covariance of the binary random field is stored as a vector depending on the spatial lag $\Delta x_{(ij)}$ together with the corresponding threshold values $\hat{d}(\bm{x}^l_i)$ and $\hat{d}(\bm{x}^l_j)$. For a certain spatial lag $\Delta x_{(ij)}$, there is a total of $(n_x-n_{\text{lags}})n_y n_z$ covariance entries with $n_x$,$n_y$,$n_z$ being the number of voxels in every spatial direction within one unit cell and $n_{\text{lags}}$ being the total number of considered lags (see Step 2 in~\cref{fig:schemeForThresholds}). The problem can be further reduced by only considering the points with intermediate probability values larger than zero and smaller than one. For voxels with probability zero or one, correlation is not defined. This filtering interval is further enlarged to remove the extreme data sets, i.e., the sets for which $\hat{\mu}_{Y}({\bm{x}}_i^l) <0.1$ and $\hat{\mu}_{Y}({\bm{x}}_i^l) >0.9$. 
	
	Assume, that the covariance data after filtering the extreme data sets has the total size $N_{\text{data}}$. Hence, the collected data set consists of the estimated covariance of the binary field $\hat{\Gamma}_{YY}(\bm{x}^l_{1m},\bm{x}^l_{2m})$ at the respective pair of spatial coordinates $(\bm{x}^l_{1m},\bm{x}^l_{2m})$ with $m=1,\ldots,N_{\text{data}}$. The set is fitted to the considered analytical Mat\'ern model by inserting~\cref{eq:correlationMatern} into~\cref{eq:separability} and then plugging into~\cref{eq:covarianceBinaryFinal2}. The optimization problem is then formulated as a standard least-square minimization:
	\begin{equation}
\min  \left(S^2(l_x,\nu_x) \right)=	\min  \left(\sum_{m=1}^{N_{\text{data}}} r_{m}^2(l_x,\nu_x) \right)=\min  \left(\sum_{m=1}^{N_{\text{data}}}  \left(\hat{\Gamma}_{YY}(\bm{x}^l_{1m},\bm{x}^l_{2m}) -{\Gamma}_{YY}(\bm{x}^l_{1m},\bm{x}^l_{2m})\right)^2\right)
	\label{eq:MinimizationFunction}
	\end{equation}
	where $S$ indicates the total formulated residual, $\hat{\Gamma}_{YY}(\bm{x}^l_{1m},\bm{x}^l_{2m})$ is the estimated covariance of the observed binary random field according to~\cref{eq:covBinaryFromImage}, and ${\Gamma}_{YY}(\bm{x}^l_{1m},\bm{x}^l_{2m})$ is the chosen model correlation computed by combining~\cref{eq:correlationMatern} with~\cref{eq:separability} and plugging in to~\cref{eq:covarianceBinaryFinal2}.

}

%% file: sections/realization/realization.tex
\section{Efficient random field simulation for large data sets}
\label{sec:numericalgeneration}
{ 
	To analyze the impact of the random microstructure on the output quantities of interest, realizations of the binary random field as in~\cref{eq:binaryDefinition} need to be generated. The latter requires the simulation of the underlying standard Gaussian random field $U(\bm{x})$. 
	This can be done by a number of methods -- a comprehensive review is given in~\cite{Liu2019}. Unlike methods based on approximate representations of the random field (e.g., Fourier representations~\cite{Shinozuka1996}, Karhunen-Lo\`eve expansion~\cite{Betz2014}), the Cholesky decomposition method generates samples from the true random field at a number of spatial locations~\cite{Wood1994}. However, this approach is rarely used as its computational cost is $\mathcal{O}(n^3)$, with $n$ denoting the number of locations, which is prohibitively expensive for large $n$. Furthermore, the necessary memory storage for the whole covariance matrix before its decomposition is exceptionally large for practical examples. The size of the problems at hand results in a simulation of more than 100 million random variables in three dimensions. This would not be feasible unless approximation methods are used or the assumption of separability as in~\cref{eq:separability} is made. If the covariance function is assumed to be separable, the exact stepwise technique based on the covariance matrix decomposition proposed in~\cite{Li2019} can be applied. This approach drastically reduces the computational costs and memory requirements for the generation of a three-dimensional Gaussian random field. In the following, the main steps of this method are recapitulated.

Recall that the Gaussian random field $U(\bm{x})$ is to be simulated at a number $n$ of spatial points ${\bm{X}}=[\bm{x}_1;\ldots;\bm{x}_{n}]$, corresponding to the voxel centers. This requires the simulation of a Gaussian random vector $\bm{U}$ with zero mean, unit variance and prescribed correlation matrix $\bm{R}=[\rho_{UU}(\bm{x}_{i},\bm{x}_{j})]_{n\times n}$. The vector $U(\bm{x})$ can be decomposed as follows:
\begin{equation}
\bm{U} = \bm{L}\bm{Z}
\label{eq:GRFGeneration}
\end{equation}
where $\bm{Z}$ is the vector of $n$-independent standard normal variables and $\bm{L}$ is the lower triangular matrix derived from the Cholesky decomposition $\bm{R}=\bm{L}\bm{L}^{\mathrm{T}}$~\cite{Fenton2008}. 
Since the correlation function is assumed to be separable as in~\cref{eq:separability}, the total correlation matrix $\bm{R}$ can be written as a Kronecker product of correlation matrices corresponding to 1D correlation functions:

\begin{equation}
\bm{R}= \bm{R}_z\otimes\bm{R}_y\otimes\bm{R}_x
\end{equation}
where $\otimes$ is the Kronecker product, $\bm{R}_x=[\rho_x(x_{i},x_{j})]_{n_x\times n_x}$, $\bm{R}_y=[\rho_y(y_{i},y_{j})]_{n_y\times n_y}$, $\bm{R}_z=[\rho_z(z_{i},z_{j})]_{n_z\times n_z}$ and $\bm{x}_i=[x_i,y_i,z_i]$. In contrast to the size of the total correlation matrix $\bm{R}$ of $(n_xn_yn_z\times n_xn_yn_z)$, the matrices corresponding to the one-dimensional correlation models have a much smaller size of  $(n_z\times n_z)$, $(n_y\times n_y)$, and $(n_x\times n_x)$. 

Using the mixed-product property of the Kroenecker product, the lower triangular matrix $\bm{L}$ can be written as the Kroenecker product of the lower triangular matrices of the Cholesky decompositions corresponding to each one-dimensional model:

\begin{equation}
\bm{L}= \bm{L}_z\otimes\bm{L}_y\otimes\bm{L}_x
\label{eq:choleskySeparable}
\end{equation}
\Cref{eq:choleskySeparable} simplifies the computation of the Cholesky decomposition of a large matrix, which is both time-consuming and prone to round-off errors due to the poor conditioning of the covariance matrix~\cite{Fenton2008}.
Inserting~\cref{eq:choleskySeparable} into~\cref{eq:GRFGeneration} gives:
\begin{equation}
\bm{U} =\left(\bm{L}_z\otimes\bm{L}_y\otimes\bm{L}_x\right)\bm{Z}
\label{eq:GRFGenerationSimplified}
\end{equation}

\Cref{eq:GRFGenerationSimplified} is still rather expensive, as a large matrix multiplication is involved. However, if matrix-array multiplication is introduced the equation can be written out in its equivalent form as follows~\cite{Li2019}:
\begin{equation}
\underline{\bm{U}} = \bm{L}_z\times_3\left[\bm{L}_y\times_2\left(\bm{L}_x\times_1\underline{\bm{Z}}\right)\right]
\label{eq:GRFGenerationSimplifiedMatrixArrayMultiplication}
\end{equation}
where $\underline{\bm{U}}$ and $\underline{\bm{Z}} $ are vectors $\bm{U},\bm{Z}$ respectively reshaped to the matrices of size $(n_x\times n_y \times n_z)$ and $\times_i$ is the matrix-array multiplication defined in the following generic way:
\begin{equation}
\begin{aligned}
c_{jpq}&=\sum_{k}a_{jk}b_{kpq} \text{ if } i=1 \\
c_{pjq}&=\sum_{k}a_{jk}b_{pkq} \text{ if } i=2 \\
c_{pqj}&=\sum_{k}a_{jk}b_{pqk} \text{ if } i=3 .
\end{aligned}
\end{equation}

This approach reduces the computational costs of the covariance decomposition to $\mathcal{O}(n_x^3+n_y^3+n_z^3)$ and of the random field generation to $\mathcal{O}\left[n_xn_yn_z(n_x+n_y+n_z)\right]$~\cite{Li2019}. Moreover, it avoids evaluation and storage of the full correlation matrix $\bm{R}$. Hence, it enables generating the large scale Gaussian random fields required in this work. The required computational times will be further addressed in~\cref{sec:numerics}.
	
}

%% file: sections/MLMC/MLMC.tex
\section{Multilevel Monte Carlo method for AM product characterization}
\label{sec:MLMC}
{
	
	The deterministic evaluation of the AM product's mechanical behavior provides an insight into the achieved quality of the final structure. The presented approach enables the evaluation of the variability of the mechanical response subject to geometrical uncertainties in the underlying microstructure. In this work, we focus on the homogenized elastic behavior of AM components; in particular, we assess the variability of the effective Young's modulus. For a given realization of the binary random field representing the material microstructure, the quantity of interest (QoI or $Q$) can be approximated numerically. In particular, the setup shown in~\cref{fig:NumericalSetup} can be used to characterize a homogenized tensile behavior of the lattices, with $u$ denoting the applied uni-axial displacement.
	The rigid body modes of this setup are then additionally fixed. Then, the homogenized Young's Modulus in the applied displacement direction can be computed by dividing the average occurring stress by the applied strain. This problem is usually solved using the Finite Element Method (FEM).
	 
	\begin{figure}[H]
		\centering
		\includegraphics[scale=0.22]{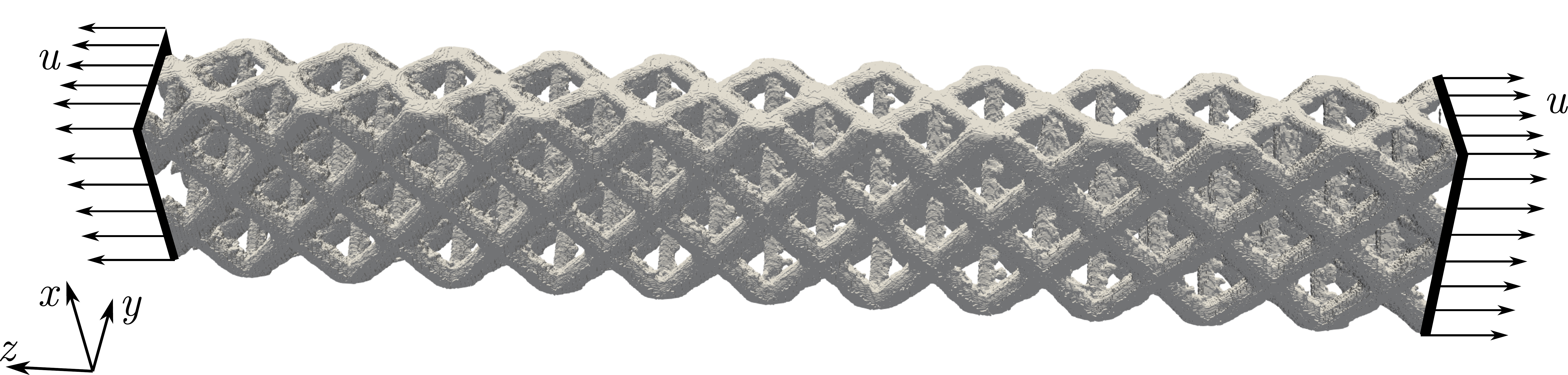} 
		\caption{Numerical setup of the tensile experiment on the lattice structure~\cite{Korshunova2020a}. }
		\label{fig:NumericalSetup}
	\end{figure}		
	
	 The effect of microstructural variability on the effective Young's modulus can be evaluated by application of the Monte Carlo (MC) method. In the crude MC method, the statistical moments of the QoI (e.g., mean and variance) are estimated based on repeated evaluations of the QoI for a number of realizations of the uncertain geometrical input. Due to the large overall size of the considered structures and the small scale of the local geometrical variations, a detailed FEM resolution is typically required to achieve an accurate approximation of the QoI. Thus, the numerical analysis for each realization with standard FEM results in significant manual labour and high computational cost. This is because every change of the underlying geometry would require a new mesh generation to resolve the generated geometry. Furthermore, the realization and the original specimens are CT images that bring an additional level of complexity to the traditional FEM for numerical analysis. In particular, before the mesh for a model can be generated, a complex geometry reconstruction must be run, which results in a large manual involvement.
	
	As an efficient alternative to FEM, we apply the Finite Cell Method, an immersed boundary method, to compute the QoI for a given geometrical model of a lattice. While we omit all details of FCM and refer to the literature (see, e.g.,~\cite{Duster2017, Korshunova2020a, Yang2012}), we summarize its significant advantages. As an immersed boundary method, the FCM embeds the geometric model in a larger, simply shaped domain. This domain can easily be divided into a coordinate aligned numerical grid. Thus, it eliminates any effort for meshing the structure. As many numerical simulations with a varied shape should be performed, the FCM allows performing the MC analysis in a fully automatic way. The true geometry is commonly retained by a specialized integration scheme for elements, which are cut by the physical boundary. As the geometrical models in this work stem from the CT images, we can apply a unique, efficient integration scheme exploiting the underlying voxel structure.  Furthermore, our implementation of this immersed boundary method can successively increase the polynomial degree of shape functions or refine the computational grid locally or globally. Thus, it is easily possible to increase the accuracy (on the cost of computational effort) of a simulation.
	
	However, the computational effort for estimating the statistical moments using a combination of the FCM with the crude MC method still remains high. We can further reduce the necessary computational cost for obtaining accurate moment estimates  by employing the multilevel Monte Carlo (MLMC) method. Instead of evaluating a large number of samples with a fixed fine discretization of the QoI, the MLMC method considers a hierarchy of $(L+1)$ numerical approximation levels. Every subsequent level has a better approximation quality. The levels can be chosen, e.g., such that every subsequent level has a finer mesh or a higher polynomial degree of the element shape functions. MLMC reduces the overall computational cost for a target accuracy of the obtained moment estimates by performing many numerical simulation runs at low approximation levels while decreasing the required runs on every subsequent level. Thus, only a few solves at high levels are necessary. 
	
    For the characterization of the variability of the homogenized Young's Modulus, we consider both the mean value $\mu_Q = \mathrm{E}[{Q}]$ and the second central moment, i.e. the variance, $\mu_{2,Q}=\mathrm{E}[({Q}-\mu_Q)^2]=\mathrm{Var}[{Q}]$. The MLMC method is well established for estimating the mean value (or other raw moments) of the QoI (e.g~\cite{Cliffe2011,Teckentrup2013}). However, the estimation of central moments is not trivial. Recently, MLMC was extended to include the estimation of central moments of the QoI of any order~\cite{Krumscheid2020}. Next we review this method, starting with a summary of the standard MLMC procedure for the mean value estimation.
    
    The MLMC estimator for the mean value $\mu_Q = \mathrm{E}[{Q}]$ can be written as follows:
	\begin{equation}
	m_1^{MLMC}=\displaystyle\sum_{l=0}^{L} \left( \widehat{Q}^l_{N_l,M_l} - \widehat{Q}^l_{N_l,M_{l-1}}\right)
	\label{eq:MLMCMean}
	\end{equation}
	where
	\begin{equation}
	    \widehat{Q}^l_{N_l,M_l} = \frac{1}{N_l}\displaystyle\sum_{i=1}^{N_l} Q_{i,M_l}
	\end{equation}
	and $\widehat{Q}^{0}_{N_{0},M_{-1}}:=0$; $\bm{Q}^l_{N_l,M_l}= \left(Q_{i,M_l}\right)_{i=1..N_l}$ is the collection of $N_l$ independent and identically distributed (i.i.d.) realizations of the quantity of interest ${Q_{M_l}}$ at numerical approximation level $l$ and $M_l$ is the corresponding number of degrees of freedom at level $l$. The estimator quantities with the same superscript $l$ are computed with the same input samples whereas estimators with different superscripts are computed with a different set of i.i.d. input samples.
	
	The accuracy of the estimator in \cref{eq:MLMCMean} can be assessed by the mean-square error, which is formulated as:
	\begin{equation}
	\mathrm{MSE}(m_1^{MLMC}) =(\mu_{{Q}_{M_L}}-\mu_Q)^2+ \sum_{l=0}^{L}\mathrm{Var}\left[ \widehat{Q}^l_{N_l,M_l} - \widehat{Q}^l_{N_l,M_{l-1}}\right]
	\label{eq:MLMCMSE}
	\end{equation}
	where $\mu_{{Q}_{M_L}}$ denotes the mean of the QoI at the final approximation level $L$. The first term in~\cref{eq:MLMCMSE} is the bias contribution and evaluates the approximation error at level $L$, while  the second term is the sampling (statistical) error. The bias contribution for the mean value can be approximated as follows:
	\begin{equation}
	\left|\mu_{{Q}_{M_L}}-\mu_Q \right| \approx \left|\widehat{Q}^l_{N_l,M_l} - \widehat{Q}^l_{N_l,M_{l-1}}\right|
	\end{equation}
	The variance terms can be estimated using the available samples through the standard variance estimator.
	
	The number of approximation levels $L$ and the number of samples at each level $N_l$ can be determined such that a small mean-square error is achieved with a low computational cost. To approach this problem, an assumption is typically made on the behavior of the bias, variance and computational cost at each level. In particular, they are assumed to follow relations:
	\begin{equation}
	\begin{aligned}
	\left|\mu_{{Q}_{M_l}}-\mu_Q\right| &\leq c_{\alpha}(M_l)^{-\alpha}\\
	V_{l} &\leq c_{\beta}(M_l)^{-\beta}\\
	\mathrm{cost}(Q_{M_l}) &\leq c_{\gamma}(M_l)^{\gamma}
	\label{eq:exponentialFunctionConstants}
	\end{aligned}
	\end{equation}
	where $M_l$ is the number of degrees of freedom for level $l$, $V_{l}:=N_l\mathrm{Var}\left[ \widehat{Q}^l_{N_l,M_l} - \widehat{Q}^l_{N_l,M_{l-1}}\right]$, $\mathrm{cost}(Q_{M_l})$ is the computational cost of evaluating the QoI at level $l$, and $c_{\alpha},\alpha, c_{\beta},\beta,c_{\gamma},\gamma$ are the problem-dependent constants.
	These constants are determined by employing a screening procedure. An initial small number of samples $N_l$ are evaluated at a few coarse levels $\bar{L}$. Then, the constants are fit via a least-squares procedure and used to extrapolate the costs and variance at the subsequent levels.
	
	 The number of degrees of freedom at the maximum level $L$, $M_L$, can be estimated for a prescribed relative tolerance $\varepsilon_r$ through:
	\begin{equation}
	M_L\geq\left(\frac{\varepsilon}{c_\alpha \sqrt{2}}\right)^{-\frac{1}{\alpha}}
	\label{eq:optimalDOFs}
	\end{equation}
	where $\varepsilon \approx \varepsilon_r \cdot m_{\mu_Q}^{MLMC}$.
	Then, by solving an optimization problem to minimize the variance of the estimator for a fixed computational cost, the optimal sample size at every level can be estimated as:
	\begin{equation}
	N_l\geq\left\lceil \frac{2}{\varepsilon^2}\sqrt{\frac{V_{l}}{\mathrm{cost}(Q_{M_l})}}\sum_{k=0}^{L}\sqrt{\mathrm{cost}(Q_{M_k})V_{k}}\right\rceil
	\label{eq:optimalNumberOfSamples}
	\end{equation}
	where the terms $V_{l}$ and $\mathrm{cost}(Q_{M_l})$ can be estimated with \cref{eq:exponentialFunctionConstants}.
	
	Having described the standard MLMC procedure for estimating the mean value of the QoI, we now discuss estimation of its central moments. We are particularly interested in estimating the second central moment, i.e., the variance of the QoI, which is a measure of its variability. It is possible to employ the standard MLMC estimator for these cases if we express the central moments in terms of raw moments. However, an approximation of the $r$-th central moments through the raw moments can lead to a large sampling error. In~\cite{Krumscheid2020} an MLMC estimator of central moments is proposed that employs h-statistics. The h-statistics is an unbiased central moment estimator, which provides a minimal variance for the level contributions. This approach allows to evaluate any central moment using closed-form expressions. In this approach, the MLMC estimator for the $r$-th central moment (with $r \geq 2$) is expressed as follows:
	\begin{equation}
	m_{r}^{MLMC}=\sum_{l=0}^{L} \Delta_lh_r=\sum_{l=0}^{L} \left(h_r(\bm{Q}^l_{N_l,M_l})-h_r(\bm{Q}^l_{N_l,M_{l-1}})\right)
	\end{equation} \label{eq:MLMCCentralMom}
	where $h_r(\bm{Q})$ is the $r$-th order h-statistic of $\bm{Q}$ and $h_r(\bm{Q}^0_{N_0,M_{-1}}):=0$. 
	
	The difference of two h-statistics between two consecutive levels for the second central moment is formulated as:
	
	\begin{equation}
	\Delta_lh_r = 
	\frac{N_lS_{1,1}^l - S_{0,1}^lS_{1,0}^l}{(N_l-1)N_l}
	\label{eq:deltaH}
	\end{equation}
	In~\cref{eq:deltaH} a power sum notation $S_{a,b}^l$ between the levels is introduced. It can be computed via the sample sum $\bm{X}_{N_l}^{l,+}$ and the sample difference $\bm{X}_{N_l}^{l,-}$ as follows:
	\begin{equation}
	S_{a,b}^l:=S_{a,b}(\bm{X}_{N_l}^{l,+},\bm{X}_{N_l}^{l,-}) = \sum_{i=1}^{N_l}(X_{i,N_l}^{l,+})^a(X_{i,N_l}^{l,-})^b
	\end{equation}
	
	with \begin{align}
		\bm{X}_{N_l}^{l,+}&=\bm{Q}^l_{N_l,M_l}+\bm{Q}^l_{N_l,M_{l-1}}\nonumber\\
		\bm{X}_{N_l}^{l,-}&=\bm{Q}^l_{N_l,M_l}-\bm{Q}^l_{N_l,M_{l-1}}\nonumber
	\end{align}
	
	The mean-square error of the MLMC estimator of~\cref{eq:MLMCCentralMom} can then be formulated:
	\begin{equation}
	\mathrm{MSE}(m_r^{MLMC}) = (\mu_{r,{Q}_{M_L}}-\mu_{r,Q})^2+ \sum_{l=0}^{L}\mathrm{Var}\left[\Delta_lh_r\right]
	\end{equation}
	where $\mu_{r,{Q}_{M_L}}$ is the $r$-th central moment of the QoI at level $L$. The expressions for the bias contribution and the variance terms for central moments of any order $r$ can be found in~\cite{Krumscheid2020}.
    
    The optimal number of the levels and the required sample number can be determined through a similar screening procedure to the one described above. In particular,~\cref{eq:exponentialFunctionConstants} can be used to fit the problem-dependent constants, the only difference being the expressions for the estimation of the bias and the variance terms.
	
}

%% file: sections/numericalExperiments/numericalExperiments.tex
\section{Numerical experiments}
\label{sec:numerics}
{ 
	In this section, the proposed workflow is applied to two additively manufactured lattice structures. The deterministic evaluation and experimental evaluation of the homogenized tensile behavior of both lattices has been examined in~\cite{Korshunova2020, Korshunova2020a}. The first example is a square lattice produced using Inconel\textregistered718 powder with a cell size of $600$ $\mu$m and the designed strut size of $96 \mu m$. The second example is an octet-truss structure with a considerably larger cell size of $4$ mm. The material of this specimen is stainless steel $SS 316L-0407$. The acquired CT images of the lattices provide the basis for the application of the proposed workflow. The parameter fitting and generation of statistically equivalent CT images are performed on a standard workstation with $i7-6500U$ CPU and 16 GB of RAM. The numerical simulations of mechanical behavior of all specimens has been performed on the Linux Cluster CoolMUC-2 at Leibniz Supercomputing Centre of Technical University of Munich.

	\subsection{Square grid lattice}
	\label{subsec:SquareLattice}
	{
		\textbf{2D model parameter identification}
		
		\noindent First, a two-dimensional slice of the square grid lattice is considered to evaluate the proposed model's applicability. The image is extracted from the $x-y$ plane of the whole specimen. The total size of the slice is $800\times400$ pixel with the pixel spacing $14.71\mu m$ (see~\cref{fig:unitCell2D}). A local unit cell size has a size of  $40\times40$ pixels with local coordinates $x^l\in[0,39]$ and $y^l\in[0,39]$. Considering the CT image's pixel spacing, the overall as-manufactured unit cell size is approximately $(603 \times 603)$ $\mu m$.
		
		\begin{figure}[H]
			\centering
			\includegraphics[scale=0.4]{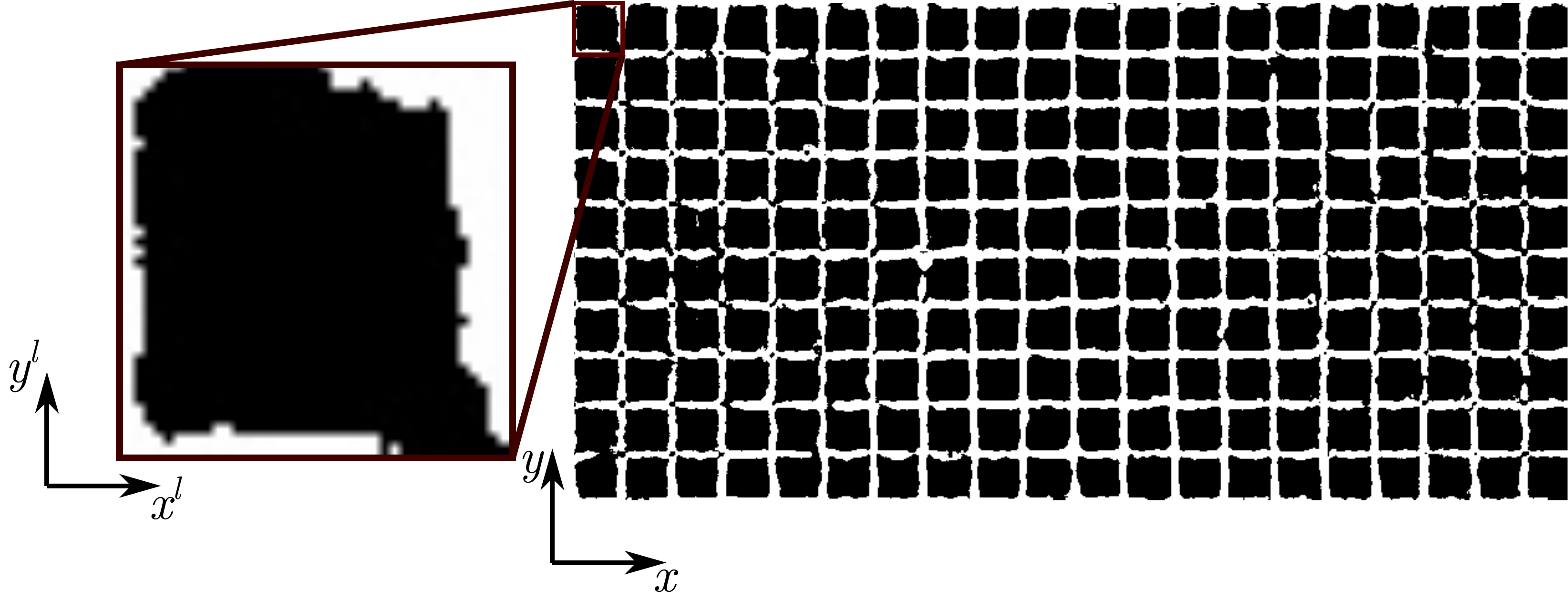}
			\caption{An example of a periodic structural cell with its local coordinate system in a 2D grid-like structure.}    
			\label{fig:unitCell2D}
		\end{figure}
		
		The local mean values at every pixel are computed according to~\cref{eq:meanBinary} and are depicted in~\cref{fig::2DLocalProbabilties}. As expected, the breakages in the corners of the square grids lower the mean value, while the middle part of the cells is precisely captured by a zero mean.
		\begin{figure}[H]
			\centering
			\includegraphics[scale=0.5]{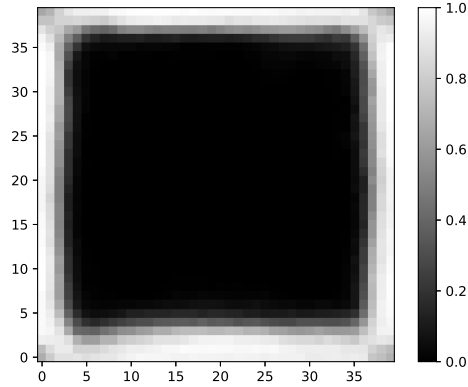}
			\caption{Mean values of the binary random field in the local coordinate system.}    
			\label{fig::2DLocalProbabilties}
		\end{figure}

		The correlation parameters are determined using the algorithm described in~\cref{sec:parameterIdentification}. To achieve the best fit of the correlation parameter, we consider $36$ lags in every spatial direction, close to the unit cell's total size. Then, only the data points with the mean value interval $\mu_Y(\bm{x})\in[0.1,0.9]$ are considered for the fitting procedure. The achieved fit of the computed binary field correlation and the Mat\'ern based correlation function is visualized in~\cref{fig:2DCovarianceFit}.  Due to a large amount of available data for every spatial lag, a specific graphical representation of the correlation fit is chosen. The sample correlation is depicted using a blue gradient indicating the density of the points within the particular region. The more points occur for a spatial lag; the darker is the color of this region. Analogously, the red gradient indicates the density of the chosen correlation model fit.
		
		\begin{figure}[H]
			\centering
			\captionsetup[subfigure]{oneside,margin={0cm,0cm},labelformat=empty}
			\hspace*{-10mm}\subfloat[]{
				\includegraphics[scale=0.38]{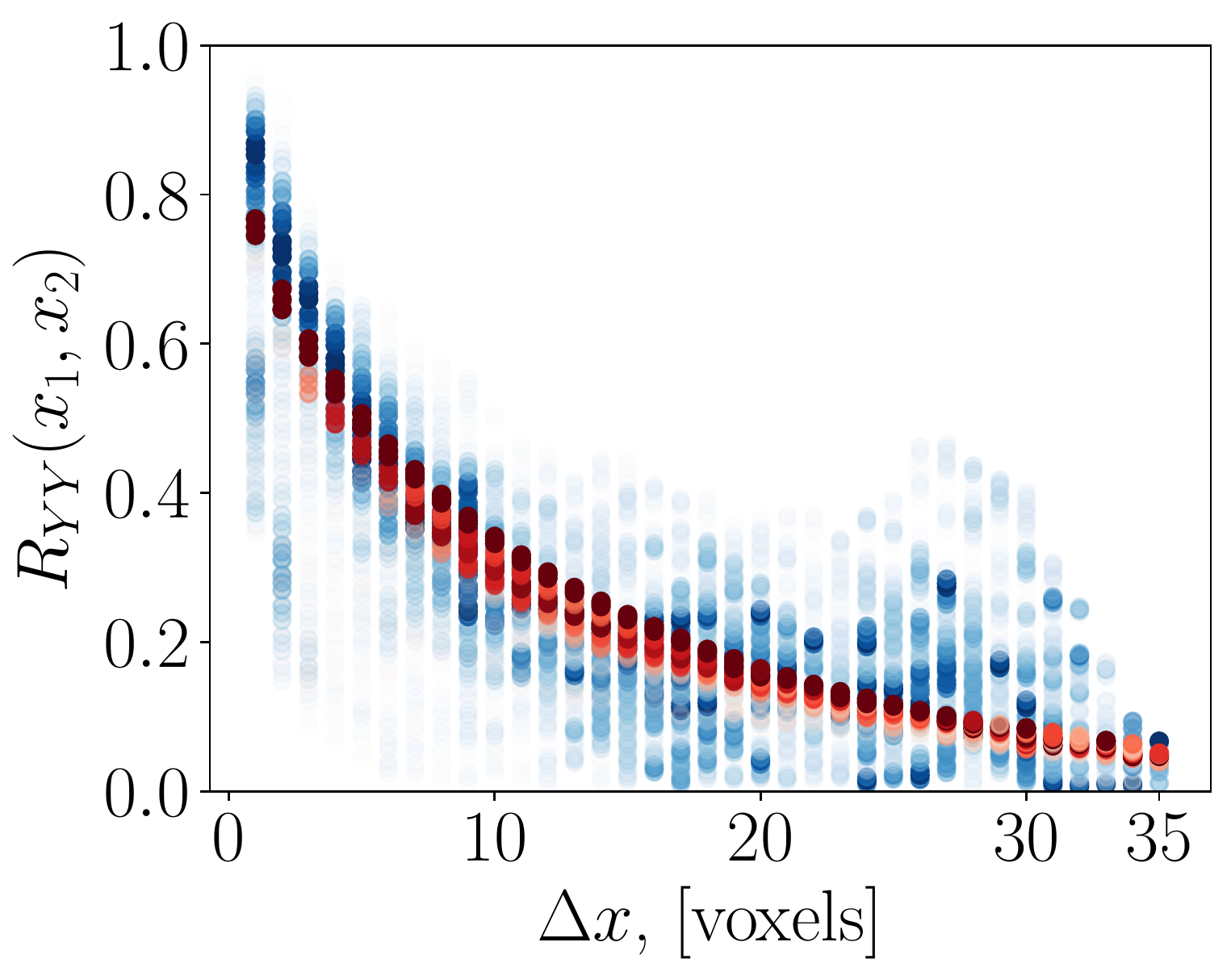} }
			\subfloat[]{
				\includegraphics[scale=0.38]{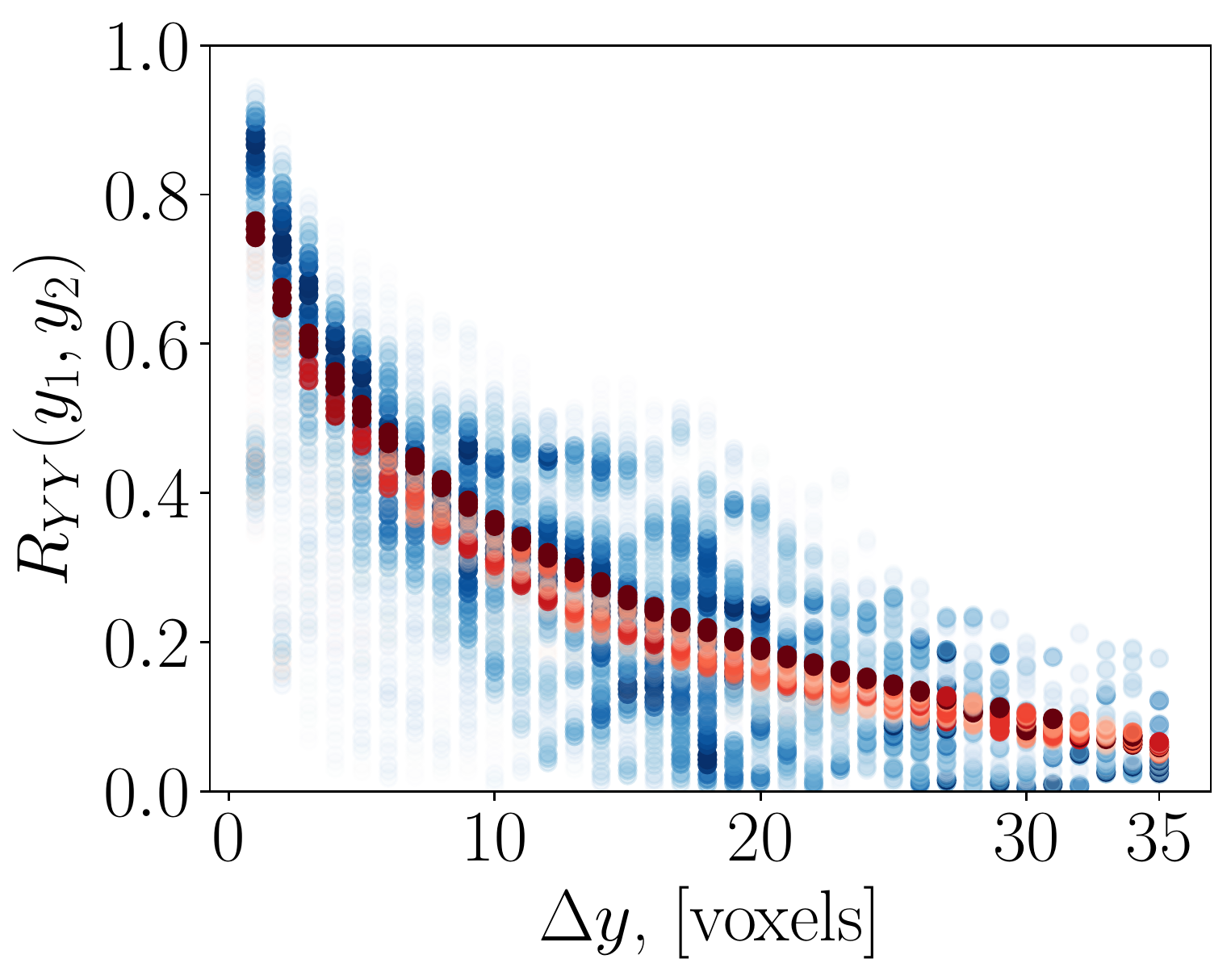} }
			\caption{Fitted and computed auto-correlation of a 2D model ( blue - sample correlation; red - Mat\'{e}rn model fit).}
			\label{fig:2DCovarianceFit}
		\end{figure}
		
		The qualitative trend of the binary correlation is captured well for all spatial lags. However, the model does not cover the whole spread of the correlation values and omits the extreme values at every lag. As this work focuses on the homogenized response of the considered lattices, we consider it a satisfactory fit as an overall homogenized correlation trend is well captured. The identified correlation model parameters are presented in~\cref{tab:parametersIdentified} together with their standard deviations. The standard deviation is computed from the performed least-square optimization procedure.
		\begin{table}[H]
			\centering
			\begin{tabular}{|c|c|c|c|c|}
				\hline
				Design parameter & $l_x,[pixels]$& $l_y,[pixels]$ &$\nu_x,[-]$ & $\nu_y,[-]$  \\\hline
				36 lags & 15.00 $\pm$ 0.23	 & 16.92 $\pm$ 0.31 & 0.52 $\pm$ 0.01 & 0.48 $\pm$ 0.01 \\\hline
			\end{tabular}
			\caption{Correlation parameter identification for a 2D slice.}
			\label{tab:parametersIdentified}
		\end{table}
		
		Next, 10 000 realizations of a 2D slice were generated using the method described in~\cref{sec:numericalgeneration}. The total generation time for all realizations is approximately 5 minutes. Three different generated slices are shown in~\cref{fig:2DRealizaiton}.
		
		\begin{figure}[H]
			\centering
			\captionsetup[subfigure]{oneside,margin={0cm,0cm},labelformat=empty}
			\subfloat[(a) Original sample]{
				\includegraphics[scale=0.28]{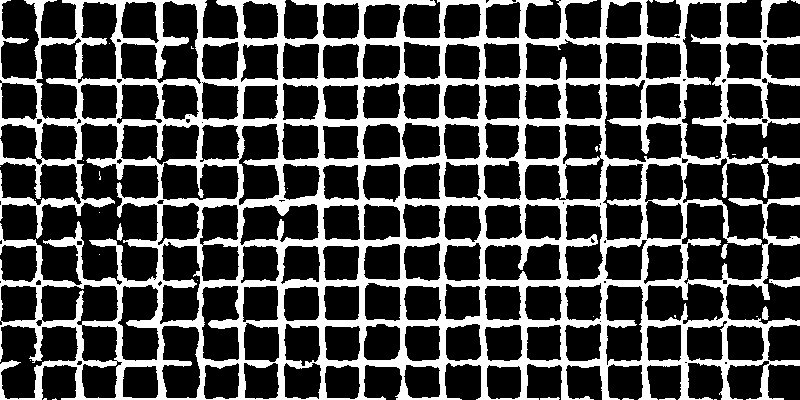} }
			\hspace*{0.2cm}
			\subfloat[{(b) Realization 1}]{
				\includegraphics[scale=0.28]{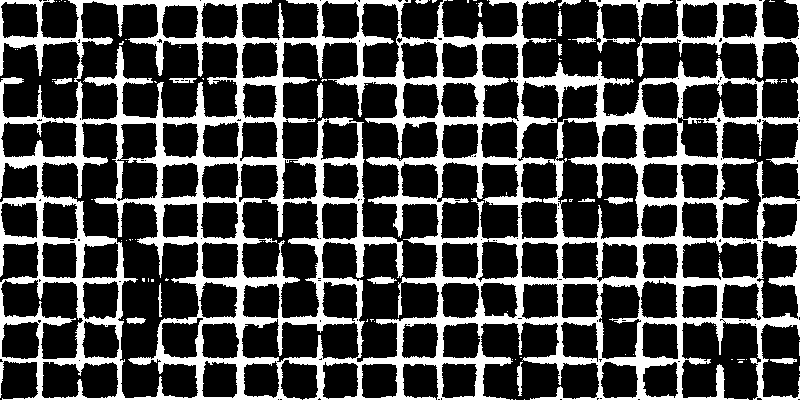} }\\
			\subfloat[{(c) Realization 2}]{
				\includegraphics[scale=0.28]{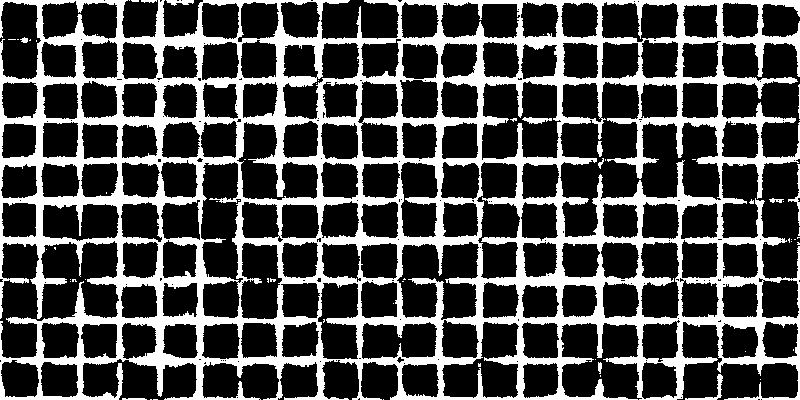}}
			\hspace*{0.2cm}
			\subfloat[{(d) Realization 3}]{
				\includegraphics[scale=0.28]{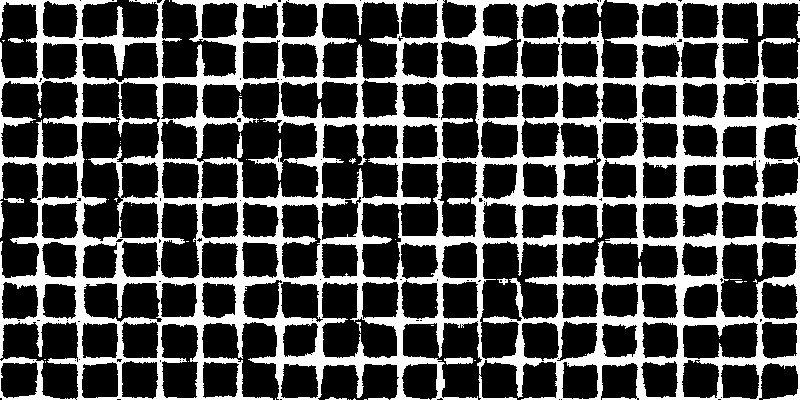} }
			\caption{Generated realizations of a 2D slice based on fitted correlation parameters.}
			\label{fig:2DRealizaiton}
		\end{figure}
		
		As the topological connectivity of a two-dimensional slice of a three-dimensional structure cannot be guaranteed, these images' mechanical analysis has not been performed. Instead, the porosity of the realizations is evaluated. The original porosity of the considered slice is $0.7237$. The variability is evaluated on 10 000 realizations and is presented in~\cref{fig:porosityDistribution}. The plot shows the normalized histogram and the normal fit. The estimated mean value is $\mu=0.726$ and the standard deviation is $\sigma=0.006$. Overall, the variability of porosity is small. The original value from the CT slice lies in the high probability mass region of the estimated distribution and is close to the estimated mean value.
		
		\begin{figure}[H]
			\centering
			\includegraphics[scale=0.7,trim={0.8cm 0.2cm 2cm 1.8cm},clip]{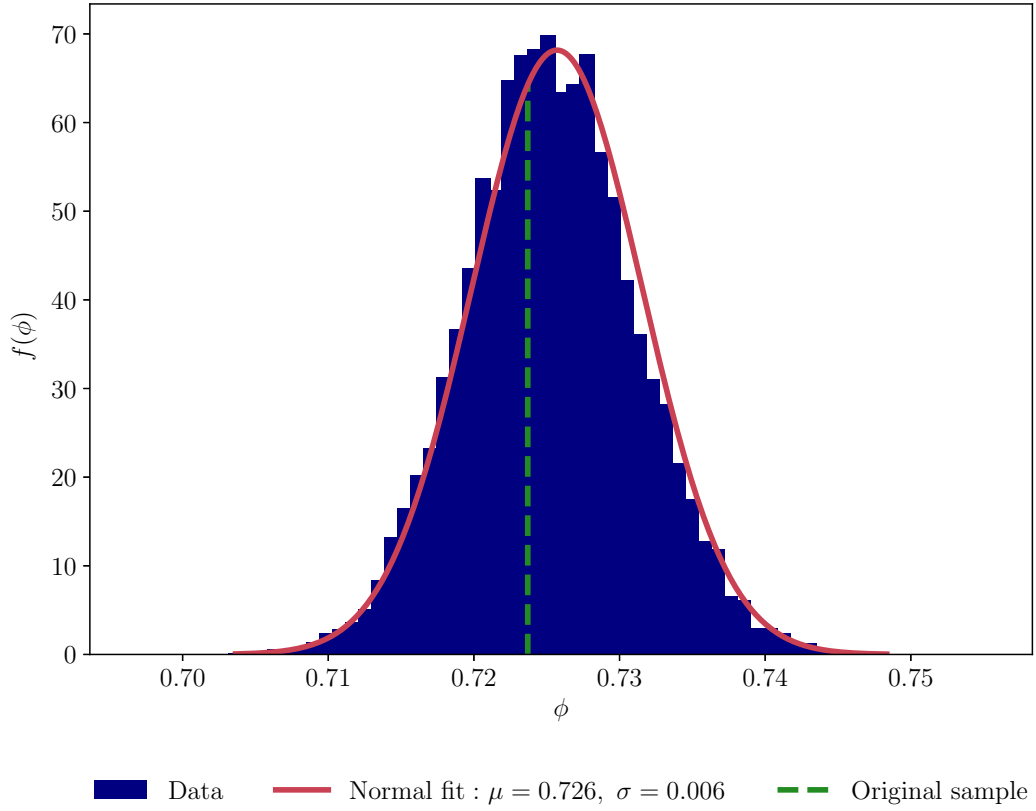}
			\caption{Porosity distribution for the 2D realizations.}    
			\label{fig:porosityDistribution}
		\end{figure}

		\textbf{3D model parameter identification}
		
		Next, the whole CT scan of the same square lattice structure is analyzed using the proposed model. The total size of the considered model is $400\times800\times368$ voxel with the same voxel spacing $14.71\mu m$. The size of the local cell is $40\times40\times 368$ voxels with local coordinates $x^l\in[0,39]$, $y^l\in[0,39]$ and $z^l\in[0,368]$ (see~\cref{fig:unitCell3D}).

		\begin{figure}[H]
			\centering
			\includegraphics[scale=0.4]{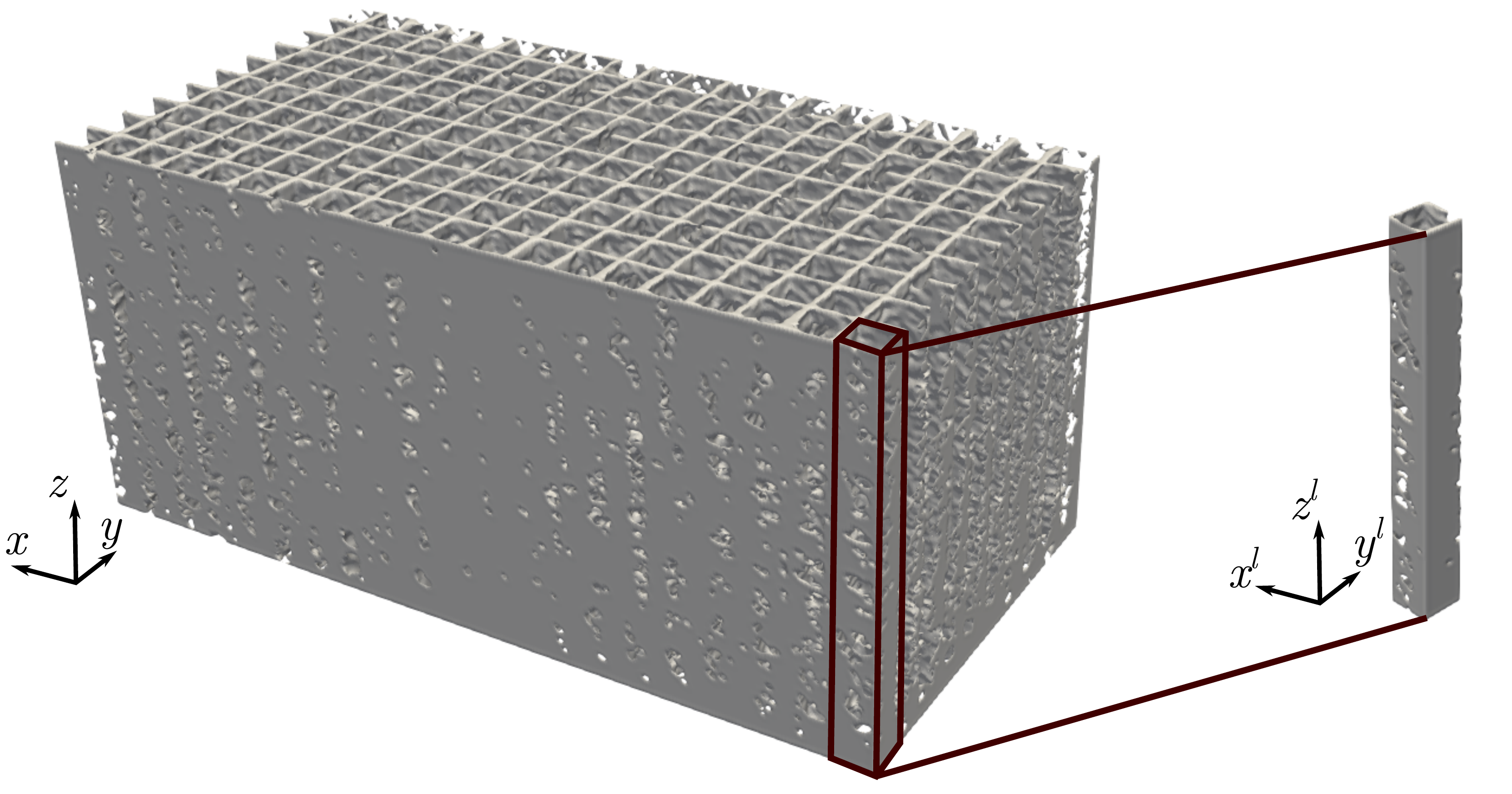}
			\caption{An example of a periodic unit cell with its local coordinate system in the 3D square lattice structure~\cite{Korshunova2020}.}    
			\label{fig:unitCell3D}
		\end{figure}

		The local probabilities are computed in a similar manner as for the 2D case considering the local unit cells extracted from the volume. As the data set already becomes very large, not all spatial lags can be considered in the fitting procedure for the parameters of the correlation model. To estimate the necessary number of spatial lags, we perform a convergence study on the correlation parameters. To this end, we increase the number of spatial lags in every direction and observe the estimated correlation parameters' behavior. 
		
		\newcommand{\graphDir}{./sections/numericalExperiments/Figures3D/Tikz/graphs}
		\newcommand{\dataDir}{./sections/numericalExperiments/Figures3D/Tikz/data}
		\begin{figure}[H]
			\centering
			\includegraphics[width=\textwidth,height=0.4\textheight]{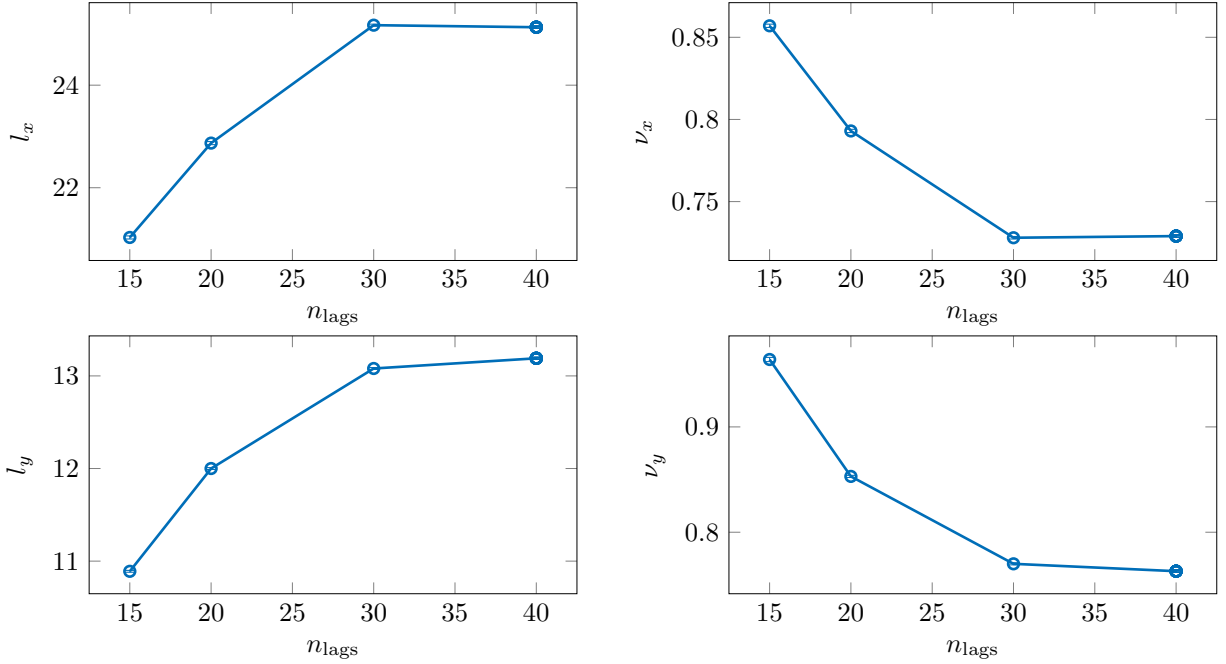}
			\caption{Convergence of the estimated correlation parameters in $x$ and $y$ direction.}    
			\label{fig:CorrelationConvergenceXYSiemens}     
		\end{figure}
		\begin{figure}[H]
			\centering
			\includegraphics[width=\textwidth,height=0.4\textheight]{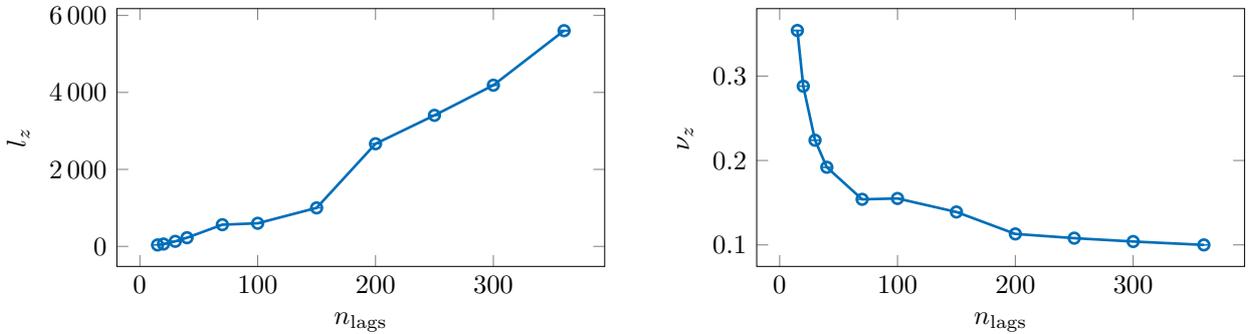}
			\caption{Convergence of the estimated correlation parameters in $z$ direction.}    
			\label{fig:CorrelationConvergenceZSiemens}     
		\end{figure}
		\Cref{fig:CorrelationConvergenceXYSiemens} indicates that after $30$ spatial lags both the correlation length and the smoothness parameters in $x$ and $y$ directions do not change significantly. Thus, for further computations we consider $30$ spatial lags.  However,~\cref{fig:CorrelationConvergenceZSiemens} shows that the correlation length in $z$ direction is constantly increasing with the number of considered lags. The smoothness parameter $\nu_z$ can be considered converged after $300$ lags. This could potentially indicate that the objective function in~\cref{eq:MinimizationFunction} is rather flat in the achieved minimum of the smoothness parameter. Thus, the change in the correlation length for the fixed smoothness parameter does not change the behavior of the residual significantly. To further investigate this, we compute the residual as in~\cref{eq:MinimizationFunction} for $300$ lags on a pre-defined grid of values for correlation length and smoothness parameter. \Cref{fig:ObjectiveFunction3DZDirection} reveals a manifold of near-optimal values of $(l_z,\nu_z)$, indicating that there is a number of combinations of the two parameters that leads to near-optimal parameter fit. The non-identifiability of unique smoothness and correlation length parameters of correlation models belonging to the Mat\'ern class has been discussed elsewhere, e.g., in \cite{DeOliveira2000}. The plot also shows that there exists a minimum for the smoothness parameter, which corresponds well to the results obtained above. However, the objective function remains almost constant once the smoothness minimum is achieved for $l_z \in (2000,10000)$. The large values of the correlation length in $z$ direction are related to the technique used to generate the AM product. In particular, the square grid structure is produced such that the in-plane grid is extruded in $z$ direction. Thus, it is natural to expect a larger correlation length in this direction. The optimal value obtained in this case is $(l_z,\nu_z)=(0.104,4187)$. For further computations we fix the number of spatial lags in $z$ direction to $300$.
 		
		\begin{figure}[H]
			\centering
			\resizebox{\textwidth}{!}{\includegraphics[scale=0.3]{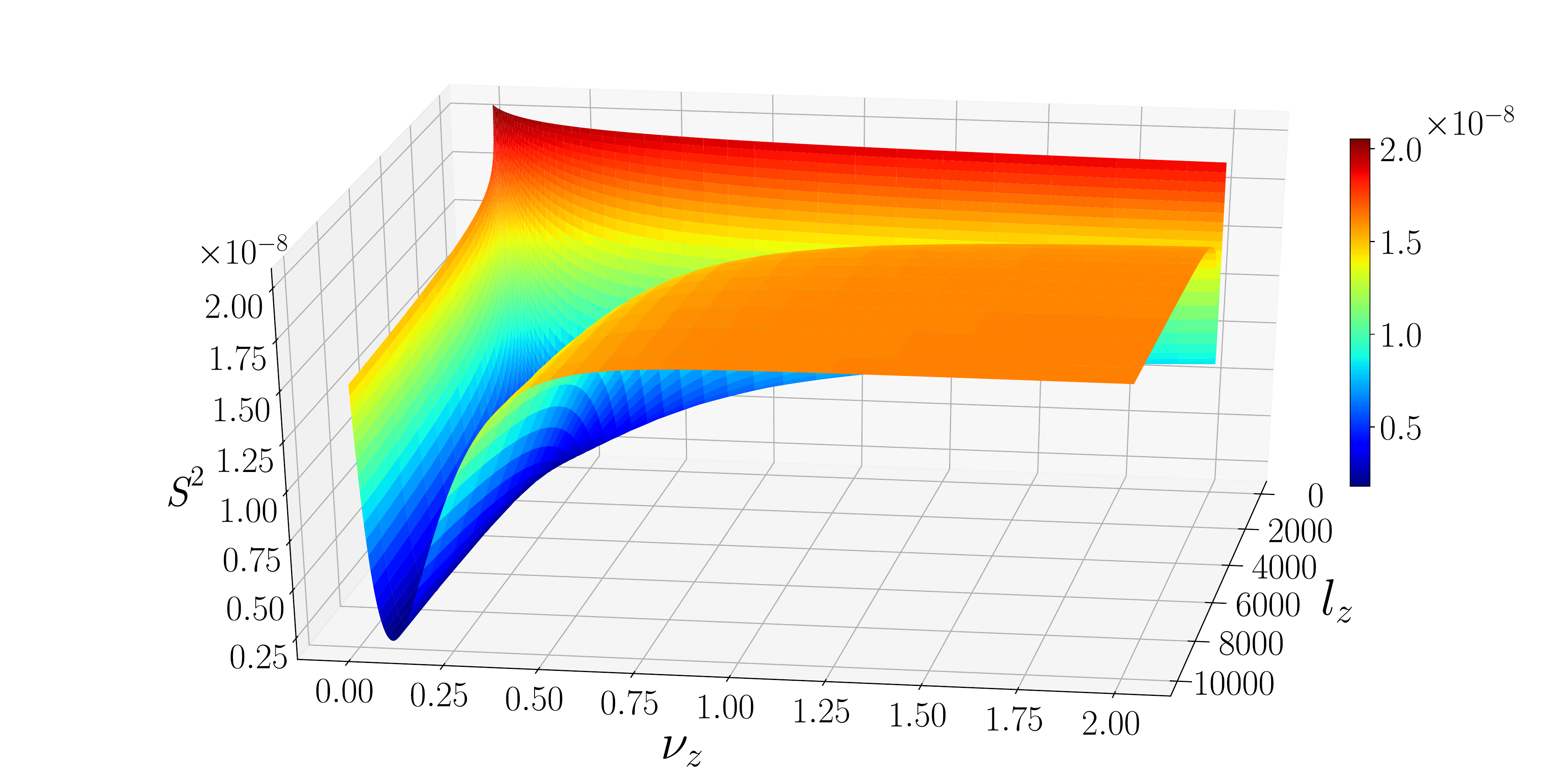}}
			\caption{Behavior of the objective function for fitting the correlation function in z-direction with $300$ lags.}    
			\label{fig:ObjectiveFunction3DZDirection}
		\end{figure}

		The achieved fit of computed covariance and the considered model of Mat\'ern covariance for $30$ spatial lags in $x$ and $y$ direction and $300$ lags in $z$ direction is shown in~\cref{fig:3DCovarianceFit300}. Similar to the 2D case, the model qualitatively follows the trend of the binary covariance. Nevertheless, the model does not capture the spread in the correlation values but rather homogenizes the solution.
		The design correlation parameters are summarized in~\cref{tab:parametersIdentified3D} together with their standard deviations.

		\begin{figure}[H]
			\centering
			\captionsetup[subfigure]{oneside,margin={0cm,0cm},labelformat=empty}
			\hspace*{-10mm}\subfloat[]{
				\includegraphics[scale=0.38]{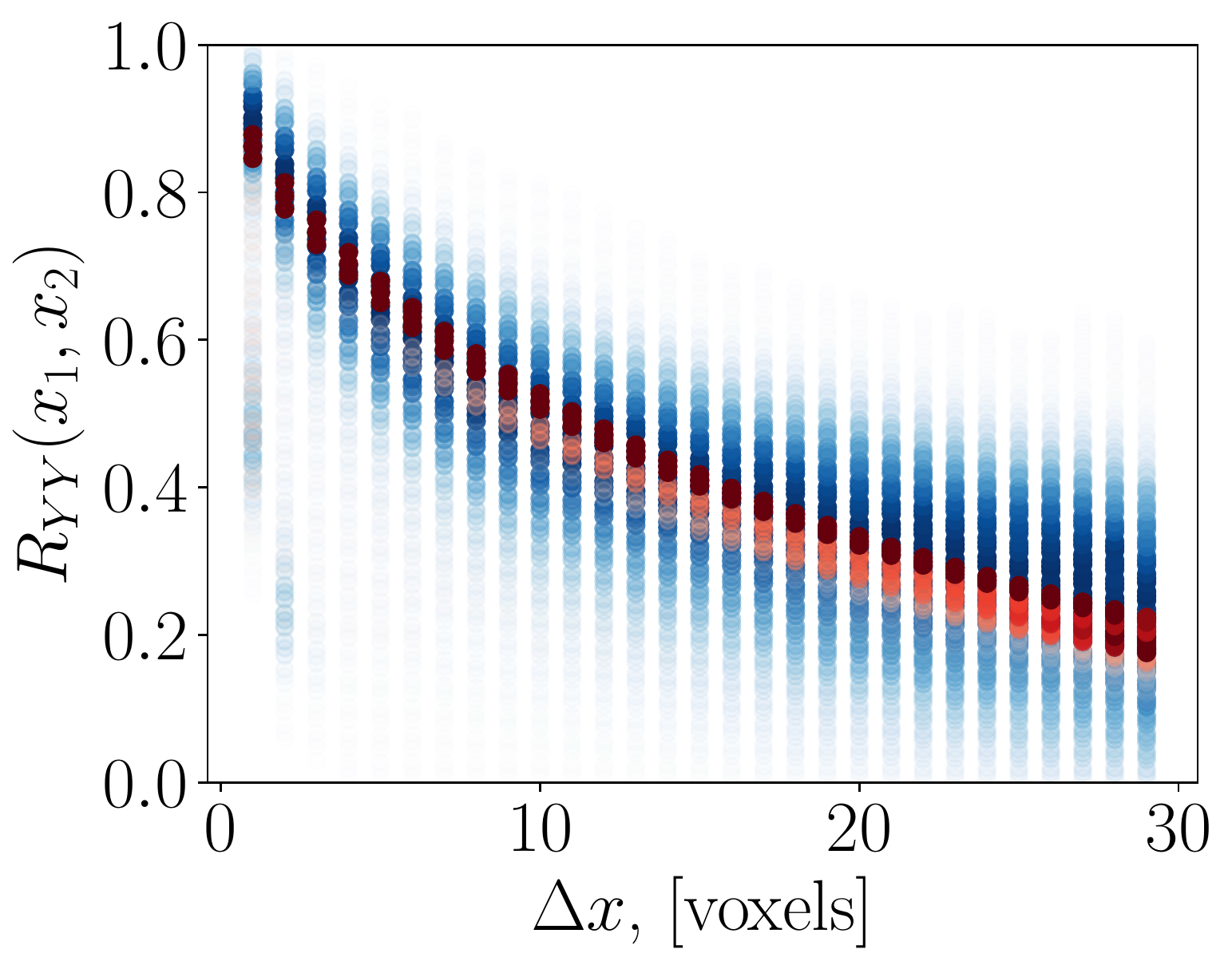} }
			\subfloat[]{
				\includegraphics[scale=0.38]{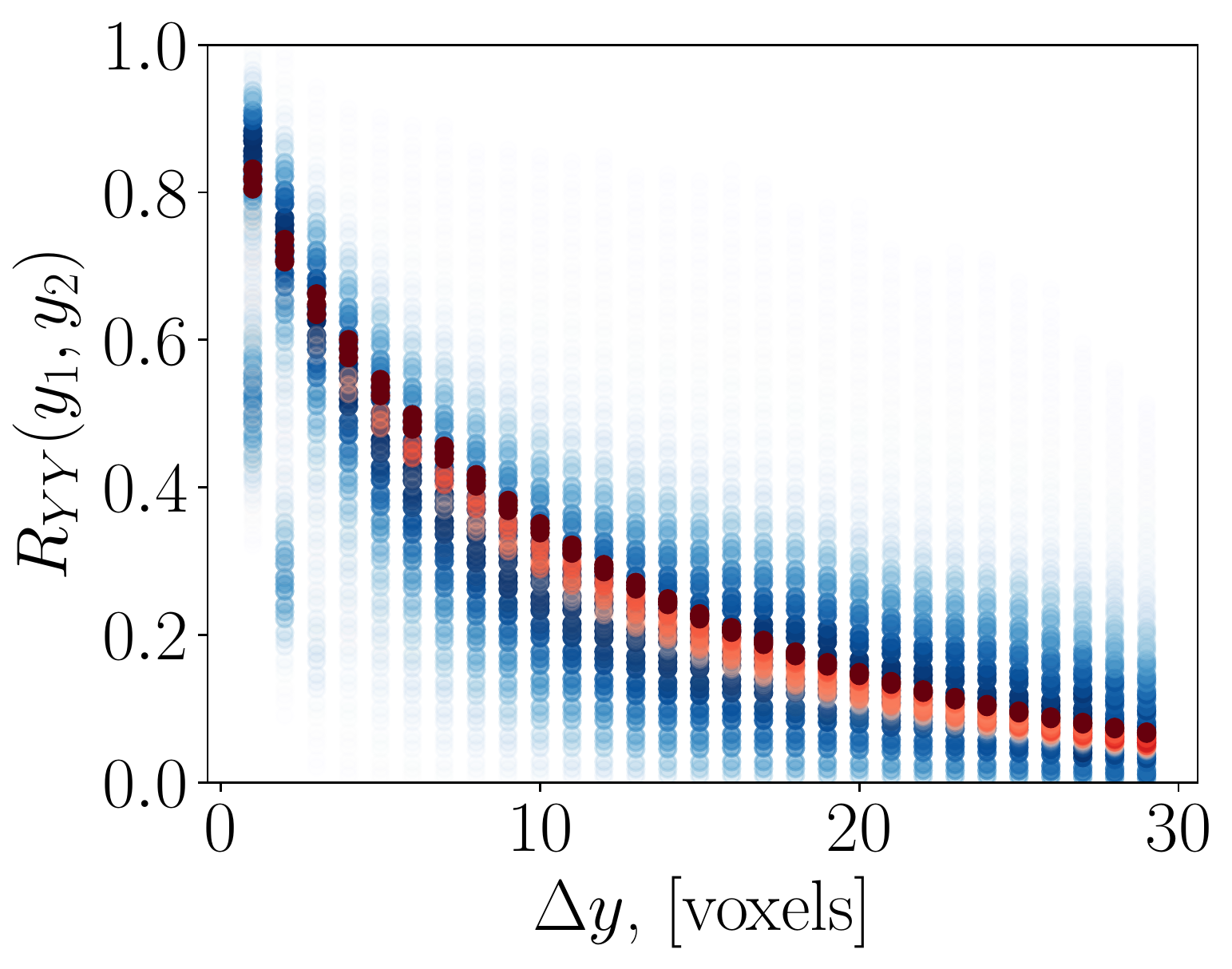} }
			\subfloat[]{
				\includegraphics[scale=0.38]{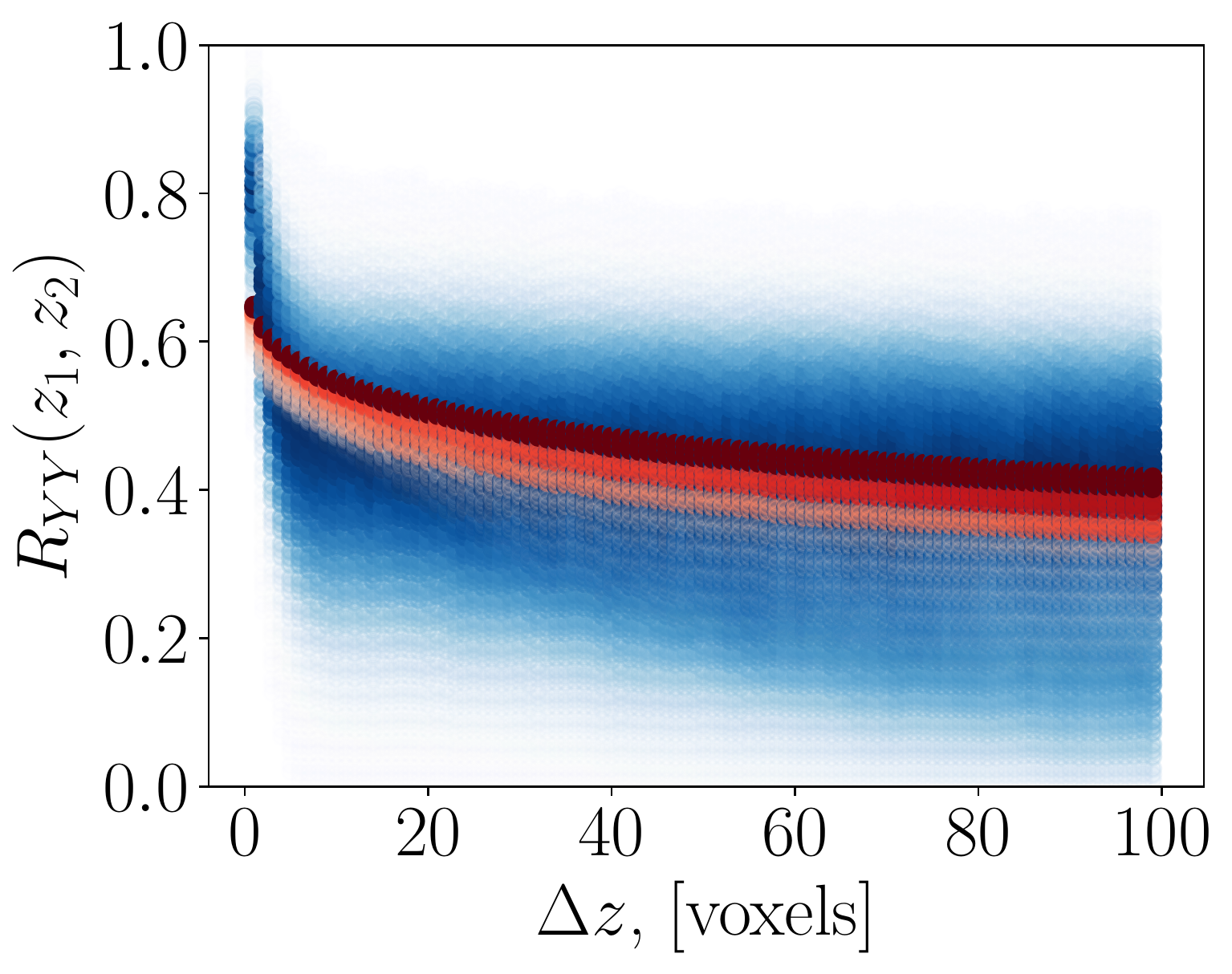}}
			\caption{Fitted and computed auto-correlation of a 3D model ( blue - sample correlation; red - Mat\'{e}rn model fit).}
			\label{fig:3DCovarianceFit300}
		\end{figure}

		\begin{table}[H]
			\centering
			\begin{tabular}{|c|c|c|c|c|c|c|}
				\hline
				$l_x, [voxels]$&$l_y, [voxels]$&$l_z, [voxels]$& $\nu_x, [-]$& $\nu_y, [-]$  & $\nu_z, [-]$ \\\hline
				$25.17 \pm 0.02$ & $13.08 \pm 0.01$ & $4186.99\pm 10.31 $ & $0.728 \pm 0.001$ & $0.770 \pm 0.001$ & $ 0.104 \pm 0.001$\\\hline
			\end{tabular}
			\caption{Correlation parameter identification for a 3D structure using 300 lags.}
			\label{tab:parametersIdentified3D}
		\end{table}
		
		Having determined the correlation parameters, the method described in~\cref{sec:numericalgeneration} is applied to generate realizations of the 3D random microstructure. The generation of every volume takes on average 6 seconds for the considered domain size. A representative realization is shown in~\cref{fig:3DCoronal,fig:3DAxial,fig:3DSagittal}. The original sample together with one realization are depicted in~\cref{fig:3DFull}.

		\begin{figure}[H]
			\centering
			\captionsetup[subfigure]{oneside,margin={0cm,0cm},labelformat=empty}
			\hspace*{-10mm}\subfloat[(a) Original sample]{
				\includegraphics[scale=0.28]{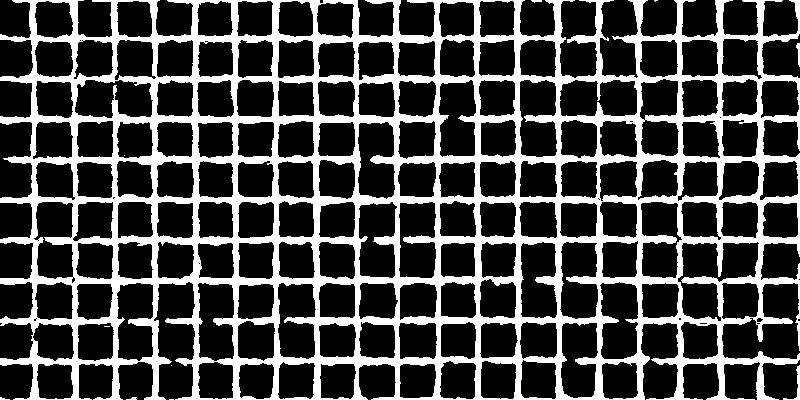} }
			\hspace*{-0.15cm}
			\subfloat[{(b) Realization}]{
				\includegraphics[scale=0.28]{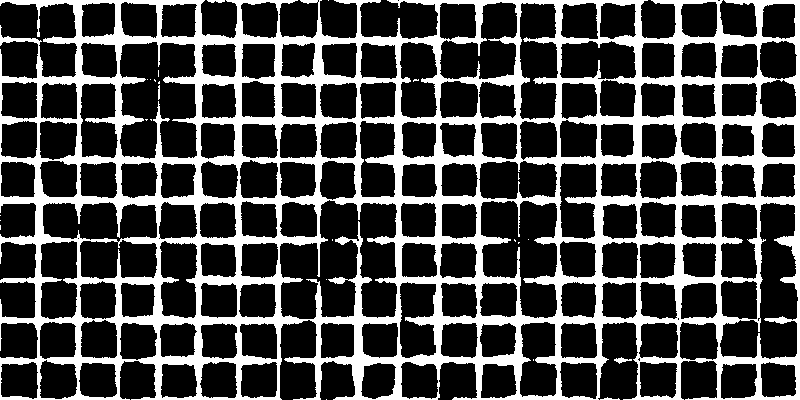} }
			\caption{An example of a realization based on fitted correlation parameters: coronal slice 71 in the 3D volume.}
			\label{fig:3DCoronal}
		\end{figure}
		\begin{figure}[H]
			\centering
			\captionsetup[subfigure]{oneside,margin={0cm,0cm},labelformat=empty}
			\subfloat[(a) Original sample]{
				\includegraphics[scale=0.28]{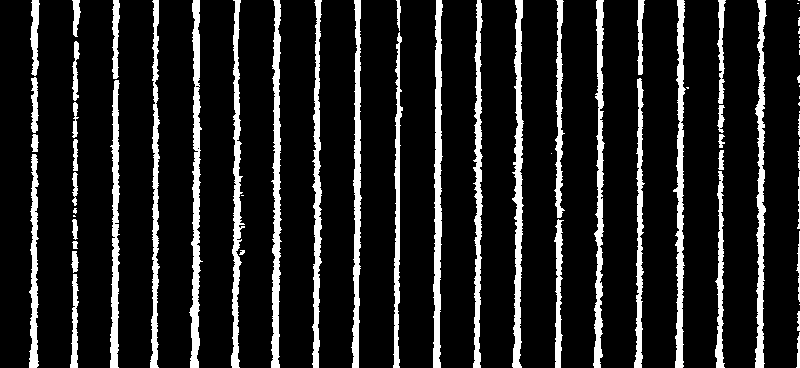} }
			\hspace*{0.2cm}
			\subfloat[{(b) Realization}]{
				\includegraphics[scale=0.28]{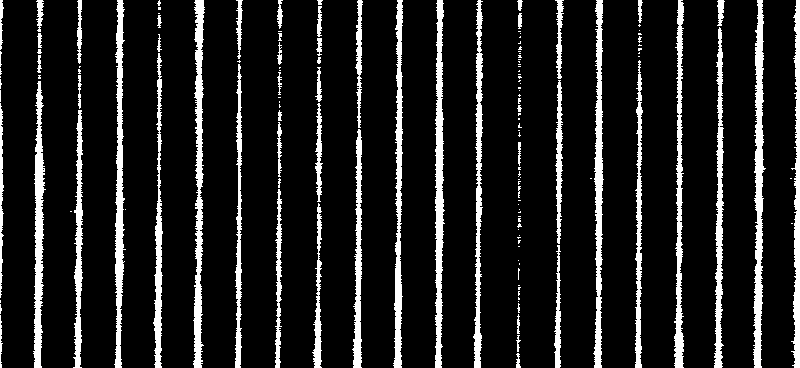} }
			\caption{An example of a realization based on fitted correlation parameters: axial slice 213 in the 3D volume.}
			\label{fig:3DAxial}
		\end{figure}
		\begin{figure}[H]
			\centering
			\captionsetup[subfigure]{oneside,margin={0cm,0cm},labelformat=empty}
			\subfloat[(a) Original sample]{
				\includegraphics[scale=0.32]{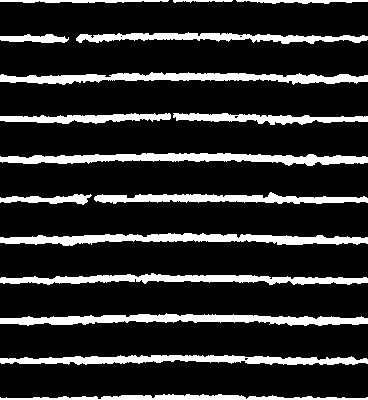} }
			\hspace*{0.2cm}
			\subfloat[(b) Realization]{
				\includegraphics[scale=0.32]{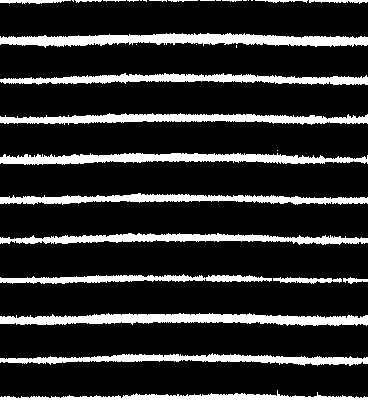} }
			\caption{An example of a realization based on fitted correlation parameters: sagittal slice 413 in the 3D volume.}
			\label{fig:3DSagittal}
		\end{figure}		
		
		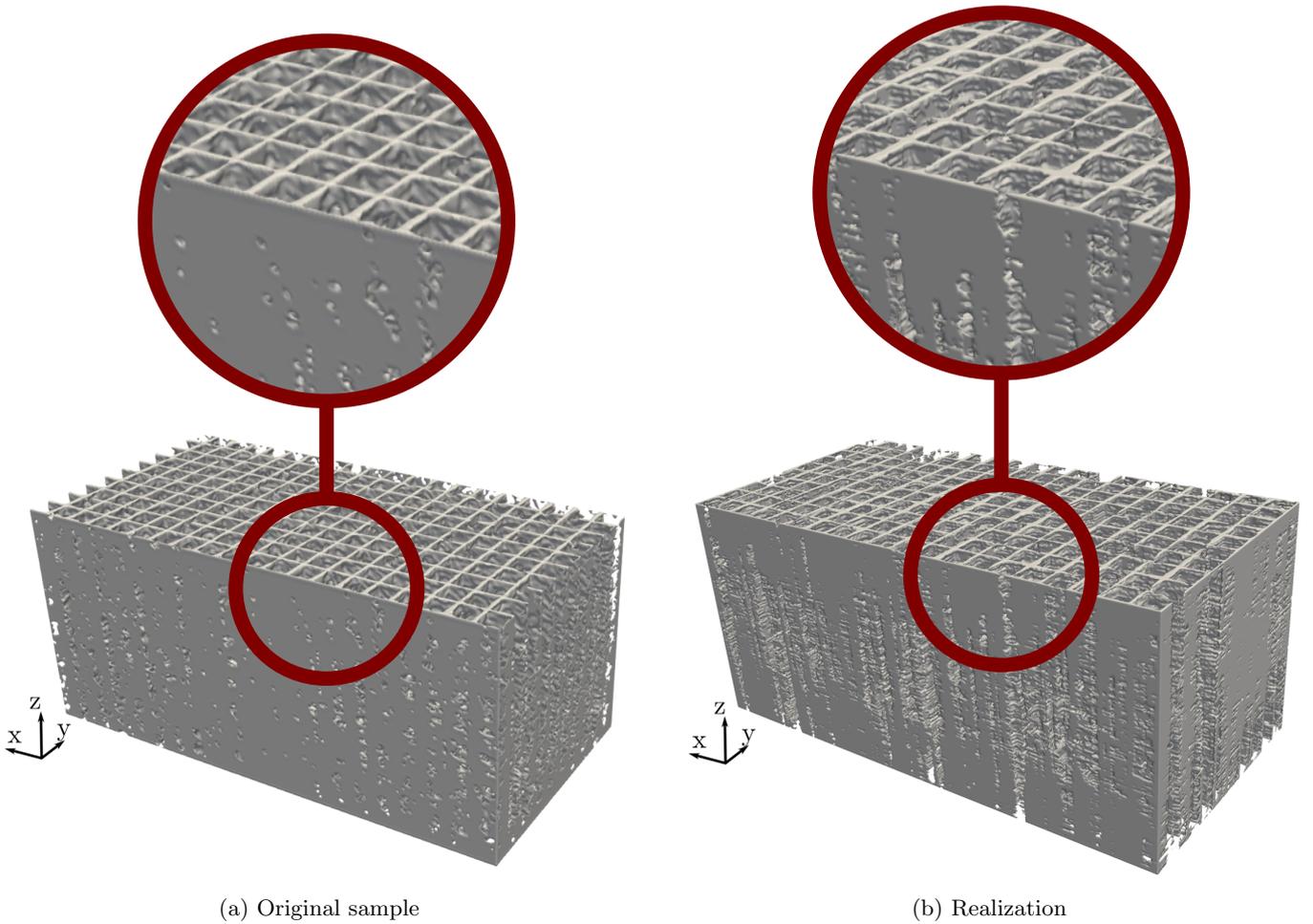
\begin{figure}[H]
			\centering
			\captionsetup[subfigure]{oneside,margin={0cm,0cm},labelformat=empty}
			\hspace*{-0.6cm}\subfloat[(a) Original sample]{
				\begin{tikzpicture}[spy using outlines={circle,red!50!black, magnification=2, connect spies}]
        		\node {\includegraphics[scale=0.35]{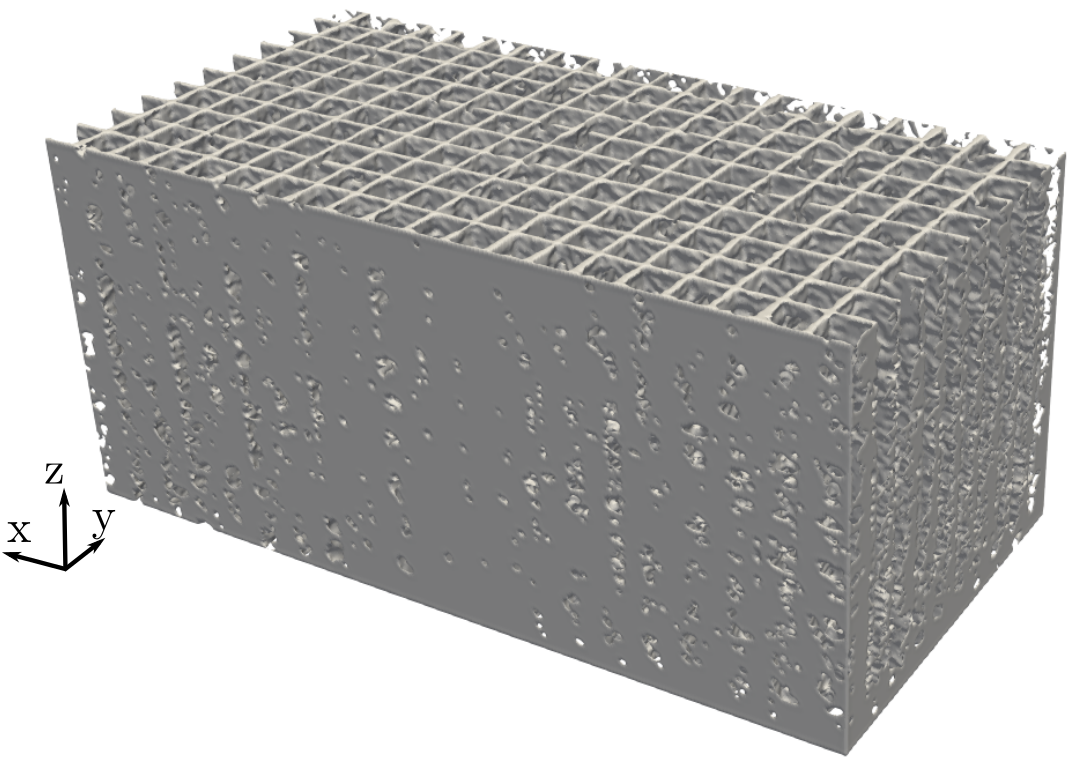} };
        		\coordinate (spypoint1) at (0.1,0.9);
        		\coordinate (magnifyglass1) at (0.1,6.0);
        		\spy [every spy on node/.append style={line width=2mm},size=5cm,spy connection path={\draw[line width=2mm,red!50!black] (tikzspyonnode) -- (tikzspyinnode);}] on (spypoint1)
        		in node[fill=white,line width=2mm] at (magnifyglass1);
        		\end{tikzpicture}
				 }
			\subfloat[{(b) Realization}]{
				\begin{tikzpicture}[spy using outlines={circle,red!50!black, magnification=2, connect spies}]
        		\node {\includegraphics[scale=0.35]{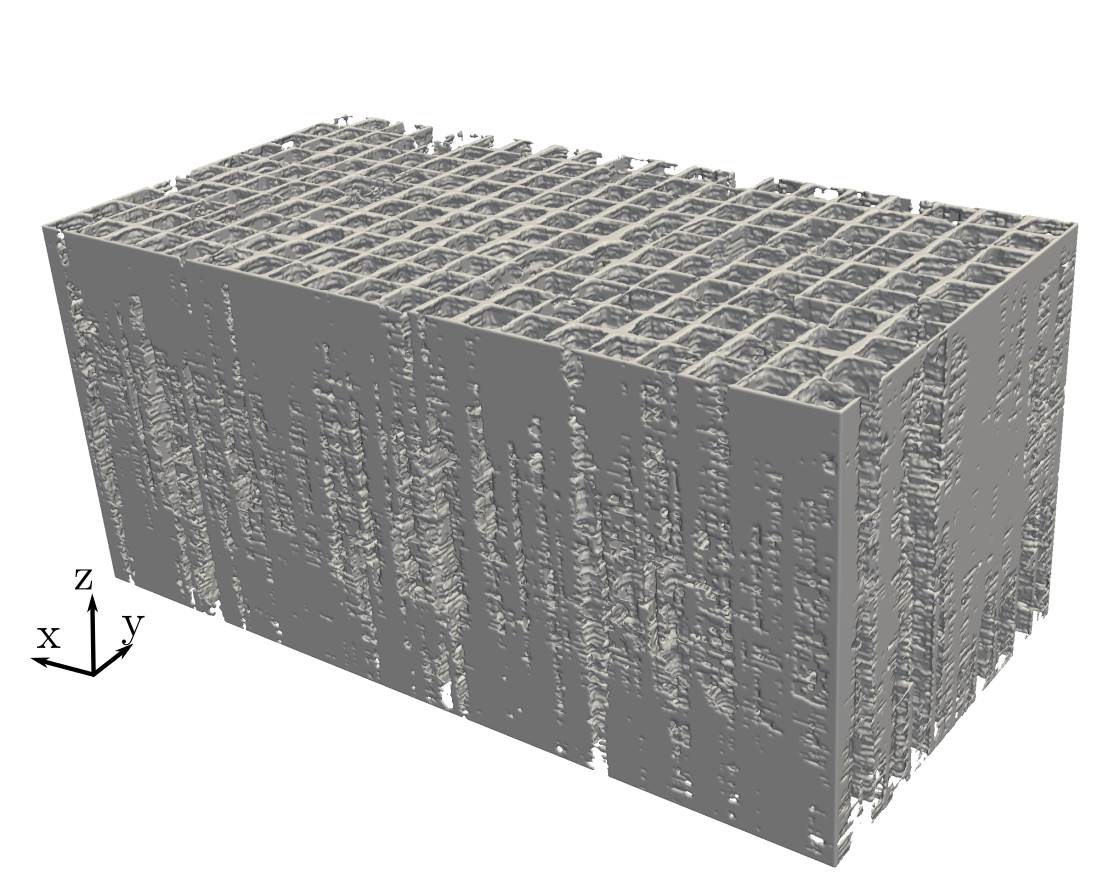} };
        		\coordinate (spypoint1) at (0.1,0.7);
        		\coordinate (magnifyglass1) at (0.1,6.0);
        		\spy [every spy on node/.append style={line width=2mm},size=5cm,spy connection path={\draw[line width=2mm,red!50!black] (tikzspyonnode) -- (tikzspyinnode);}] on (spypoint1)
        		in node[fill=white,line width=2mm] at (magnifyglass1);
        		\end{tikzpicture}}
			\caption{An example of a model realization based on fitted correlation parameters: full 3D model.}
			\label{fig:3DFull}
		\end{figure}
		Visually, the overall geometry is similar to the original specimen. The underlying structure is well preserved, while the small features are varied. The distribution of the porosity computed with 5000 samples is shown in~\cref{fig:porosityDistribution3D}. The porosity of the original specimen is $\phi=0.7293$, which is close to the mean value computed with samples from the proposed model. The computed standard deviation again appears to be relatively small. However, the macroscopic porosity is not the only determining factor for the macroscopic mechanical behavior. The topological features, such as the strut connectivity, can have a significant influence on the response of the final product.
		
		\begin{figure}[H]
			\centering
			\includegraphics[scale=0.7,trim={0.8cm 0.2cm 2cm 1.8cm},clip]{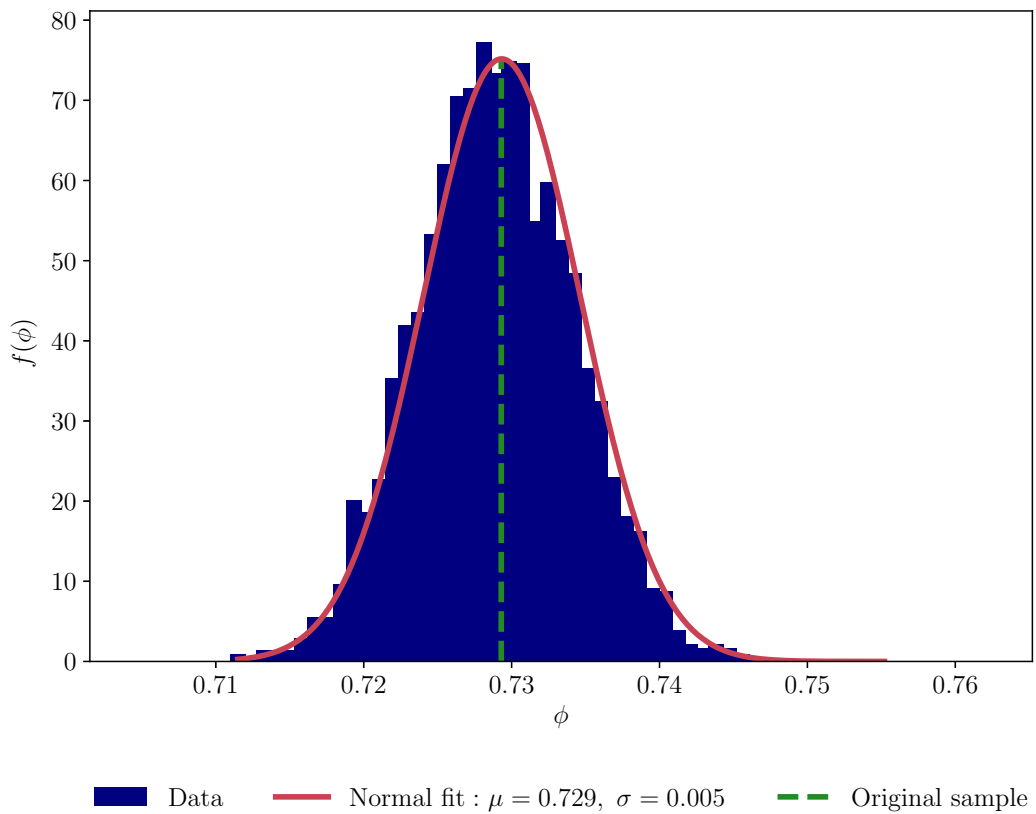}
			\caption{Porosity distribution for 3D square lattice realizations.}    
			\label{fig:porosityDistribution3D}
		\end{figure}

	\textbf{Multilevel Monte Carlo analysis of the homogenized mechanical behavior}
	  
	 The considered square lattice specimen was analyzed experimentally and numerically. The results are discussed in details in~\cite{Korshunova2020}. Here, the achieved results are briefly recapitulated.

The homogenized Young's modulus of three square lattice specimens (indicated as $600$ L1-L3 in~\cref{tab::ExperimentalResults600}) was determined experimentally. All specimen were printed with the same geometrical input, the same process parameters, and the same manufacturing plate. As the scale of the square lattice is only $96\mu m$, a considerable variation was observed. In particular, the three specimens showed a spread of the homogenized Young's modulus from $15\,339$ MPa to $26\,731$ MPa. The specimen that was CT scanned before the experimental testing and analyzed numerically, is referred to as $600$ L2 in~\cref{tab::ExperimentalResults600}. This specific specimen in the experiment showed an estimated Young's modulus lays between $20 \, 851$ and $25\,915$ MPa. 

\begin{table}[H]
	\centering
	\begin{tabular}{| c | c |}
		\hline        
		Specimens & $E_{xx}$, [MPa] \\    \hline
		600 L1-L3 & 15 339...26 731  \\\hline
		600 L2 & 20 851...25 915  \\\hline
	\end{tabular}
	\caption{Specimen 600: Experimental results of a tensile test on three specimens~\cite{Korshunova2020}.}
	\label{tab::ExperimentalResults600}
\end{table}       

The numerical analysis of the CT scan of the $600$ L2 specimen was performed through applying the FCM directly on the provided image. The coarsest discretization level $L=0$ consists of $100\times46\times100$ finite cells with the polynomial degree $p=1$. At the level $L=1$ the $h-$refinement is performed to obtain $200\times46\times100$ finite cells with polynomial degree $p=1$. With every higher level, $L>2$, the finite cells' polynomial degree is raised from $p=1$ to $p=5$. The deterministic evaluation of the homogenized Young's modulus of the considered specimen is shown in~\cref{fig::ConvergenceDNS600}. The relative error is computed with respect to an overkill solution obtained with $800\times368\times400$ cells of polynomial degree $p=4$. The achieved results agree well with the experimental data providing a reliable estimate of the homogenized Young's modulus that lies within the two experimental estimates.

\renewcommand{\graphDir}{./sections/numericalExperiments/TensileGrid/graphs}
\renewcommand{\dataDir}{./sections/numericalExperiments/TensileGrid/data}

\begin{figure}[H]
	\centering
	\vspace*{-0.01cm}
	\includegraphics[height=0.37\textheight,width=0.70\textwidth]{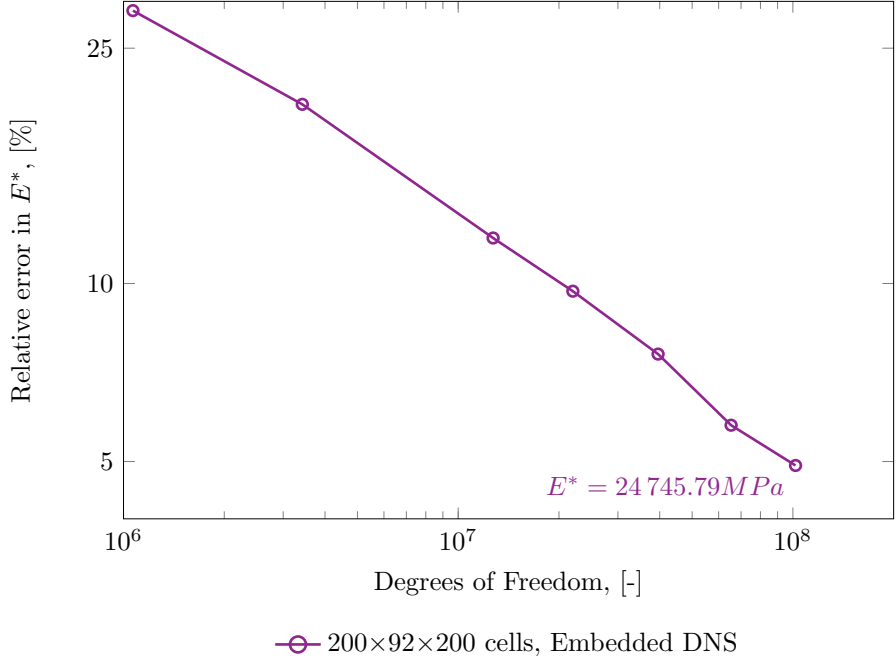}
	\caption{Specimen 600: Convergence of the directional Young's modulus $E^*$~\cite{Korshunova2020}.}    
	\label{fig::ConvergenceDNS600}
\end{figure}

 Next, we evaluate the influence of the geometrical variability on the homogenized Young's modulus. The mechanical analysis of the 3D realizations with the correlation parameters determined above is performed using the same Finite Cell discretization levels as for the original CT image. As the constants $c$ in~\cref{eq:exponentialFunctionConstants} are not known a priori for this problem, the screening procedure is held to evaluate the optimal hierarchy parameters.   

For the screening procedure, a few samples on the levels $\overline{L}=\{0,1,2, 3, 4, 5\}$ are evaluated. The fit of the constants for $\Delta_Lh_p, V_{l,p}$ and $\mathrm{cost}(Q_{M_l})$ are performed assuming an exponential dependence as in~\cref{eq:exponentialFunctionConstants} and then extrapolated to the higher levels. Using~\cref{eq:optimalDOFs,eq:optimalNumberOfSamples} the optimal number of hierarchy levels together with the optimal number of samples at each level are estimated for different relative tolerances $\varepsilon_r$.

\begin{figure}[H]
	\centering
	\captionsetup[subfigure]{oneside,margin={0cm,0cm},labelformat=empty}
	\subfloat[(a) Mean estimation]{
		\includegraphics[scale=0.5,trim={2.4cm 0.1cm 3.6cm 2.1cm},clip]{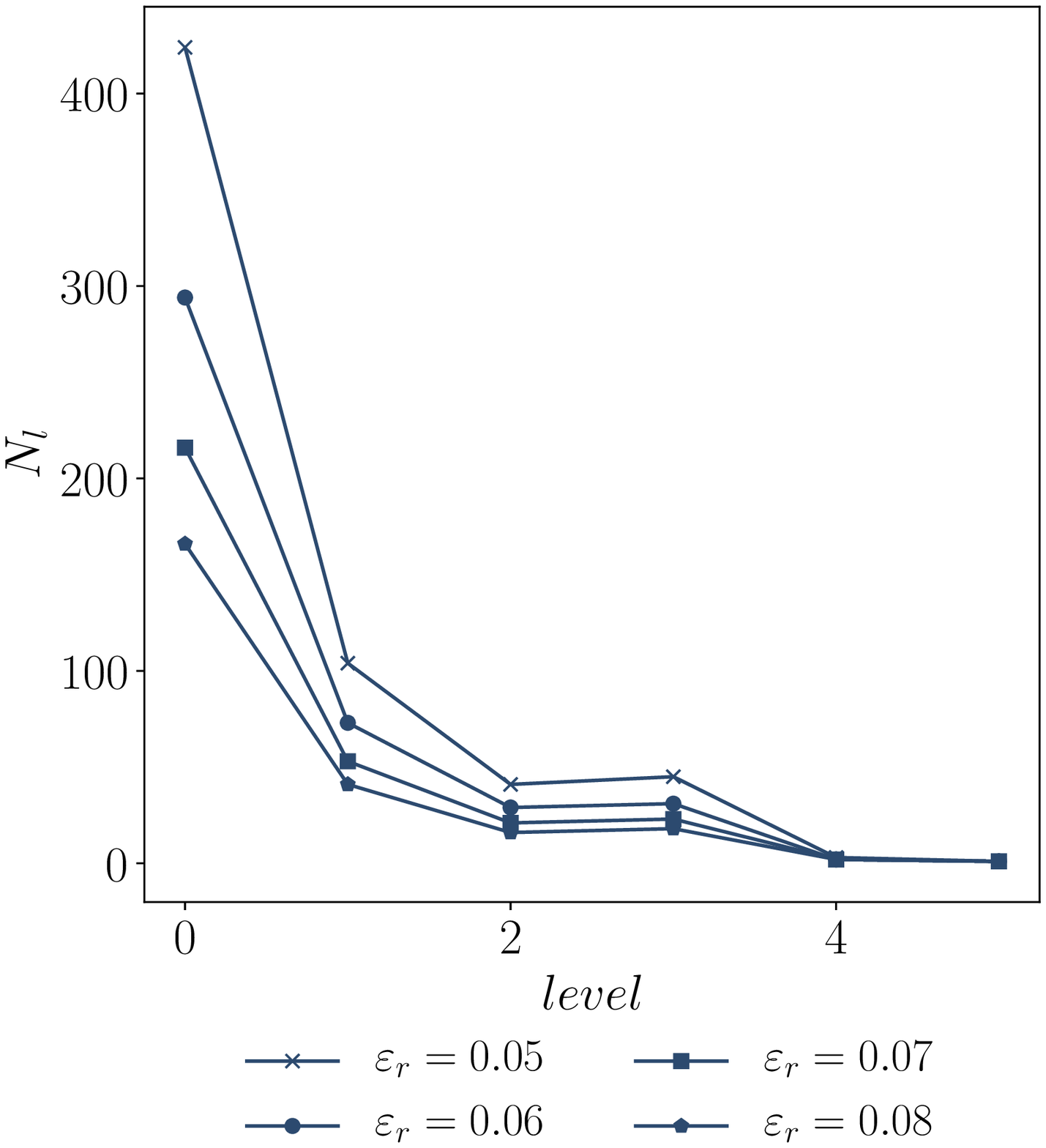} }
	\hspace*{0.2cm}
	\subfloat[(b) Variance estimation]{
		\includegraphics[scale=0.5,trim={2.4cm 0.1cm 3.6cm 2.1cm},clip]{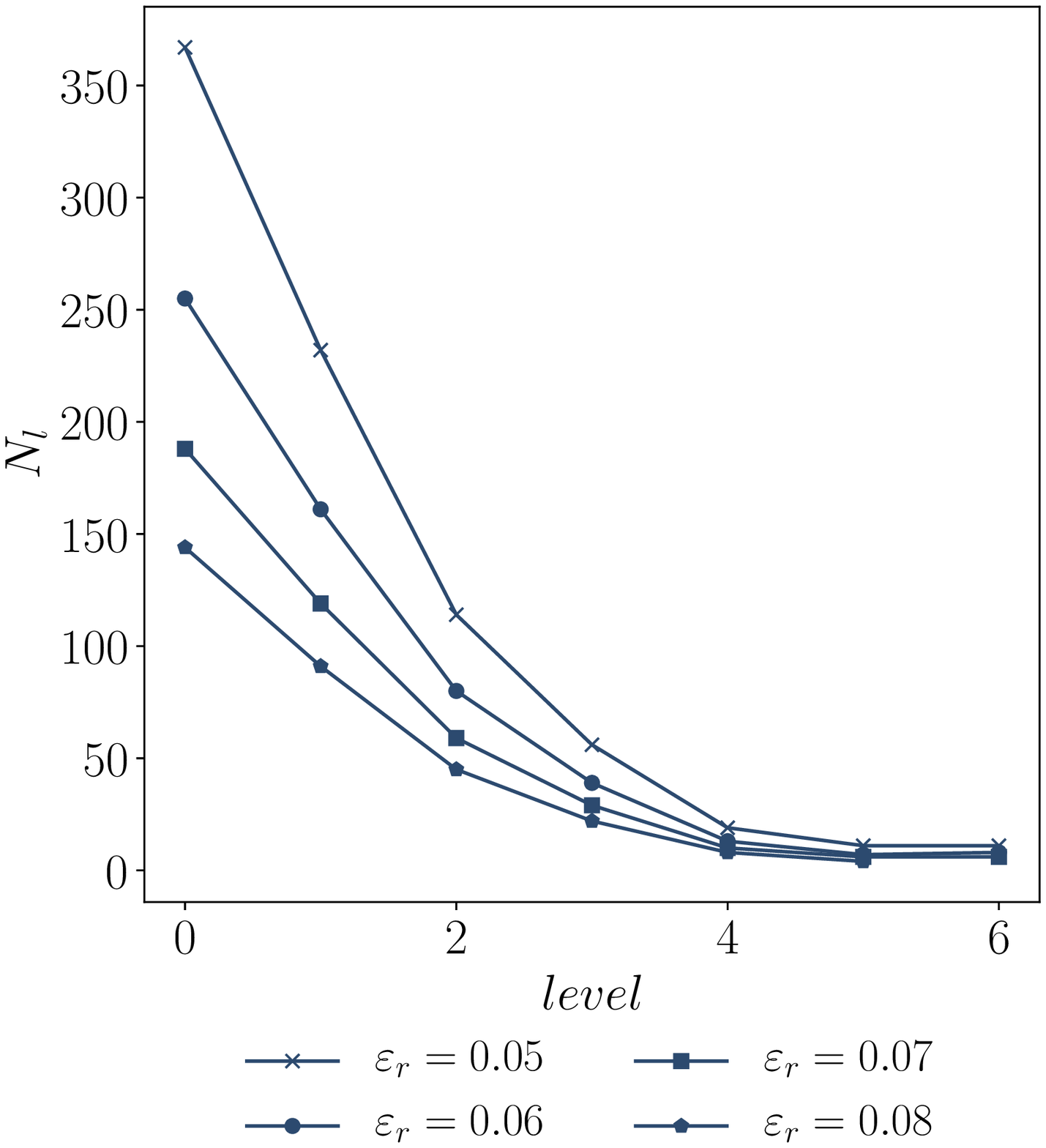} }
	\caption{Results of MLMC screening procedure for the square lattice.}
	\label{fig:screeningMLMC}
\end{figure}

\Cref{fig:screeningMLMC} depicts the number of necessary samples for the mean and variance estimation of the homogenized Young's modulus. The mean value estimation requires a lower value of optimal levels than the variance estimation. The screening procedure shows that seven hierarchy levels with sample numbers larger than $N_l\geq( 367, 232, 114, 56, 19, 11, 11 )$ are required to achieve a relative error of $0.05$ for the estimated variance. The mean value can be evaluated with the same achieved error with six levels and sample numbers larger than $N_l\geq( 424, 104, 41, 45, 3, 1 )$. The final number of samples used to obtain the estimates of the moments is updated to achieve the target accuracy of $\varepsilon_r=0.05$ giving $N_l=( 621, 250, 146, 99,  38,  19,  15 )$. Although the screening procedure is performed for the first two moments, the third and the fourth moments are evaluated with the provided number of samples to get an idea about the final distribution.

The results of the MLMC algorithm are summarized in~\cref{tab:MLMCResultGrid}.

\begin{table}[H]
	\centering
	\begin{tabular}{|c|c|c|c|}
		\hline
		$\mu$, [MPa]  & $\sigma$, [MPa]   & $\gamma$, [-]  & $\kappa$, [-]\\\hline
		$20\,625$ &  $1\,619$ & $-0.20$ & 2.68  \\\hline
	\end{tabular}
	\caption{The estimated moments from the MLMC procedure for the square grid lattice ($\gamma$ is skewness, whereas $\kappa$ is kurtosis).}
	\label{tab:MLMCResultGrid}
\end{table}

In~\cref{fig:resultMLMC2}, a normal fit of the probability density function of the homogenized Young's modulus based on the estimated mean and variance together with the intervals that cover the $90\%$, $95\%$, $98\%$, and $99\%$ of the probability mass. The determined skewness $\gamma$ and kurtosis $\kappa$ as in~\cref{tab:MLMCResultGrid} are close to the one expected for the normal distribution ($0$ and $3$ respectively). However, these estimates have significant uncertainty. Considering the sampling uncertainty related to the moment estimates, the plot shows a conservative visualization of the distribution and intervals. 

The Young's modulus evaluated with the original sample falls into the $99\%$ interval (shown as the dashed line in~\cref{fig:resultMLMC2}). Overall, the performed MLMC procedure seems to capture the observed spread in the experimentally determined homogenized Young's modulus. Due to the additive manufacturing process's underlying complex physics, there are many geometrical and topological variations to be expected for such microscale lattices. Undoubtedly, these defects play a significant role in the final part's mechanical behavior, which is supported by this analysis. 
	 
	 \begin{figure}[H]
	 	\centering
	 	\includegraphics[width=0.9\textwidth, height=0.5\textwidth,trim={0.0cm 0.1cm 0.1cm 0.5cm},clip]{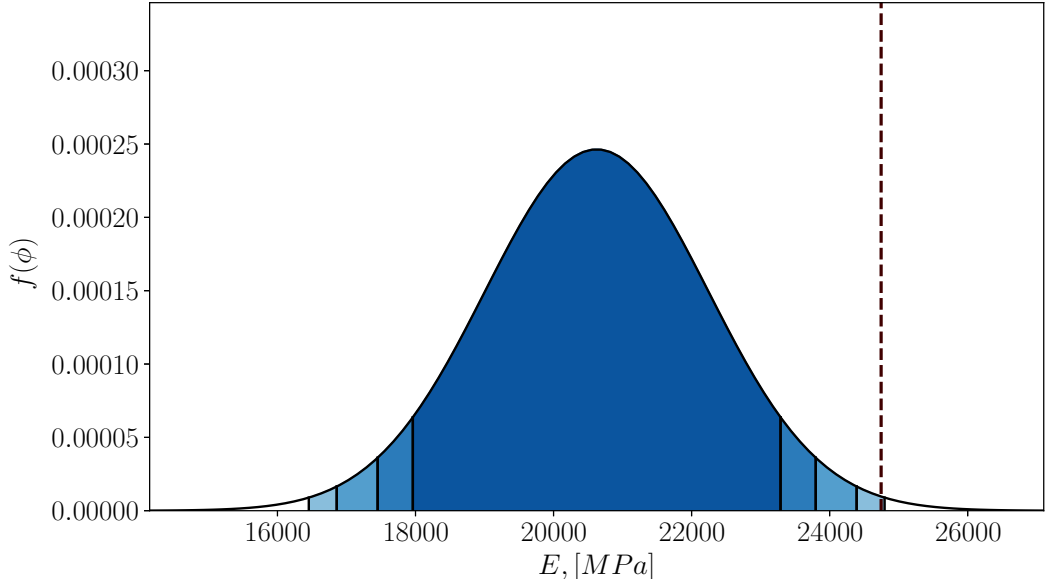}
	 \caption{Results of MLMC on the homogenized Young's modulus for the square lattice. Normal fit based on the estimates of the first two moments. The shaded areas plot the intervals covering the $90\%$, $95\%$, $98\%$, and $99\%$ of the probability mass.}
	 \label{fig:resultMLMC2}
 \end{figure}	
	 
	}

\subsection{Octet-truss lattice}
{
	To gain further confidence in the proposed workflow, we consider an octet-truss lattice structure. The scale of the printed lattice is considerably different from the example above. The overall unit cell size is $4$ mm, while the diagonal struts' size is $0.2$ mm. At this scale, the manufacturing process provides results with a significantly better reproducibility than those of example in~\cref{subsec:SquareLattice}. Due to many process-induced defects the as-manufactured geometry is different from the as-designed one. However, when multiple specimens are printed, similar porosity values are achieved. The CT scan was performed on one of such tensile specimens as shown in~\cref{fig:unitCell3DOctetTruss}~\cite{Korshunova2020a}. The size of the considered example is $292\times176\times1768$ voxels with the spacing of $2.69\mu m$. 
\begin{figure}[H]
	\centering
	\def\svgwidth{0.3\textwidth}
	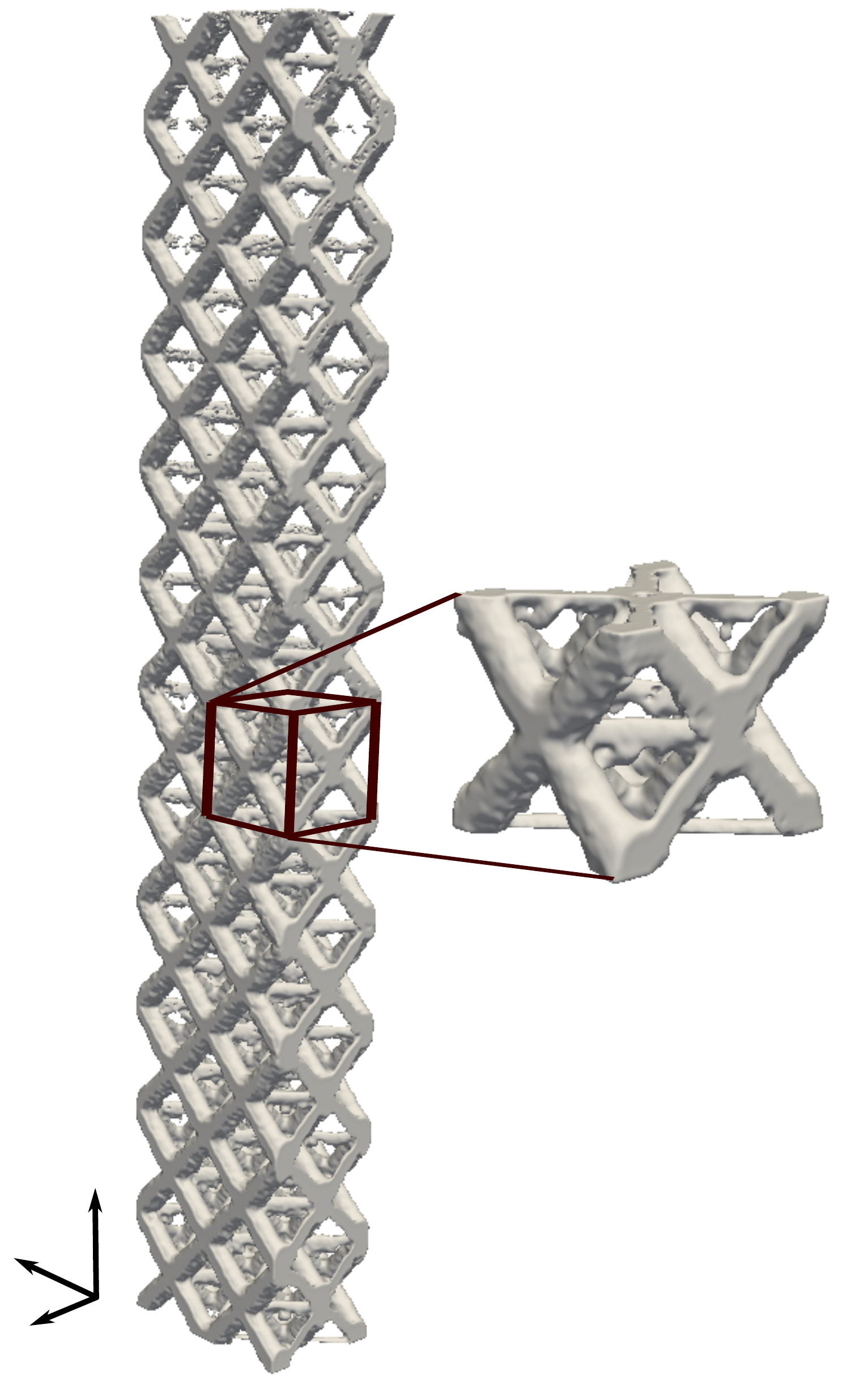
	\caption{An example of a periodic unit cell with its local coordinate system in the 3D octet-truss lattice~\cite{Korshunova2020a}.}    
	\label{fig:unitCell3DOctetTruss}
\end{figure}

\textbf{3D model parameter identification}

First, the local probabilities of the underlying local cell are evaluated. The size of the unit cell is $146\times176\times146 $ voxels as indicated in~\cref{fig:unitCell3DOctetTruss}. Overall, $20$ cells are extracted to evaluate the probabilities at every local coordinate according to to~\cref{eq:meanBinary}. 

Second, the correlation parameters are determined. In this case, convergence was not achieved for any number of considered spatial lags. We observe that with the increase of the number of lags, the smoothness parameter $\nu$ is continuously increasing. Such behavior was observed in all directions. Thus, to gain further insight into the behavior of the residual in~\cref{eq:MinimizationFunction}, we evaluate the residual on the grid of correlation length and smoothness values.  The objective function's behavior in all directions is somewhat similar. \Cref{fig:3DOctetObjectiveFunction} shows an apparent plateau in the smoothness parameter, explaining the lack of convergence. As the residual appears to decrease with increase of the smoothness parameter, we choose to fit the Gaussian model, which is a particular case of Mat\'ern correlation function when $\nu \to \infty$.

\begin{figure}[H]
	\centering
	\resizebox{\textwidth}{!}{\includegraphics[scale=0.3]{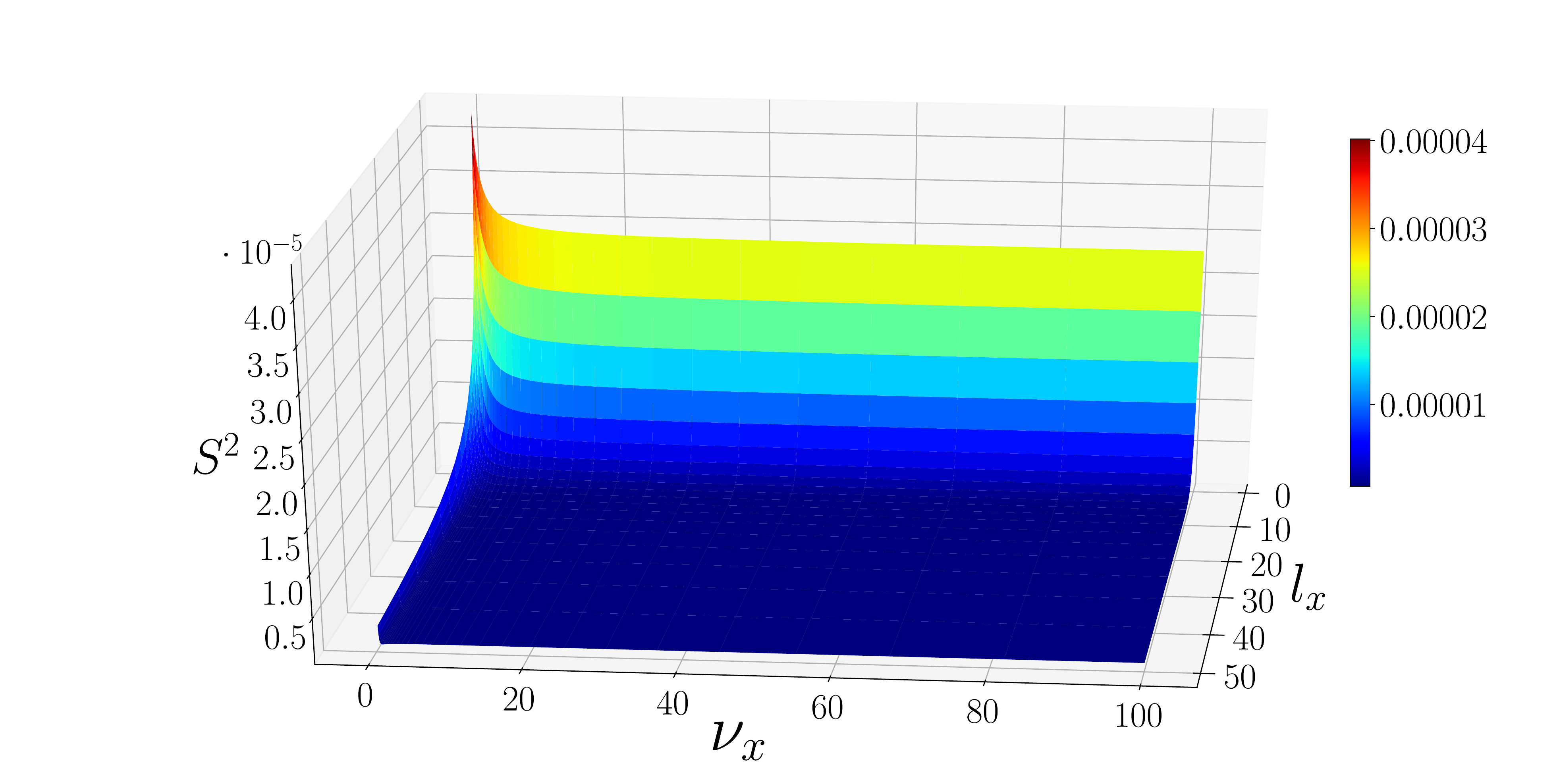}}
	\caption{Objective function for an octet-truss lattice in x-direction with $100$ lags.}    
	\label{fig:3DOctetObjectiveFunction}
\end{figure}

Having fixed the smoothness parameter, we perform the convergence study on the achieved correlation length. \Cref{fig:CorrelationConvergenceOctet} shows that the convergence in $x$ direction can be achieved relatively quickly as the correlation length parameter changes only slightly with the increasing number of lags. However, there is a significant change in other directions. The results converge after $100$ spatial lags.

\newcommand{\graphDir}{./sections/numericalExperiments/Figures3D/Tikz/graphs}
\newcommand{\dataDir}{./sections/numericalExperiments/Figures3D/Tikz/data}
\begin{figure}[H]
	\centering
	\hspace*{-1cm}
	\includegraphics[width=\textwidth,height=0.4\textheight]{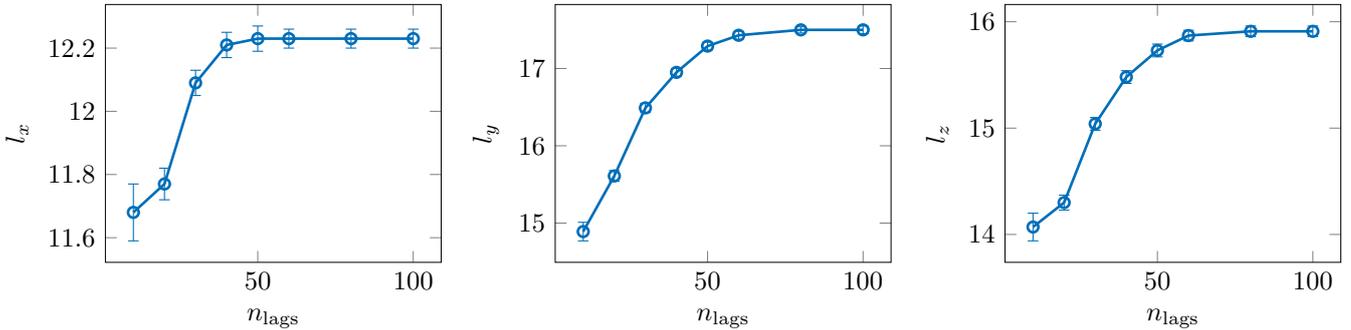}
	\caption{Convergence of the estimated correlation parameters for the octet-truss lattice.}    
	\label{fig:CorrelationConvergenceOctet}     
\end{figure}

The qualitative fitting as shown in~\cref{fig:3DCovarianceFitOctet} shows that the trend is only captured on average. The extreme values of the observed binary correlation are not well contained in the proposed structure. It is important to note, that due to a very large size of the data set, it is hard to visualize it. The computed correlation coefficients for every lag follow a distribution, i.e. large amount of data points are concentrated around a value. The fitting procedure captures this value, or homogenizing the spread of the possible correlation parameters. Quantitatively, the results presented in~\cref{tab:parametersIdentified3DOctet} seem to have a relatively small standard deviation indicating a good fit.

\begin{figure}[H]
	\centering
	\captionsetup[subfigure]{oneside,margin={0cm,0cm},labelformat=empty}
	\hspace*{-10mm}\subfloat[]{
		\includegraphics[scale=0.38]{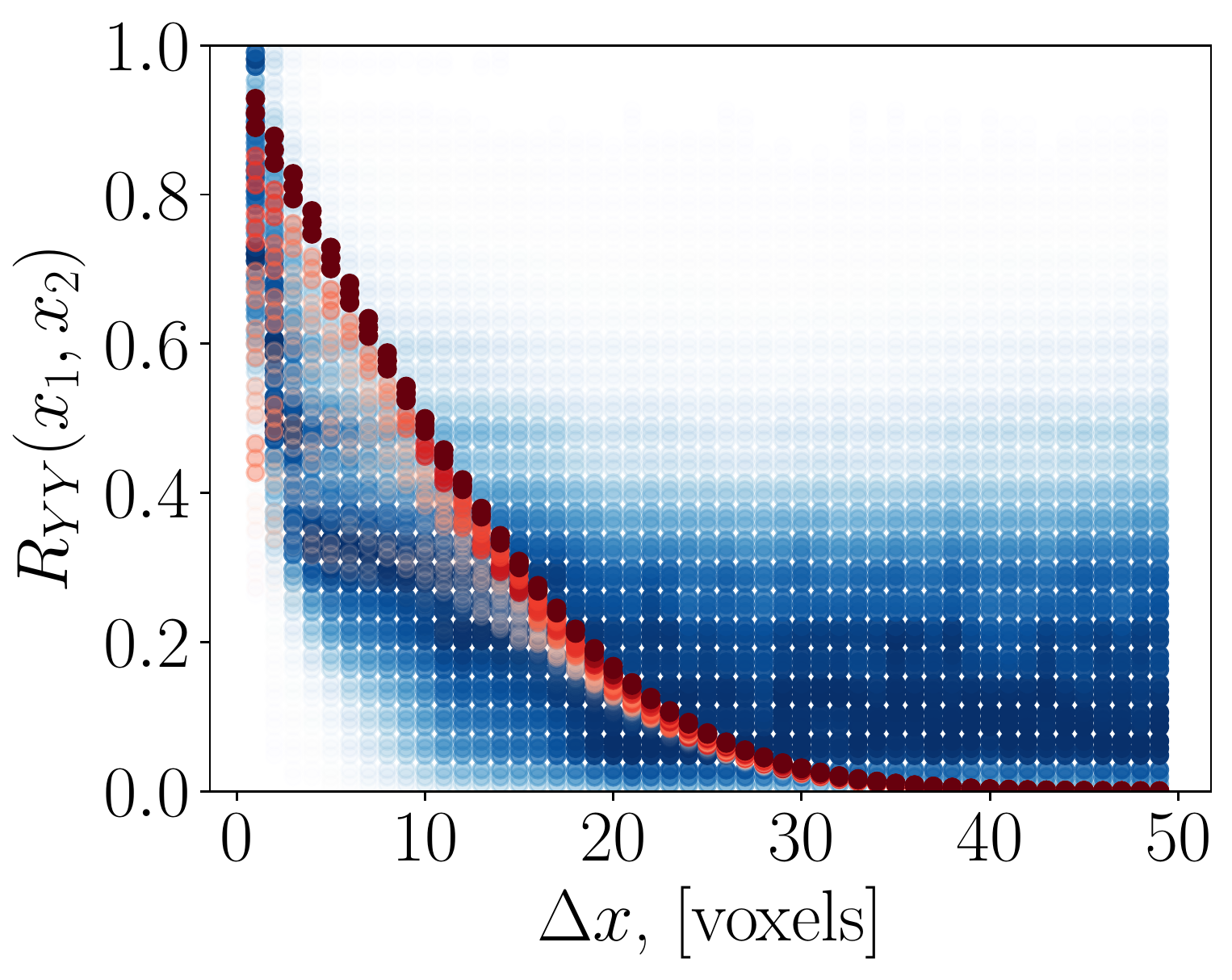} }
	\subfloat[]{
		\includegraphics[scale=0.38]{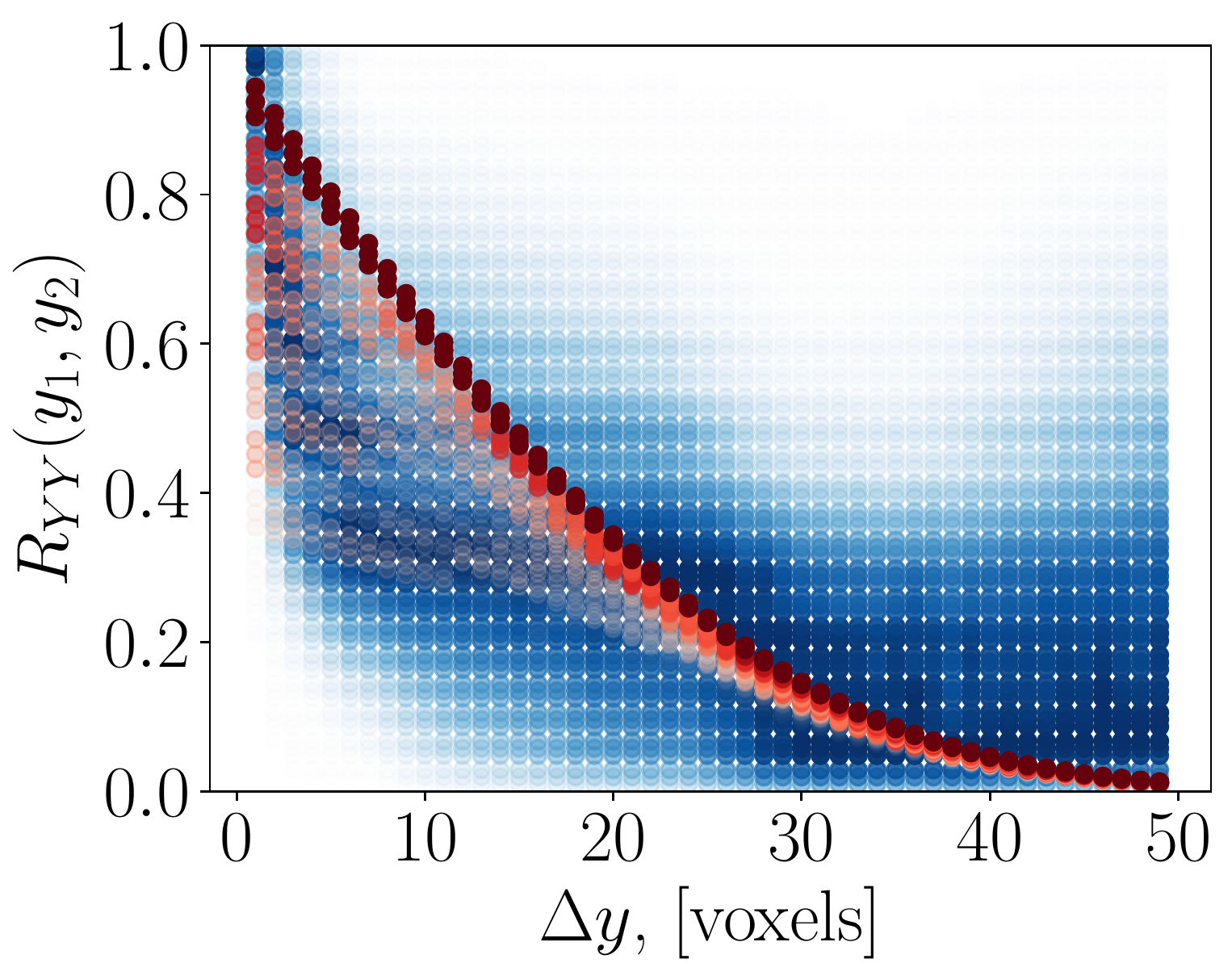} }
	\subfloat[]{
		\includegraphics[scale=0.38]{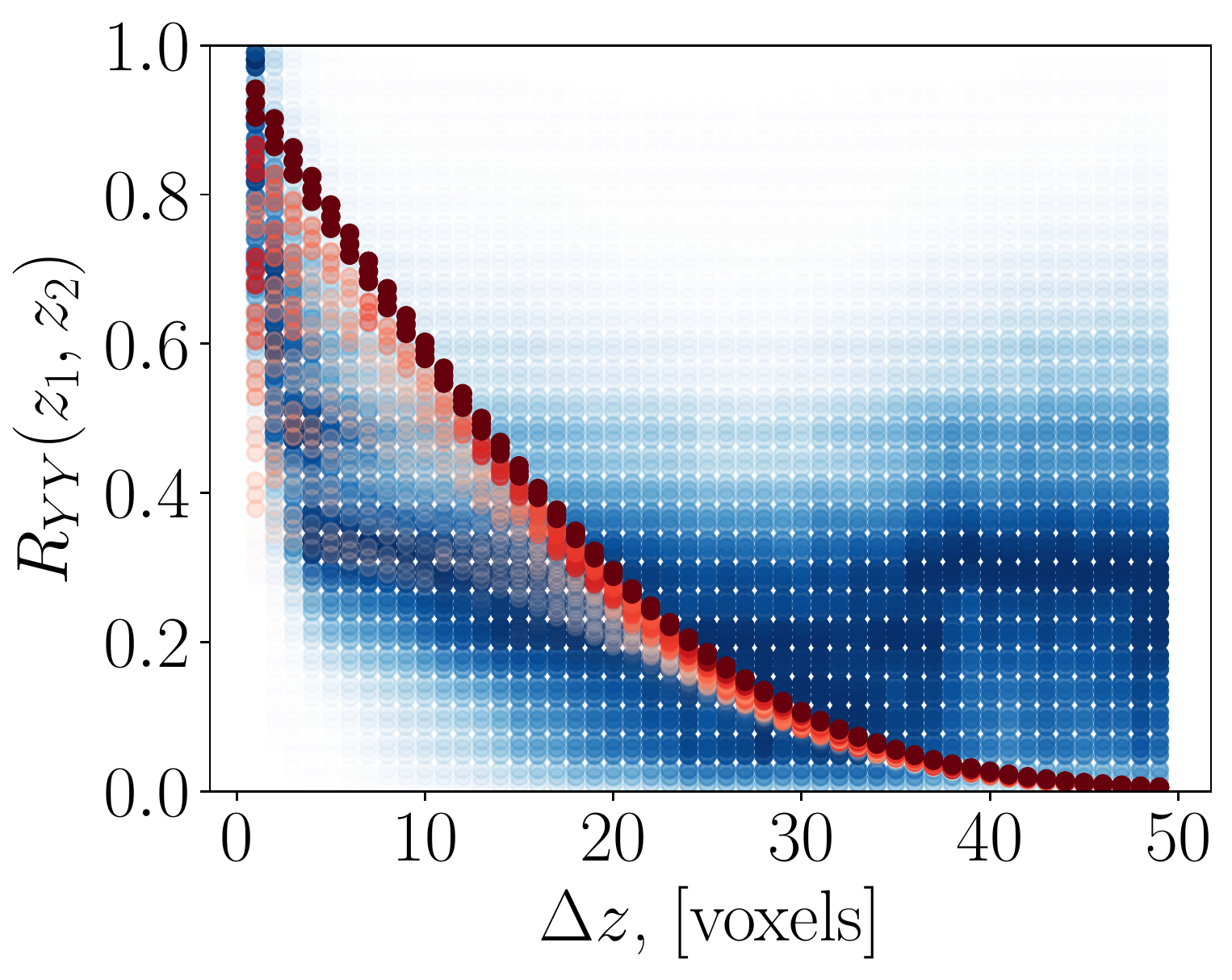}}
	\caption{Fitted and computed auto-covariance of the octet-truss structure ( blue - sample correlation; red - squared exponential model fit).}
	\label{fig:3DCovarianceFitOctet}
\end{figure}

\begin{table}[H]
	\centering
	\begin{tabular}{|c|c|c|c|}
		\hline
		Considered lags &$l_x, [voxels]$&$l_y, [voxels]$&$l_z, [voxels]$ \\\hline
		100 lags& 12.23 $\pm$ 0.03 & 17.50 $\pm$ 0.05 & 15.91 $\pm$ 0.05\\\hline
	\end{tabular}
	\caption{Correlation parameter identification for the octet-truss lattice.}
	\label{tab:parametersIdentified3DOctet}
\end{table}

Based on the fitted parameters, 3D samples of the microstructure are generated according to the procedure described in~\cref{sec:numericalgeneration}. One sample takes about 10 seconds to be generated. A representative realization is compared to the original image in three orthogonal views in~\cref{fig:3DCoronalOctet,fig:3DAxialOctet,fig:3DSagittalOctet}. Visually, the structures are similar to the original image and the periodic structure of the underlying octets is well-captured.

\begin{figure}[H]
	\centering
	\captionsetup[subfigure]{oneside,margin={0cm,0cm},labelformat=empty}
	\subfloat[(a) Original sample]{
		\includegraphics[scale=0.28, angle =90]{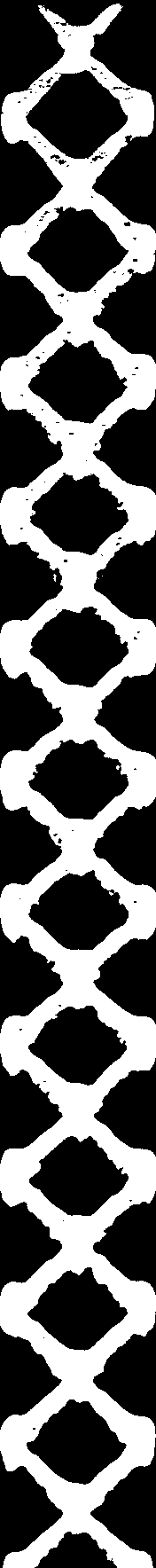} }\\%
	\subfloat[{(b) Realization }]{
		\includegraphics[scale=0.28, angle =90]{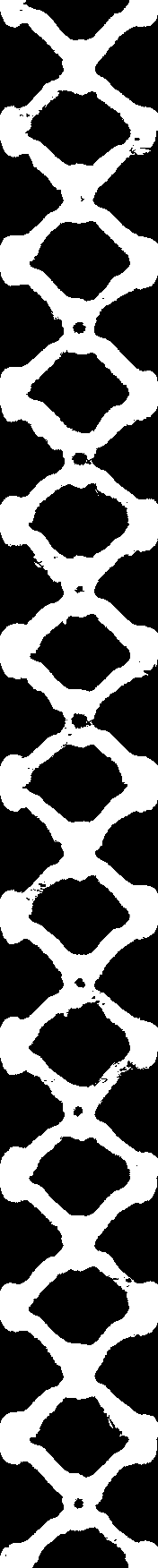} }
	\caption{An example of a realization based on fitted correlation parameters: coronal slice 290 of the octet-truss lattice.}
	\label{fig:3DCoronalOctet}
\end{figure}
\begin{figure}[H]
	\centering
	\captionsetup[subfigure]{oneside,margin={0cm,0cm},labelformat=empty}
	\subfloat[(a) Original sample]{
		\includegraphics[scale=0.28, angle =90]{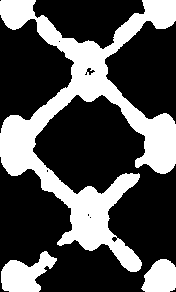} }
	\hspace*{0.2cm}
	\subfloat[{(b) Realization}]{
		\includegraphics[scale=0.28, angle =90]{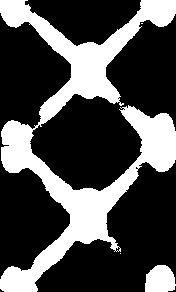} }
	\caption{An example of a realization based on fitted correlation parameters: axial slice 872 of the octet-truss lattice.}
	\label{fig:3DAxialOctet}
\end{figure}
\begin{figure}[H]
	\centering
	\captionsetup[subfigure]{oneside,margin={0cm,0cm},labelformat=empty}
	\subfloat[(a) Original sample]{
		\includegraphics[scale=0.25, angle =90]{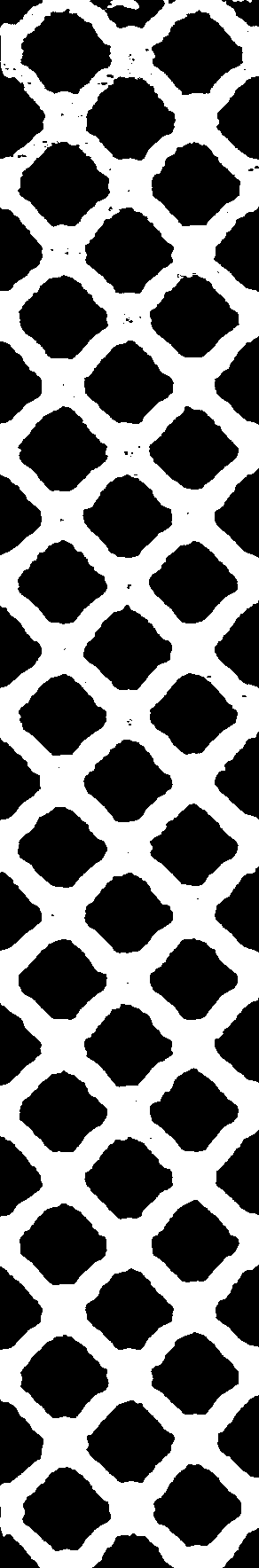} }\\%
	\subfloat[(b) Realization]{
		\includegraphics[scale=0.25, angle =90]{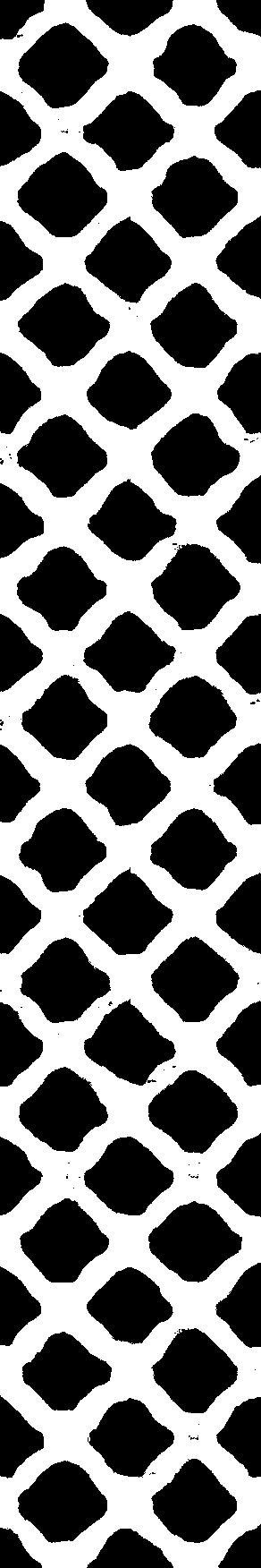} }
	\caption{An example of a realization based on fitted correlation parameters: axial slice 77 of the octet-truss lattice.}
	\label{fig:3DSagittalOctet}
\end{figure}	
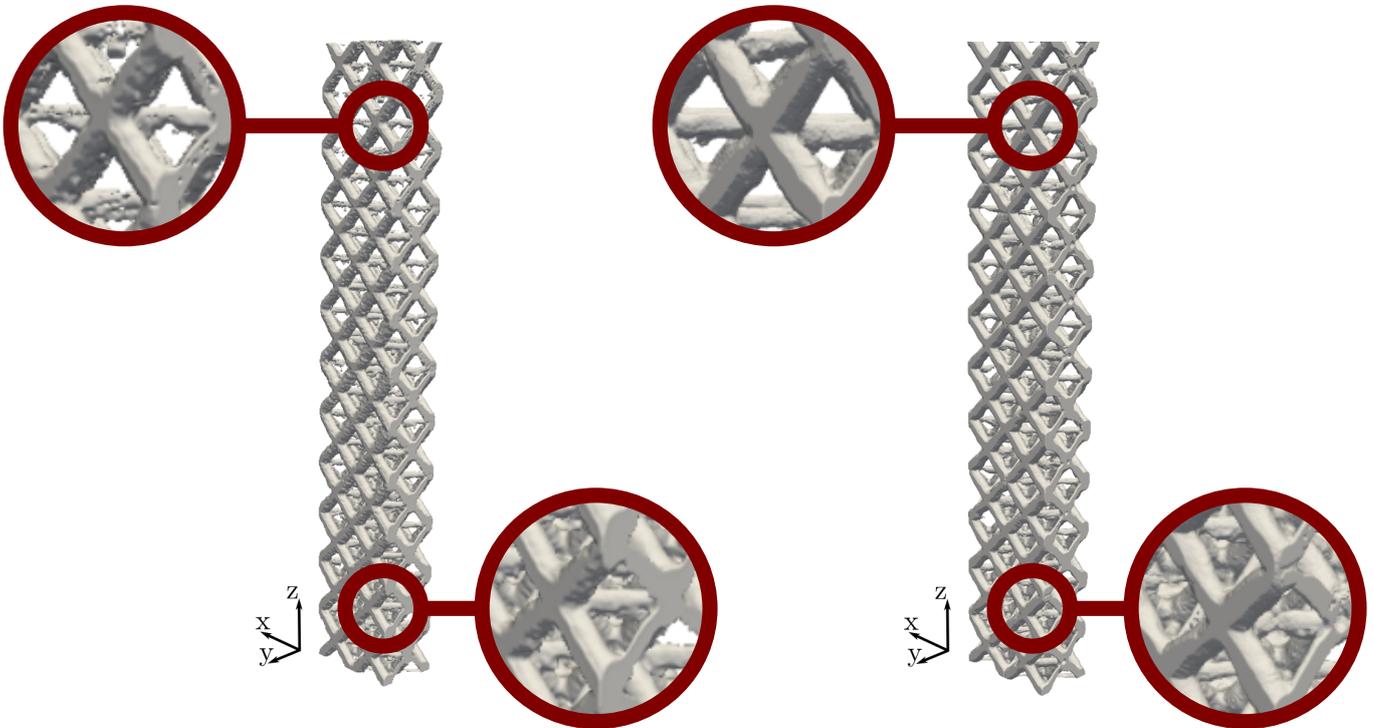
\begin{figure}[H]
\centering
\captionsetup[subfigure]{oneside,margin={0cm,0cm},labelformat=empty}
\hspace*{-1.2cm}\subfloat[(a) Original sample]{
				\begin{tikzpicture}[spy using outlines={circle,red!50!black, magnification=3, connect spies}]
        		\node {\includegraphics[scale=0.35]{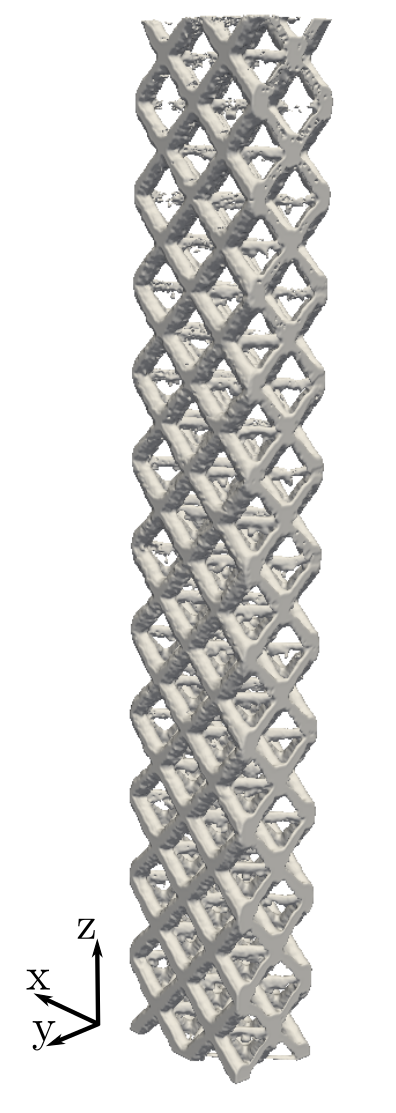} };
        		\coordinate (spypoint1) at (0.3,3.2);
        		\coordinate (magnifyglass1) at (-3.1,3.2);
        		\coordinate (spypoint2) at (0.3,-3.2);
        		\coordinate (magnifyglass2) at (3.1,-3.2);
        		\spy [every spy on node/.append style={line width=2mm},size=3cm,spy connection path={\draw[line width=2mm,red!50!black] (tikzspyonnode) -- (tikzspyinnode);}] on (spypoint1)
        		in node[fill=white,line width=2mm] at (magnifyglass1);
        		\spy [every spy on node/.append style={line width=2mm},size=3cm,spy connection path={\draw[line width=2mm,red!50!black] (tikzspyonnode) -- (tikzspyinnode);}] on (spypoint2)
        		in node[fill=white,line width=2mm] at (magnifyglass2);
        		\end{tikzpicture}
	 }
\hspace*{-1.2cm}
\subfloat[{(b) Realization}]{
				\begin{tikzpicture}[spy using outlines={circle,red!50!black, magnification=3, connect spies}]
        		\node {\includegraphics[scale=0.35]{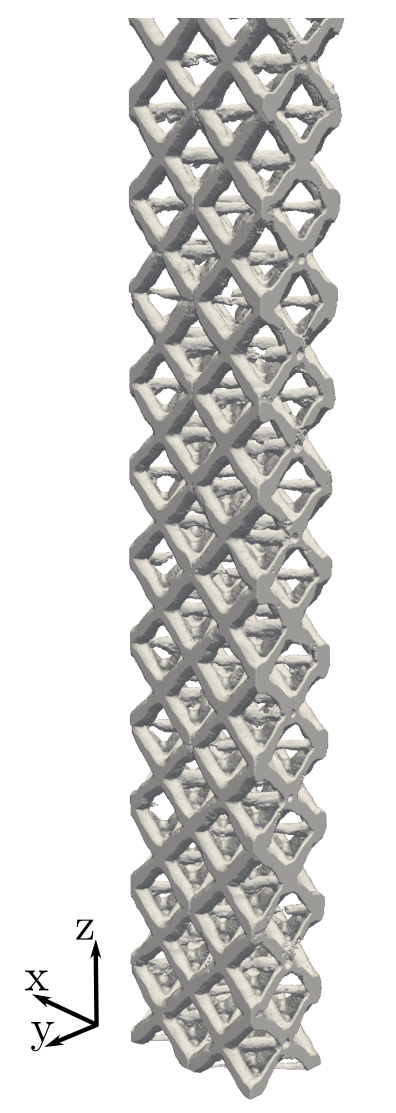}  };
        		\coordinate (spypoint1) at (0.3,3.2);
        		\coordinate (magnifyglass1) at (-3.1,3.2);
        		\coordinate (spypoint2) at (0.3,-3.2);
        		\coordinate (magnifyglass2) at (3.1,-3.2);
        		\spy [every spy on node/.append style={line width=2mm},size=3cm,spy connection path={\draw[line width=2mm,red!50!black] (tikzspyonnode) -- (tikzspyinnode);}] on (spypoint1)
        		in node[fill=white,line width=2mm] at (magnifyglass1);
        		\spy [every spy on node/.append style={line width=2mm},size=3cm,spy connection path={\draw[line width=2mm,red!50!black] (tikzspyonnode) -- (tikzspyinnode);}] on (spypoint2)
        		in node[fill=white,line width=2mm] at (magnifyglass2);
        		\end{tikzpicture}
	}
\caption{An example of a realization based on fitted correlation parameters: full 3D model of the octet-truss lattice.}
\label{fig:3DFullOctet}
\end{figure}

\textbf{Multilevel Monte Carlo analysis of the homogenized mechanical behavior}

The original octet-truss lattice is analyzed both experimentally and numerically. The results can be found in greater detail in~\cite{Korshunova2020a}. 

In this case, only one tensile experiment was performed on the specimen. 
The numerical and experimental homogenized Young's modulus is shown in~\cref{tab::ExperimentalResultsOctet}. The standard deviation is determined solely based on the instrumentation error.

\begin{table}[H]
	\centering
	\begin{tabular}{|c|c|c|c|}
		\hline
		Specimen &  Experimental $E$, [MPa]&CT-based $E$, [MPa]\\\hline
		Tensile octet-truss & 12 533 $\pm$ 751  & 13 081  \\\hline
	\end{tabular}
	\caption{Comparison of experimentally and numerically determined Young's modulus of the octet-truss specimen~\cite{Korshunova2020a}.}
	\label{tab::ExperimentalResultsOctet}
\end{table}

The numerical analysis of the CT scan of the octet-truss lattice was performed with fewer hierarchy levels than the example above. The coarsest discretization level $L=0$ consists of $73\times44\times442$ finite cells with the polynomial degree $p=1$. At the level $L=1$ $h-$refinement is performed to obtain $146\times88\times884$ finite cells with the polynomial degree $p=1$. With every higher level, $L>2$, the finite cells' polynomial degree is raised from $p=1$ to $p=3$. The results achieved based on the analysis of the original CT scan are presented in~\cref{fig::ConvergenceDNSOctet}.

\renewcommand{\graphDir}{./sections/numericalExperiments/TensileGrid/graphs}
\renewcommand{\dataDir}{./sections/numericalExperiments/TensileGrid/data}

\begin{figure}[H]
	\centering
	\vspace*{-0.01cm}
	\includegraphics[height=0.37\textheight,width=0.70\textwidth]{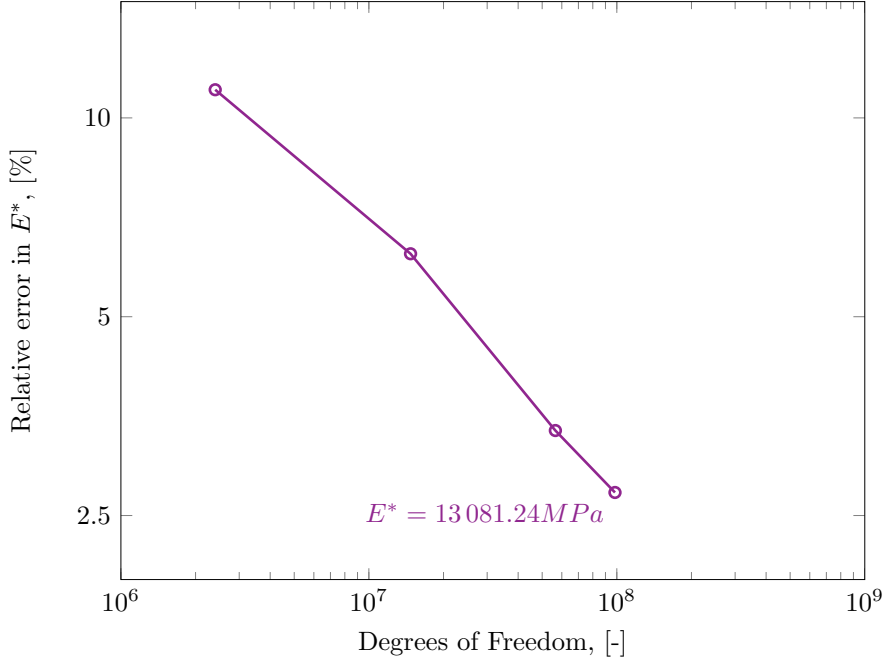}
	\caption{Octet-truss lattice: Convergence of the directional Young's modulus $E^*$~\cite{Korshunova2020a}.}    
	\label{fig::ConvergenceDNSOctet}
\end{figure}

To evaluate the number of necessary samples for the MLMC, the screening was performed (see~\cref{fig:screeningMLMCOctet}). The procedure showed that four hierarchy levels with the sample numbers larger than $N_l\geq$  $( 5413, 298,$ $145, 27 )$ are required to achieve a relative error of $2.5\%$ for the estimated variance. The mean value can be evaluated with the same relative error with four levels and sample numbers larger than $N_l\geq(70, 3, 1, 1 )$. The number of samples for the mean estimation is much smaller as a fast convergence for provided discretizaiton is expected. Again, we combine both estimates and update the number of samples during the computation procedure to achieve a relative accuracy of $\varepsilon_r=0.025$ for both the mean and the variance, which results in the number of samples $N_l=(5479, 396, 166, 43)$ meaning $5479$ simulations with the coarsest and $43$ simulations with the finest discretization. To get a better understanding of the final distribution, the third and the fourth moments are estimated without evaluating the optimal sample number for these central moments.

\begin{figure}[H]
	\centering
	\captionsetup[subfigure]{oneside,margin={0cm,0cm},labelformat=empty}
	\subfloat[(a) Mean estimation]{
		\includegraphics[scale=0.5,trim={2.4cm 0.1cm 3.6cm 2.1cm},clip]{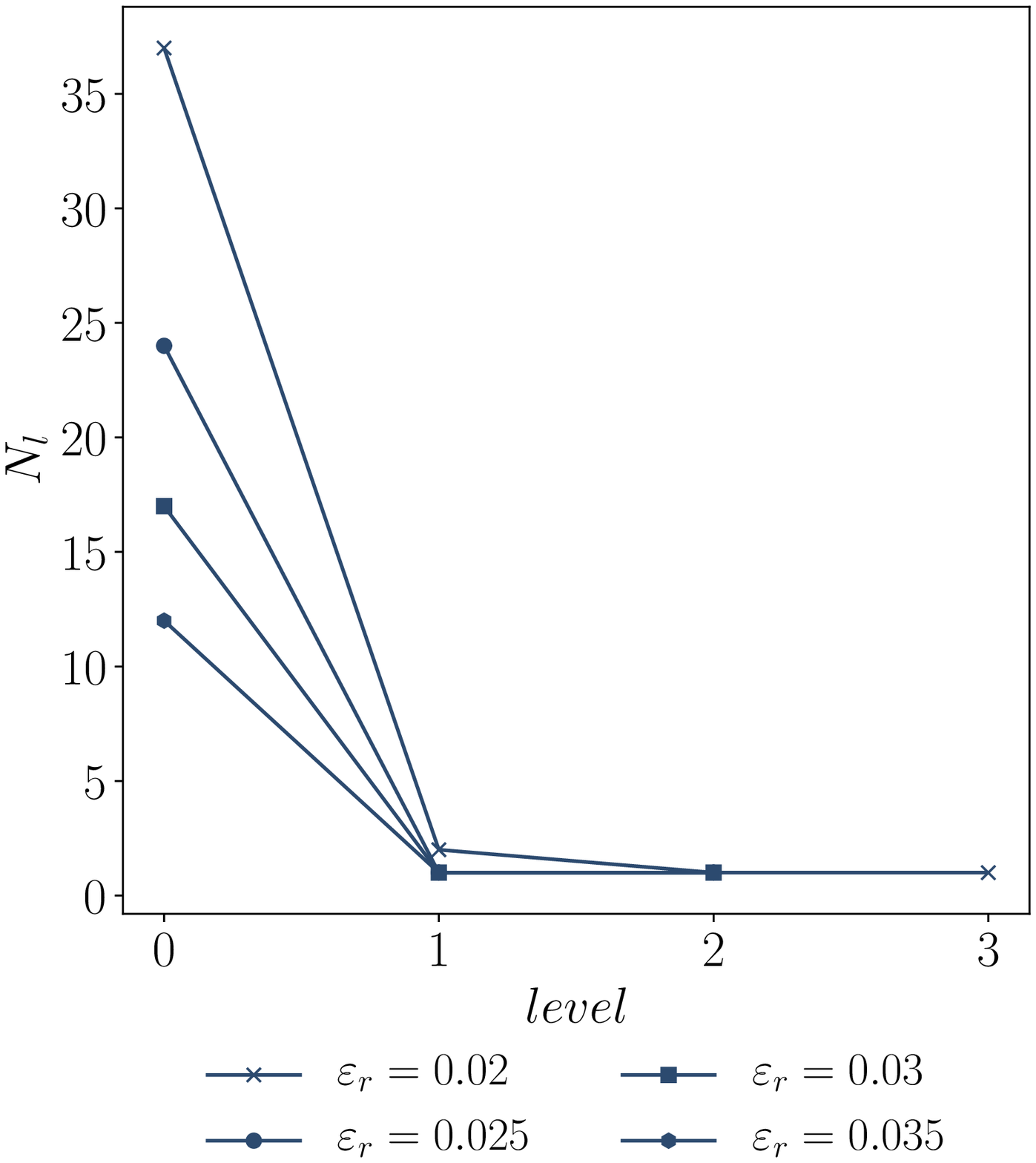} }
	\hspace*{0.2cm}
	\subfloat[(b) Variance estimation]{
		\includegraphics[scale=0.5,trim={2.4cm 0.1cm 3.6cm 2.1cm},clip]{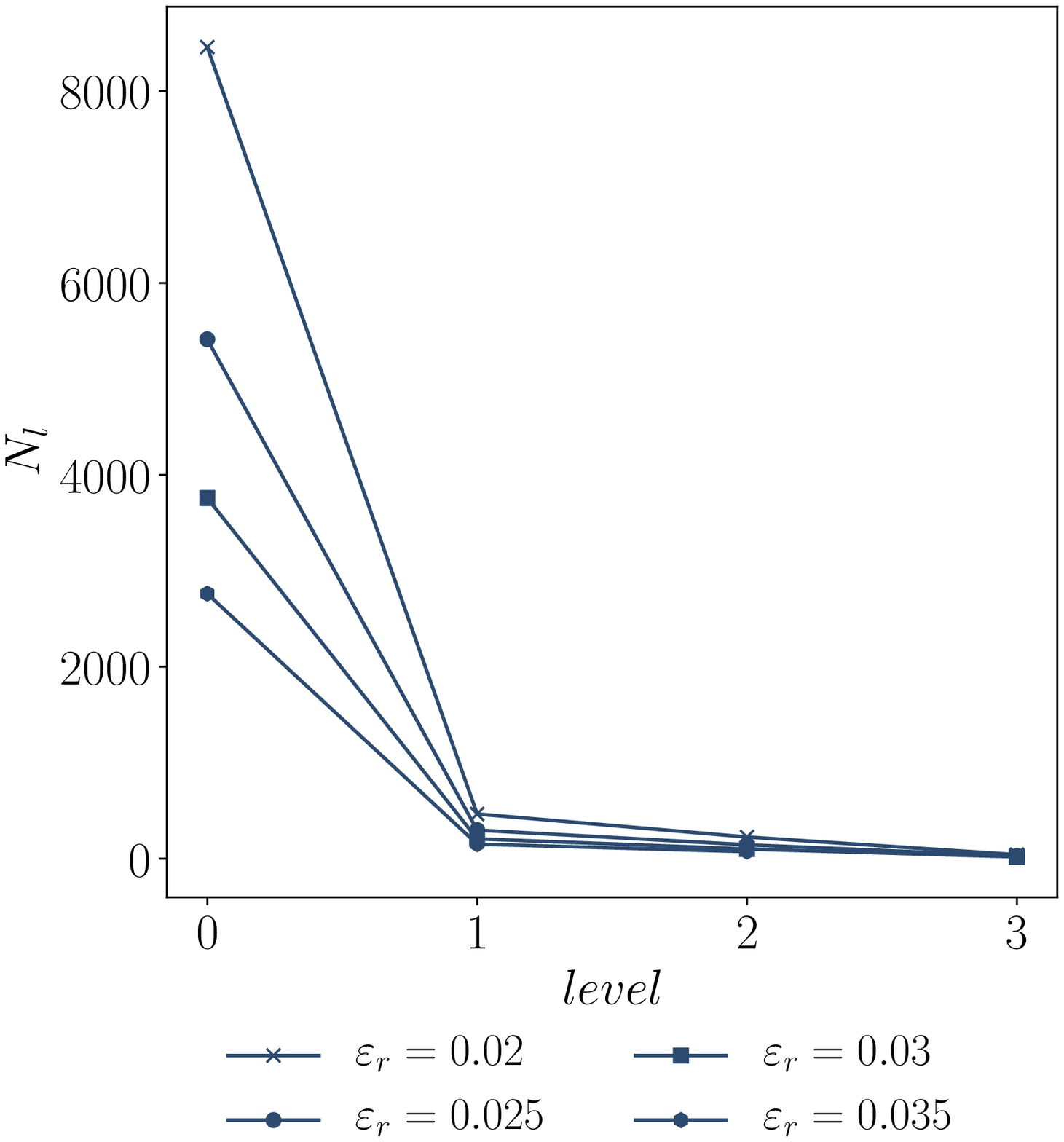} }
	\caption{Results of MLMC screening procedure for the octet-truss lattice.}
	\label{fig:screeningMLMCOctet}
\end{figure}

\Cref{tab:MLMCResultOctetTruss} summarized the achieved results. To get a visual overview of the estimated distribution, we approximate it with the normal distribution based on the estimated first and second moments. The estimated skewness and kurtosis as indicated in~\cref{tab:MLMCResultOctetTruss} are again close to the values for the normal distribution.

\begin{table}[H]
	\centering
	\begin{tabular}{|c|c|c|c|}
		\hline
		$\mu$, [MPa] & $\sigma$, [MPa]  & $\gamma$,[-] & $\kappa$,[-] \\\hline
		$12\,807$ &  $164$   &  $-0.16$  &  $3.66$ \\\hline
	\end{tabular}
	\caption{The estimated moments from the MLMC procedure for the octet-truss lattice ($\gamma$ indicates skewness, whereas $\kappa$ kurtosis).}
	\label{tab:MLMCResultOctetTruss}
\end{table}

\begin{figure}[H]
	\centering
	\includegraphics[width=0.9\textwidth, height=0.5\textwidth,trim={0.0cm 0.1cm 0.1cm 0.5cm},clip]{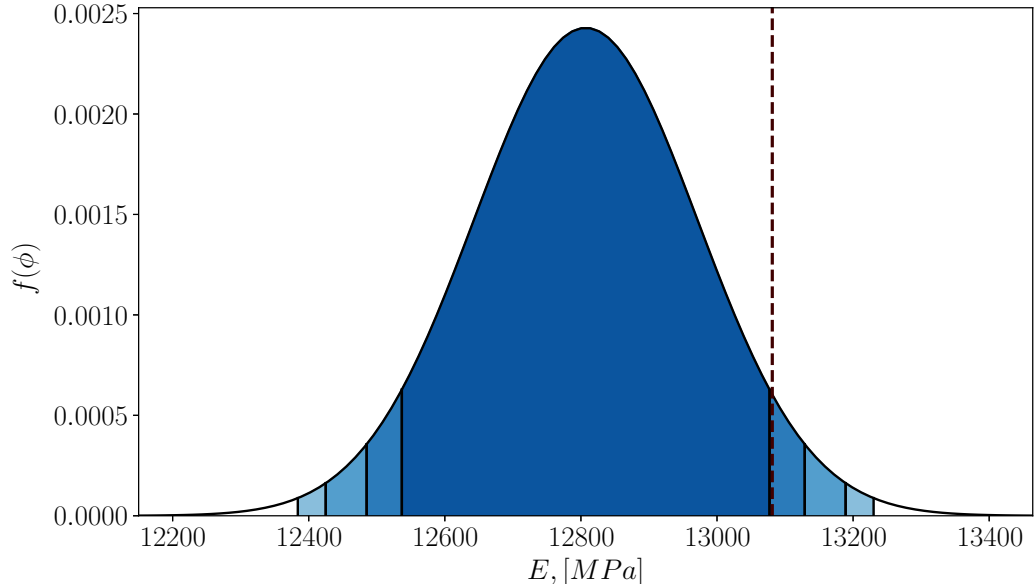}
	\caption{Results of MLMC on the homogenized Young's modulus for the octet-truss lattice. Normal fit based on the estimates of the first two moments. The shaded areas plot the intervals covering the $90\%$, $95\%$, $98\%$, and $99\%$ of the probability mass.}
	\label{fig:resultMLMCOctet}
\end{figure}	

For this example, the Young's modulus evaluated numerically with the original specimen is at the border of the estimated $90\%$ interval. The estimated mean value agrees well with the experimentally obtained values. However, the standard deviation appears to be relatively small $\sigma=164$. This phenomenon points once more to the nature of the additive manufacturing process. As indicated, the octet-truss lattice is produced at a much larger scale than the square one. Thus, the as-manufactured geometries are expected to exhibit smaller relative variability than for teh square grid lattice structure. This behavior is mirrored in the results of the MLMC procedure. We observe a much smaller variability in the final homogenized Young's modulus due to the present geometrical variations. The considered structure is much smoother with a smaller number of geometrical and topological defects.

	}

}

%% file: sections/numericalExperiments/Figures3D/3DOctetWithUnitCell.pdf_tex
\begingroup%
  \makeatletter%
  \providecommand\color[2][]{%
    \errmessage{(Inkscape) Color is used for the text in Inkscape, but the package 'color.sty' is not loaded}%
    \renewcommand\color[2][]{}%
  }%
  \providecommand\transparent[1]{%
    \errmessage{(Inkscape) Transparency is used (non-zero) for the text in Inkscape, but the package 'transparent.sty' is not loaded}%
    \renewcommand\transparent[1]{}%
  }%
  \providecommand\rotatebox[2]{#2}%
  \newcommand*\fsize{\dimexpr\f@size pt\relax}%
  \newcommand*\lineheight[1]{\fontsize{\fsize}{#1\fsize}\selectfont}%
  \ifx\svgwidth\undefined%
    \setlength{\unitlength}{441.09754944bp}%
    \ifx\svgscale\undefined%
      \relax%
    \else%
      \setlength{\unitlength}{\unitlength * \real{\svgscale}}%
    \fi%
  \else%
    \setlength{\unitlength}{\svgwidth}%
  \fi%
  \global\let\svgwidth\undefined%
  \global\let\svgscale\undefined%
  \makeatother%
  \begin{picture}(1,1.60025718)%
    \lineheight{1}%
    \setlength\tabcolsep{0pt}%
    \put(0,0){\includegraphics[width=\unitlength,page=1]{3DOctetWithUnitCell.pdf}}%
    \put(0.00997828,0.157095){\color[rgb]{0,0,0}\makebox(0,0)[lt]{\lineheight{1.25}\smash{\begin{tabular}[t]{l}x\end{tabular}}}}%
    \put(-0.00202576,0.06926962){\color[rgb]{0,0,0}\makebox(0,0)[lt]{\lineheight{1.25}\smash{\begin{tabular}[t]{l}y\end{tabular}}}}%
    \put(0.07838591,0.23144124){\color[rgb]{0,0,0}\makebox(0,0)[lt]{\lineheight{1.25}\smash{\begin{tabular}[t]{l}z\end{tabular}}}}%
  \end{picture}%
\endgroup%

%% file: sections/conclusion/conclusion.tex
\section{Concluding remarks}
\label{sec:conclusions}
{ 
	
	Process-induced defects in additive manufacturing can have a significant impact on the mechanical behavior of final parts. Therefore, the evaluation of the influence of the microstructural variability is crucial for the quality assessment of AM products. In this contribution, we propose a binary random field model to generate statistically equivalent CT images of the as-manufactured lattices. The necessary design parameters can be entirely deduced from a single CT image of the original part. In this way, it includes all occurring geometrical and topological process-induced defects up to the acquired scan resolution. Furthermore, the presented approach removes the limitation of existing approaches that model a specific set of geometric deviations, thus providing a flexible tool to account for multiple geometrical features at once.

Additionally, we demonstrate a technique for efficient evaluation of the influence of the microstructural variability on the homogenized mechanical behavior of the lattice structures, which combines the Finite Cell Method with the multilevel Monte Carlo method. In particular, the immersed Finite Cell Method enables to directly process the CT images generated by the presented random field model in a natural and efficient image-to-material-characterization workflow. The square and octet-truss lattice structure analyses have shown a good agreement with the experimental and numerical results. The inherent process effects at two different manufacturing scales appear to be reflected in the estimated variations of the homogenized Young's Modulus. The production at smaller scales causes more geometrical and topological variability, leading to a larger spread in the final mechanical values.

From the presented results, one can conclude that the CT-based binary random field model in combination with the immersed Finite Cell Method is able to efficiently characterize the influence of the process-induced defects on the homogenized behavior of metal lattices. Moreover, it allows gaining invaluable insight into the geometrical and topological variations as well as to estimate the quality of the produced parts. Although this study has gone some way towards enhancing the possibilities of numerical analysis in the field of AM product simulations, it also opens possibilities for future research. The proposed model assumes the Mat\'ern correlation model for the underlying Gaussian random field. A potential future direction would be to investigate alternative correlation models, e.g., models that introduce a periodic correlation structure. Additionally, a Bayesian approach could be applied for estimating the parameters of the correlation model. Thereby, a challenge that would need to be addressed is the increased computational cost for parameter fitting. The current random field model enables modeling the intra-specimen variability in the construction of AM products. An important future step would be to incorporate multiple CT images of similar parts produced with different process parameters into the proposed model. This would enable modeling additional sources of uncertainty that introduce inter-specimen variability.
	
}

%% file: sections/appendix.tex
\section{Covariance function of the binary random field}
\label{sec:appendix}
{

In this Appendix, we derive the expression for the auto-covariance function of the binary random field $Y(\bm{x})$ defined in \cref{eq:binaryDefinition}. The second-order PMF of $Y(\bm{x})$ is given as (cf. \cite{Vanmarcke2007})
\begin{equation}
p_{YY} (y_1,\bm{x}_1;y_2,\bm{x}_2) = 
\begin{cases} F_{UU}\left(d(\bm{x}_1),\bm{x}_1;d(\bm{x}_2),\bm{x}_2 \right)
& \text{ for } y_1=y_2 = 0\\ \Phi(d(\bm{x}_2)) -
F_{UU}\left(d(\bm{x}_1),\bm{x}_1;d(\bm{x}_2),\bm{x}_2 \right) & \text{ for } y_1=1,y_2 = 0 \\
\Phi(d(\bm{x}_1)) -
F_{UU}\left(d(\bm{x}_1),\bm{x}_1;d(\bm{x}_2),\bm{x}_2 \right) & \text{ for } y_1=0,y_2 = 1 \\
1-\Phi(d(\bm{x}_2))-\Phi(d(\bm{x}_1)) +
F_{UU}\left(d(\bm{x}_1),\bm{x}_1;d(\bm{x}_2),\bm{x}_2 \right) & \text{ for } y_1=y_2 = 1 
\label{eq:jointPMF}
\end{cases}
\end{equation}
where $F_{UU}(u_1,\bm{x}_1;u_2,\bm{x}_2)=\Phi_2 \left( u_1,u_2,\rho_{UU}(\bm{x}_1,\bm{x}_2) \right)$ is the second-order CDF of $U(\bm{x})$ and $\Phi_2(\cdot,\cdot,r)$ is the joint CDF of the bivariate standard normal distribution with correlation parameter $r$.
Hence, the mean-of-product function $\mathrm{E}[Y(\bm{x}_1)Y(\bm{x}_2)]$ is given as:
\begin{equation}
\begin{aligned}
\mathrm{E}[Y(\bm{x}_1)Y(\bm{x}_2)] &= \sum_{i=1}^2\sum_{j=1}^2 y_1 y_2 p_{YY} (y_1,\bm{x}_1;y_2,\bm{x}_2) \\ &=1-\Phi(d(\bm{x}_2))-\Phi(d(\bm{x}_1)) +
\Phi_2 \left( d(\bm{x}_1),d(\bm{x}_2),\rho_{UU}(\bm{x}_1,\bm{x}_2) \right)
\end{aligned}
\label{eq:meanOfProduct}
\end{equation}
For the auto-covariance function of $Y(\bm{x})$, $\Gamma_{YY}(\bm{x}_1, \bm{x}_2)=\mathrm{Cov}(Y(\bm{x}_1),Y(\bm{x}_1))$, it holds:
\begin{equation}
\Gamma_{YY}(\bm{x}_1, \bm{x}_2)=\mathrm{E}[Y(\bm{x}_1)Y(\bm{x}_2)] - \mu_{Y}(\bm{x}_1) \mu_{Y}(\bm{x}_2)
\label{eq:covarianceBinary}
\end{equation}
Plugging in~\cref{eq:meanOfProduct} and~\cref{eq:meanDefinition} into~\cref{eq:covarianceBinary} the following result is obtained:
\begin{equation}
\Gamma_{YY}(\bm{x}_1, \bm{x}_2)=\Phi_2 \left( d(\bm{x}_1),d(\bm{x}_2),\rho_{UU}(\bm{x}_1,\bm{x}_2) \right) - \Phi(d(\bm{x}_1)) \Phi\left(d(\bm{x}_2)\right)
\label{eq:covarianceBinarySimplified}
\end{equation}
The bivariate joint CDF can be expressed in terms of a single-fold integral, as follows \cite{Owen1956}:
\begin{equation}
\Phi_2 \left( u_1,u_2,r \right)
=\int_{0}^{r}\frac{1}{2\pi \sqrt{1-z^2}}\exp\left[-\frac{u_1^2 + u_2^2 - 2u_1u_2z}{2(1-z^2)}\right]dz + \Phi(u_1) \Phi(u_1)
\label{eq:bivariateJointPDF}
\end{equation}
Then, combining~\cref{eq:bivariateJointPDF} with~\cref{eq:covarianceBinarySimplified} leads to the final equation for the covariance of the binary random field:
\begin{equation}
\Gamma_{YY}(\bm{x}_1, \bm{x}_2)=\int_{0}^{\rho_{UU}(\bm{x}_1,\bm{x}_2)}\frac{1}{2\pi \sqrt{1-z^2}}\exp\left[-\frac{d(\bm{x}_1)^2 + d(\bm{x}_2)^2 - 2d(\bm{x}_1)d(\bm{x}_2)z}{2(1-z^2)}\right]dz
\label{eq:covarianceBinaryFinal}
\end{equation}
}

%% file: template/acknowledgements.tex
We gratefully acknowledge the support of Deutsche Forschungsgemeinschaft (DFG) through the project 414265976 – TRR 277 C-01 and TUM International Graduate School of Science and Engineering (IGSSE), GSC 81. The authors also gratefully acknowledge the Gauss Centre for Supercomputing e.V. (www.gauss-centre.eu) for funding this project by providing computing time on the Linux Cluster CoolMUC-2 and on the GCS Supercomputer SuperMUC-NG at Leibniz Supercomputing Centre (www.lrz.de). We gratefully acknowledge the support of Siemens AG, in particular, Dr. Daniel Reznik, sponsoring this research. We also kindly acknowledge the Department of Civil Engineering and Architecture of the University of Pavia, in particular Gianluca Alaimo and Massimo Carraturo, for providing facilities for additive manufacturing and holding the experimental testing on the octet-truss lattices (http://www-4.unipv.it/3d/laboratories/3dmetalunipv/). Finally, the authors gratefully acknowledge Giorgio Vattasso from LaborMet Due (http://www.labormetdue.it/) for his technical support in obtaining CT scan images.